\documentclass[aos,preprint]{imsart}
\setattribute{journal}{name}{}

\RequirePackage[round]{natbib}
\RequirePackage[colorlinks,citecolor=blue,urlcolor=blue, linkcolor=blue]{hyperref}

\usepackage{bbm}
\usepackage{comment}
\usepackage{amsmath, amssymb}
\usepackage{algpseudocode,algorithm}
\usepackage{graphicx}
\usepackage{subfigure}
\usepackage{lineno}
\usepackage[usenames,dvipsnames]{color}
\usepackage{multirow}
\usepackage{booktabs}

\usepackage{xr}
\let\hat\widehat

\newtheorem{theorem}{Theorem}
\newtheorem{theorem1}{Theorem}
\newtheorem{lemma}[theorem]{Lemma}
\newtheorem{proposition}[theorem]{Proposition}
\newtheorem{example}[theorem1]{Example}

\newtheorem{remark}[theorem1]{Remark}

\let\hat\widehat
\let\tilde\widetilde

\newcommand{\R}{\mathbb R}
\newcommand{\EIF}{{\sf EIF}}
\newcommand{\E}{\mathbb E}

\newcommand{\mcM}{\mathcal M}

\catcode`@=11
\newskip\beforeproofvskip
\newskip\afterproofvskip
\beforeproofvskip=\medskipamount
\afterproofvskip=\bigskipamount

\def\prooftag{Proof}
\def\proofskip{\enspace}

\def\proof{\@ifnextchar[{\@@proof}{\@proof}}  
\def\@startproof{\par\vskip\beforeproofvskip\leavevmode}
\def\@proof{\@startproof{\scshape\prooftag.}\proofskip}
\def\@@proof[#1]{\@startproof {\scshape\prooftag #1.}\proofskip}

\catcode`@=12

\let\hat\widehat
\let\tilde\widetilde

\begin{document}

\begin{frontmatter}

\title{Pattern graphs: a graphical approach to nonmonotone missing data}
\runtitle{Pattern graphs}

%

{
\center
Yen-Chi Chen\footnote{yenchic@uw.edu}\\
Department of Statistics\\University of Washington
}

\begin{abstract}
We introduce the concept of pattern graphs--directed acyclic graphs representing how response patterns are associated. 
A pattern graph represents an identifying restriction that is nonparametrically identified/saturated and is often a missing not at random restriction.
We introduce a selection model and a pattern mixture model formulations using the pattern graphs and show that they are equivalent. 
A pattern graph leads to an inverse probability weighting estimator as well as an imputation-based estimator.
We also study the semi-parametric efficiency theory and derive a multiply-robust estimator using pattern graphs.
\end{abstract}



\end{frontmatter}

\section{Introduction}


Missing data problems are prevalent in modern scientific research \citep{LittleRubin02,molenberghs2014handbook}.
Based on the intrinsic constraints of missing/response patterns,
these problems can be categorized into 
monotone and nonmonotone missing data problems.
In the case of monotone missing data,
the missingness of variables is ordered in such a way
that if a variable is missing, all following variables are missing.
This occurs in a scenario in which individuals drop out of a study, which
is common in longitudinal studies \citep{diggle2002analysis}.

In the case of nonmonotone missing data, the missingness is not necessarily monotone, and
the missingness of one variable does not necessarily place constraints on the missingness of any other
variables. 
There have been several attempts to
use the missing at random (MAR) restriction/assumption 
in this case \citep{Robins97,robins1997non, sun2018inverse}.
However, 
the resulting inverse probability weighting (IPW) estimator
may not be stable \citep{sun2018inverse},
and the MAR restriction is not easy to interpret in nonmonotone cases \citep{robins1997non, Linero17}.
Therefore,  several attempts have  been made to use missing not at random (MNAR) restrictions which are interpretable. 
For instance, 
\cite{shpitser2016consistent,sadinle2017itemwise, malinsky2019semiparametric}
proposed a non-self-censoring/itemwise conditionally independent nonresponse
restriction,
\cite{little1993pattern} and \cite{tchetgen2018discrete}
considered a complete-case missing value (CCMV) restriction, and
\cite{Linero17} introduced the transformed-observed-data restriction. 
However, each study proposed only one MNAR restriction to handle data,
and it remains unclear how to construct a general class of identifying restrictions for nonmonotone missing data.

\begin{figure}
\center
\includegraphics[width=1.5in]{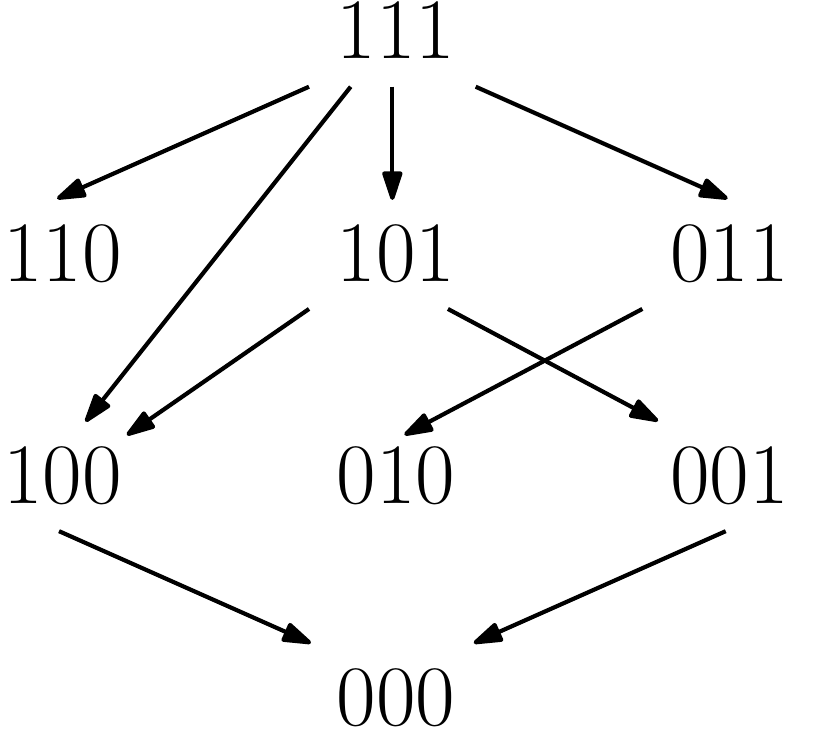}\quad
\includegraphics[width=1.5in]{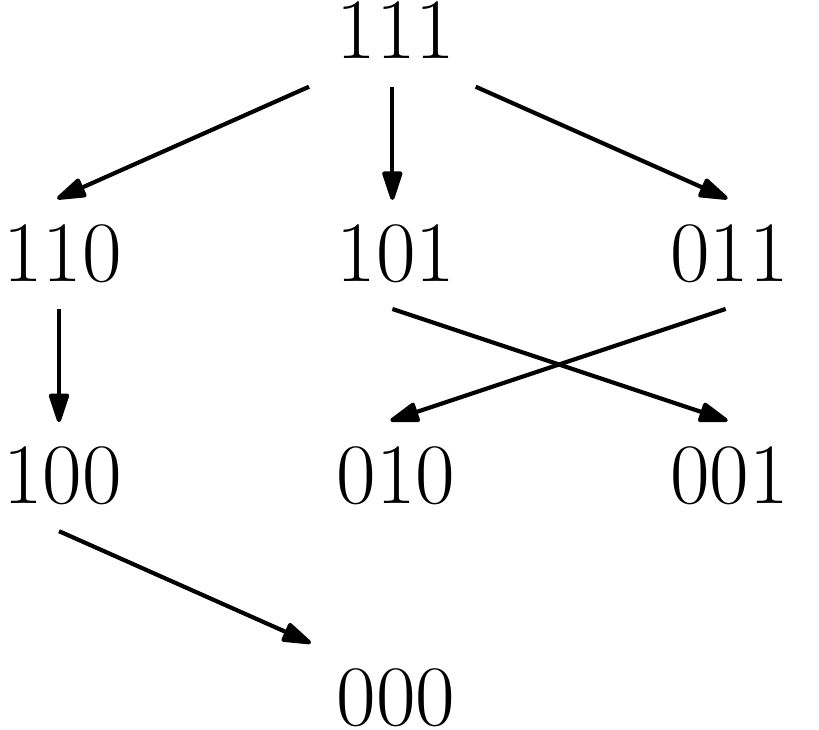}\quad
\includegraphics[width=1.5in]{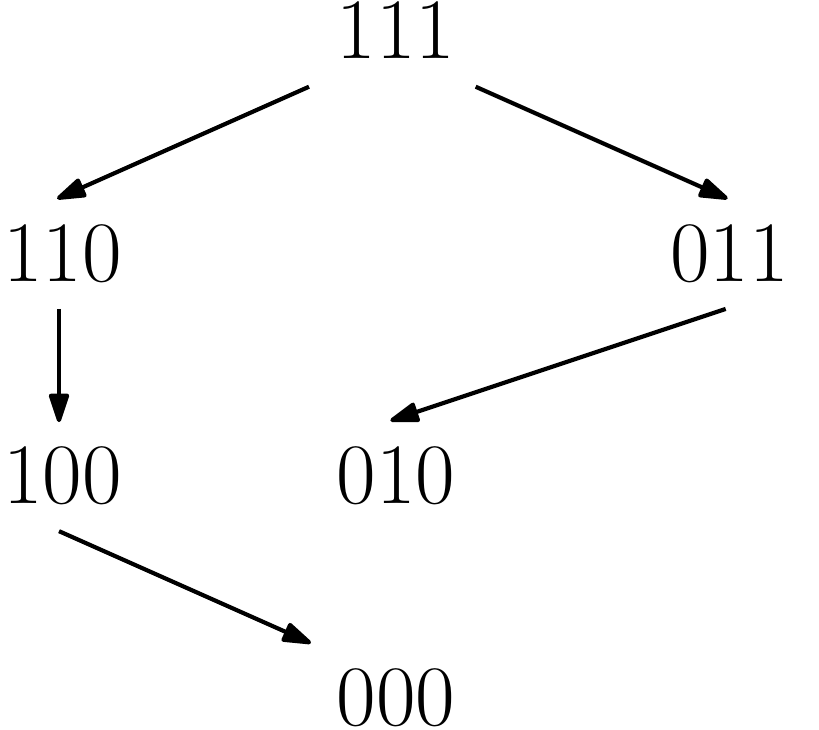}
\caption{Regular pattern graphs in the case of three potentially missing variables. The binary vector indicates the response patterns, e.g., 
$101$ signifies that the first and the third variables are observed while the second variable is missing.
The left and middle panels display examples of regular pattern graphs
when all response patterns are possible.
The right panel shows a regular pattern graph where there are only six possible response patterns (this occurs
when $P(R=101) = P(R=001) = 0$). 
}
\label{fig::ex02}
\end{figure}

In this paper, we introduce a graphical approach to constructing identifying restrictions for nonmonotone missing data problems.
This graphical approach
defines an identifying restriction
using a graph of response patterns; thus, the resulting graph is called a pattern graph.
Formally, a pattern graph is a directed graph where nodes
are possible response patterns
and whose edges/arrows represent the relationship between
the selection probability of patterns (also known as the missing data mechanism in \citealt{LittleRubin02}).
A pattern graph represents an identifying restriction 
placing conditions on the unobserved part of  data,
and is always nonparametrically identified/saturated (Theorem~\ref{thm::PMM1}; \citealt{robins2000sensitivity}); that is, it
does not contradict  the observed data.
In general, the identifying restriction of a pattern graph is an MNAR restriction.
Figure~\ref{fig::ex02} provides examples of pattern graphs when three variables 
may be missing, and a response pattern is described by a binary vector (e.g., $110$ signifies
that for a variable $L = (L_1,L_2,L_3)$, $L_1$ and $L_2$ are observed and $L_3$ is missing).
Different pattern graphs correspond to different identifying restrictions,
so pattern graphs define  a large class of identifying restrictions.
It should be emphasized  that 
\emph{a pattern graph is not a conventional graphical model}. 

\emph{Main results.}
The main results of this paper can be summarized as follows:
\begin{enumerate}
\item We introduce the concept of pattern graphs (Section~\ref{sec::ID})
and derive a graphical criterion leading to 
an identifiable full-data distribution using selection odds model and pattern mixture model formulations
(Theorem~\ref{thm::Codds} and \ref{thm::PMM1}).

\item We demonstrate that  the selection odds model and  the pattern mixture
model are equivalent	 (Theorem~\ref{thm::PMM2}).

\item We introduce an IPW estimator and study its statistical properties (Theorem~\ref{thm::IPW}).

\item We propose a regression adjustment estimator and derive its asymptotic normality (Theorem~\ref{thm::RA}).

\item We study the semi-parametric theory of the pattern graph (Theorem~\ref{thm::EIF}) 
and propose a multiply robust  estimator by augmenting the IPW estimator (Theorem~\ref{thm::MR}). 

%

\end{enumerate}

\emph{Related work.}
The CCMV restriction \citep{little1993pattern,tchetgen2018discrete}
can be represented by a  pattern graph.
In monotone missing data problems,
the available-case missing value restriction \citep{ACMV} and the neighboring-case missing value restriction \citep{Thijs}
and some donor-based identifying restrictions \citep{chen2019nonparametric}
can also be represented by pattern graphs.
There have been studies that utilize graphs to
analyze missing data. 
\cite{mohan2013graphical,mohan2014graphical,tian2015missing,mohan2018graphical, bhattacharya2020causal, nabi2020full}
proposed methods to test missing data assumptions
under graphical model frameworks.
\cite{shpitser2015missing,shpitser2016consistent,sadinle2017itemwise, malinsky2019semiparametric}
proposed a non-self censoring graph that leads to 
an identifying restriction under the MNAR scenario. 
However, it should again be emphasized that pattern graphs are different from graphical models;
thus, our graphical approach is very different from the above-mentioned studies.


\emph{Outline.}
In Section~\ref{sec::ID}, we formally introduce the concept
of (regular) pattern graphs and describe how they represent an identifying restriction.
We discuss strategies for constructing an estimator
under a pattern graph in Section~\ref{sec::estimator}.
We discuss potential future work in Section~\ref{sec::discussion}.
In  the supplementary materials \citep{chen2020supp},
we present a sensitivity procedure in Appendix ~\ref{sec::sensitivity}, 
a study on the equivalence class in Appendix \ref{sec::GPG},
and an application to a real data in Appendix \ref{sec::data}. 
Technical assumptions and proofs are provided in Appendix \ref{sec::assumptions} and \ref{sec::proof}.


 
 
 
 
 

\section{Pattern graph and identification}	\label{sec::ID}

Let $L\in\R^d$ be a vector of the study variables  of interest
and $R\in\{0,1\}^d$ be a binary vector representing the response pattern. 
Variable $R_j = 1$ signifies that variable $L_j $ is observed. 
Let $1_d = (1,1,\cdots, 1)$ be the pattern corresponding to 
the completely observed case
and $\bar r = 1_d - r$ be the reverse (flipping $0 $ and $1$) of pattern $r$.
We use the notation $L_r = (L_j: r_j=1)$. 
For example, suppose that $L = (L_1,\cdots, L_4)$, then 
$L_{1010} = (L_1,L_3)$, $L_{1100} = (L_1,L_2)$ and $L_{\overline{1100}} = L_{0011} = (L_3,L_4)$.
Table~\ref{tab::001} presents an example of data with missing entries and the corresponding pattern indicator $R$.
Both $L$ and $R$ are random vectors from a joint distribution $F(\ell,r)$ with a probability density function (PDF) $p(\ell,r)$, and
we denote $\mathbb{S}_r$ as the support of random variable $L_r$.
For a binary vector $r$, we use $|r| = \sum_{j}r_j$ to denote  the number of
non-zero elements.

\begin{table}
\center
\begin{tabular}{rrrr|c}
ID                       & $L_1$ & $L_2$  & $L_3$  &$R$\\ \hline
\multicolumn{1}{r|}{001} & 5     & 1.3   &  *  &110 \\ \hline
\multicolumn{1}{r|}{002} & 6     & *     &  1.1  &101 \\ \hline
\multicolumn{1}{r|}{003} & *     & *     &  1.0  &001 \\ \hline
\multicolumn{1}{r|}{004} & 5     & *     &  *    &100 \\ \hline
\multicolumn{1}{r|}{005} & 2     & 2.1   & 0.8  &111\\ \hline
\multicolumn{1}{r|}{$\vdots$}&$\vdots$&$\vdots$&$\vdots$&$\vdots$
\end{tabular}
\caption{Example of a hypothetical dataset with missing entries. 
Variable $L=(L_1,\cdots, L_3)$ represents the study variable and variable $R\in\{0,1\}^3$
represents the response pattern. The star symbol ($*$) indicates a missing entry.
}
\label{tab::001}
\end{table}

Let $\mathcal{R}\subset \{0,1\}^d$ be the collection of all possible
response patterns, i.e., $P(R\in\mathcal{R}) = 1$. 
A pattern graph is a directed graph $G = (V,E)$,
where each vertex represents a response pattern (vertex/node set $V = \mathcal{R}$),
and the directed edge represents associations
of the distribution of $(L,R)$ across different patterns. 
Figure~\ref{fig::ex02} provides examples of pattern graphs.
Later we will give a precise definition of how a pattern graph
factorizes the underlying distribution. 
The joint distribution of $(L,R)$ is called the full-data distribution and 
identifying the full-data distribution is a key topic in missing data problems.

When we equip the pattern set $\mathcal{R}$
with a graph $G$,
we can define the notion of parents and children in the graph. 
For two patterns $r_1,r_2\in\mathcal{R}$,
if there is an arrow $r_1\rightarrow r_2$,
we say that $r_1$ is a parent of $r_2$
and $r_2$ is a child of $r_1$. 
Let ${\sf PA}_{r} = \{s: s\rightarrow r\}$ denote the parents 
of pattern/node $r$. 
A pattern/node is called a source if it has no parent.

For two patterns $s,r\in \mathcal{R}$,
we say that $s>r$ if $s_j\geq r_j$ for all $j$ and there is at least one element $k$
such that $s_k>r_k$.
For instance, $110>100$ and $110>010$; however, $110$ cannot be compared with $011$ or $001$.
An immediate result from the above ordering is that 
when $s>r$, the observed variables in pattern $r$ are also observed in pattern $s$.







A pattern graph $G$ is called a {\bf regular pattern graph} if 
it satisfies the following conditions:
\begin{itemize}
\item[\bf(G1)] Pattern $1_d = (1,1,\cdots,1)$ is the only source in $G$.
\item[\bf(G2)] If there is an arrow from pattern $s$ to $r$ (i.e., $s\rightarrow r$),
then $s>r$.
\end{itemize}


Figure~\ref{fig::ex02} presents three examples of regular pattern graphs
when there are three variables subject to missingness. 
The first two panels are regular pattern graphs when all eight response patterns are possible,
and the last panel displays a regular pattern graph when only six patterns are possible.

A regular pattern graph has several interesting properties. 
(G1) implies that the fully observed pattern $R=1_d$ is the only common ancestor
of all patterns except for $R=1_d$. 
Moreover, if $s$ is a parent of $r$, then observed variables in $r$ must be observed in $s$ (due to (G2)). 
In a sense, this means that a parent pattern is more informative than its child.
Condition (G2) implies the following condition:
\begin{itemize}
\item[\bf(DAG)] $G$ is a directed acyclic graph (DAG).
\end{itemize}
Namely, a regular pattern graph is a DAG.
In Appendix~\ref{sec::GPG},
we demonstrate that replacing (G2) with (DAG) still leads to an identifiable full-data distribution.


\subsection{Pattern graph and selection odds models}	\label{sec::Codds}


A common approach for the missing data problems is
the selection model \citep{LittleRubin02}, in which we factorize the  full-data density function as 
$$
p(\ell,r) = P(R=r|\ell) p(\ell),
$$
and  attempt to identify both quantities. 
Here, we focus on modeling the selection probability $P(R=r|\ell)$ 
due to its role in constructing an IPW estimator.
To illustrate this, suppose that we are interested in estimating a parameter of interest $\theta_0$ 
that is defined by a mean function, i.e.,
$
\theta_0 = \E(\theta(L)). 
$
Using simple algebra, it can be shown that 
$$
\theta_0 = \E(\theta(L)) = \E\left(\frac{\theta(L)I(R=1_d)}{P(R=1_d|L)}\right),
$$
which suggests that we can construct an IPW estimator if we know the propensity score 
$\pi(\ell) = P(R=1_d|\ell)$. 

To associate a pattern graph with the missing data mechanism, 
we consider the selection odds \citep{robins2000sensitivity} between a pattern $r$ against its parents ${\sf PA}_r$: $\frac{P(R=r|\ell)}{P(R\in{\sf PA}_r|\ell)}$.
Formally, {\bf the selection odds model of $(L, R)$  factorizes} with respect to pattern graph $G$
if 
\begin{equation}
\frac{P(R=r|\ell)}{P(R\in {\sf PA}_r|\ell)} = \frac{P(R=r|\ell_r)}{P(R\in {\sf PA}_r|\ell_r)}.
\label{eq::Codds}
\end{equation}
Namely, we assume that the (conditional) odds of a pattern $r$
against its parents depend only on the observed entries. 
Note that assumption (G2) in the regular pattern graph assumption
implies that for any parent nodes of $r$, variable $L_r$ is observed. 
Thus, factorization in terms of the selection odds implies that
the selection odds are identifiable.
From equation \eqref{eq::Codds}, it can be seen that
the corresponding restriction is an MNAR restriction in general. 
Equation \eqref{eq::Codds} is related to the MAR restriction
in a more involved way (see Section~\ref{sec::discussion} for a detailed discussion).

Let $O_r(\ell_r) = \frac{P(R=r|\ell_r)}{P(R\in {\sf PA}_r|\ell_r)}$
be the odds based on the variable $\ell_r$. 
Equation \eqref{eq::Codds} can be written 
as
\begin{equation}
P(R=r|\ell) = P(R\in{\sf PA}_r|\ell) \cdot O_r(\ell_r) = \sum_{s\in {\sf PA}_r} P(R=s|\ell)\cdot O_r(\ell_r).
\label{eq::Codds2}
\end{equation}
Namely, the probability of observing pattern $R=r$
is the summation of the probability of observing any of its parents multiplied by the observable odds. 
Later in Proposition \ref{lem::pathG}, we provide another interpretation of equation \eqref{eq::Codds} using the path selection.
A useful property of graph factorization 
is that the propensity score is identifiable, as described in the following theorem.

\begin{theorem}
Assume that the selection odds model of $(L, R)$ factorizes with respect to a regular pattern graph $G$.
Define $$
Q_r(\ell ) = \frac{P(R=r|L=\ell)}{P(R=1_d|L= \ell)},
$$
for each $r$ and $Q_{1_d}(\ell) = 1$.
Then 
$\pi(\ell) \equiv P(R=1_d|\ell)$ is identifiable
and has the following recursive-form:
$$
\pi(\ell) = \frac{1}{\sum_{r}Q_r(\ell)},\quad Q_r(\ell)= O_r(\ell_r) \sum_{s\in {\sf PA}_r} Q_s(\ell).
$$
\label{thm::Codds}
\end{theorem}


The identifiability follows from the induction. $Q_{1_d}=1$ is clearly identifiable,
and we recursively deduce the identifiability of $Q_r$ from $|r|=d-1,d-2,d-3,\cdots, 0$. 
Assumption (G2) guarantees that this recursive procedure is possible.
Note that with an identifiable $\pi(\ell)$, we can identify $P(R=r|\ell) = Q_r(\ell)\pi(\ell)$ 
and $p(\ell) = \frac{p(\ell,R=1_d)}{P(R=1_d|\ell)} = \frac{p(\ell,R=1_d)}{\pi(\ell)}$.
Thus, the full-data density $p(\ell,r) = P(R=r|\ell) p(\ell)$ is identifiable.

\begin{figure}
\center
\includegraphics[height=1.5in]{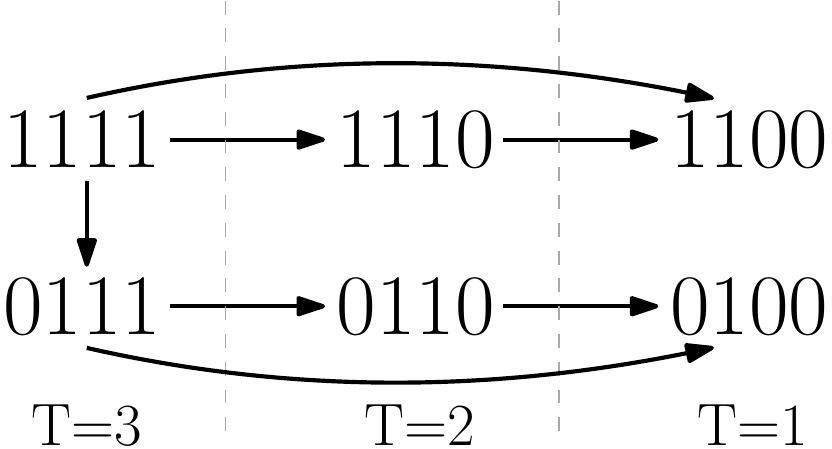}
\qquad\qquad
\includegraphics[height=1.5in]{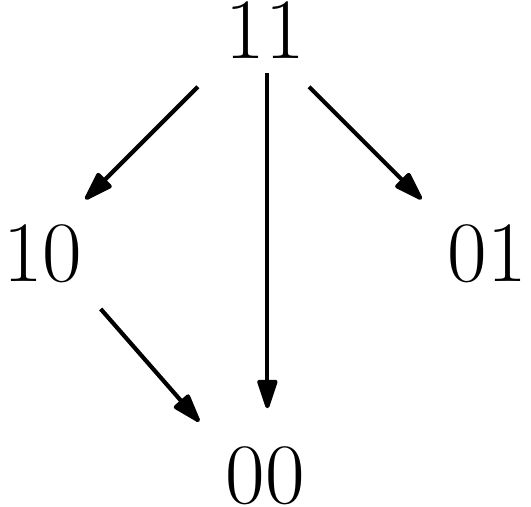}
\caption{Example of regular pattern graphs.
{\bf Left:} The regular pattern graph used in Example~\ref{ex::ex02}, where we have a longitudinal variable with three time points $Y = (Y_1,Y_2,Y_3)$
and a regular variable $Z$ where both are subject to missingness. The missingness of $Y$
is monotone. 
Note that this pattern graph leads to conditional missing at random of $Y$ given $Z$ being observed or not.
See Example~\ref{ex::ex02} for further discussion.
{\bf Right:} The regular pattern graph used in Example~\ref{ex::ex01}.
}
\label{fig::ex01}
\end{figure}

\begin{example}[Conditional MAR]	\label{ex::ex02}
Consider the scenario in which we have a longitudinal variable $Y$ with three time points, i.e., $Y=(Y_1,Y_2,Y_3). $
In addition, we have another study variable $Z$ that is observed once at the baseline. 
The total study variable $L = (Z,Y) = (Z,Y_1,Y_2,Y_3)$.
Variable $Y$ is subject to monotone missingness (dropout), and variable $Z$ may also be missing. 
There are a total of six possible patterns in this case, as illustrated in the left panel of Figure~\ref{fig::ex01}. 
We use the variable $T = R_2+R_3+R_4$ to denote the dropout time and $R_z = R_1$ to denote the response indicator
of variable $Z$. 
Suppose that we use the regular pattern graph as in the left panel of Figure~\ref{fig::ex01}. 
This graph implies the following assumptions on $T$ and $R_z$
(see Appendix~\ref{sec::ex02} in \citealt{chen2020supp} for the derivation):
\begin{align*}
P(T=t|R_z=1, L) & = P(T=t|R_z=1, Z,Y_1,\cdots, Y_t),\qquad t=1,2,3\\
P(T=t|R_z=0, L) & = P(T=t|R_z=0,Y_1,\cdots, Y_t), \qquad t=1,2,3\\
P(R_z=0|T=3, L)&= P(R_z=1|T=3, L) \cdot  \frac{P(R_z=0|T=3,Y_1,Y_2,Y_3)}{P(R_z=1|T=3,Y_1,Y_2,Y_3)}
\end{align*}
The first two equations present the conditional MAR restriction, i.e.,
we have MAR of $Y$ given $R_z$ and the observed $Z$.
The third equation describes how the missing data mechanism of $Z$ occurs.  
The graph provides a simple way to jointly model the dropout time and the missingness of variable $Z$.
%
\end{example}

Selection odds factorization  provides an alternative interpretation
of the missing data mechanism using the concept of path selection.
A (directed) path
$\Xi = \{r_0,\cdots, r_m\}$, is the collection of ordered patterns 
$$
r_0>r_1>r_2\cdots>r_m
$$
such that there is an arrow from $r_i$ to $r_{i+1}$ in the graph. 
A path from $s$ to $r$ refers to a path where initial node $r_0 = s$ and the end node $r_m = r$.
Let 
$$
\Pi_r = \{\mbox{all paths from $1_d$ to $r$}\},\quad \Pi = \cup_r \Pi_r,
$$
and
operationally define $\Pi_{1_d} = \{11\rightarrow 11\}$. 
If there exists a path from $s$ to $r$, we call $s$ an ancestor (pattern) of $r$. 
With the above notation, we have the following decomposition. 

\begin{proposition}
Assume  that the selection odds model of $(L,R)$ factorizes with respect to a regular pattern graph $G$.
Then 
\begin{equation}
\begin{aligned}
1& =  \sum_{\Xi\in \Pi} \pi(L)\prod_{s\in\Xi}O_s(L_s),\\
P(R=r|L) & = \sum_{\Xi\in \Pi_r} \pi(L)\prod_{s\in\Xi}O_s(L_s).
\end{aligned}
\label{eq::naive}
\end{equation}
\label{lem::pathG}
\end{proposition}


Proposition~\ref{lem::pathG} implies
\begin{equation}
\pi(L) = \frac{1}{\sum_{\Xi\in \Pi} \prod_{s\in\Xi}O_s(L_s)},
\label{eq::path::factor}
\end{equation}
which is a closed form of the propensity score $\pi(L)$.


Proposition \ref{lem::pathG} presents an interesting interpretation of the selection odds model.
Define $\kappa(\Xi|L) = \pi(L) \prod_{s\in \Xi} O_s(L_s)$
to be a path-specific score. 
It can be seen that $\kappa(\Xi|L)\geq 0$
and $\sum_{\Xi\in\Pi}\kappa(\Xi|L) = 1$ by the first equality in Proposition \ref{lem::pathG}.
Thus,
$\kappa(\Xi|L)$ can be interpreted as \emph{the probability of selecting path $\Xi$ from $\Pi$.}
The second equality can be written as 
$$
P(R=r|L) = \sum_{\Xi\in \Pi_r} \pi(L)\prod_{s\in\Xi}O_s(L_s) =  \sum_{\Xi\in \Pi_r} \kappa(\Xi|L),
$$
which implies that the probability of observing pattern $r$ 
is the summation of 
all path-specific probabilities corresponding to paths ending at $r$.

Because every path starts from $1_d$, 
a path can be interpreted as a scenario in which the missingness occurs (from a fully observed case).
A path $\Xi$ is randomly selected  with a probability of $ \kappa(\Xi|L) $,
and missingness occurs sequentially as the elements in $\Xi$. 
So the last element in $\Xi$ is the observed pattern.
Therefore, the probability of observing a particular pattern $r$
is the summation of the probabilities of all possible paths that end at $r$.
The choice of a graph is a means of incorporating our scientific knowledge of the
underlying missing data mechanism; in Section~\ref{sec::data},
we provide a data example to illustrate this concept.

\begin{example}	\label{ex::ex01}
Consider the pattern graph in the right panel of Figure~\ref{fig::ex01}, where
it is generated by two variables and four patterns $11,10,01,00$ and
has four arrows $11\rightarrow10\rightarrow00$, $11\rightarrow00$ and $11\rightarrow 10$.
There are five paths (including $11\rightarrow 11$): 
$$
11\rightarrow 11,\quad 11\rightarrow 10,\quad 11\rightarrow 01,\quad 11\rightarrow 00, \quad11\rightarrow10\rightarrow 00
$$
and each corresponds to probability
\begin{align*}
\kappa(11\rightarrow 11|L)&= \pi(L),\\ 
\kappa(11\rightarrow 10|L)&= \pi(L)O_{10}(L_{10}),\\
\kappa(11\rightarrow 01|L)&= \pi(L)O_{01}(L_{01}),\\
\kappa(11\rightarrow 00|L)&= \pi(L)O_{00}(L_{00}),\\ 
\kappa(11\rightarrow 10\rightarrow 00|L)&= \pi(L)O_{10}(L_{10})O_{00}(L_{00}).
\end{align*}
Each path represents a possible scenario that generates the response pattern. 
Since the probability must sum to $1$, we obtain 
$$
\pi(L) = \frac{1}{1+O_{10}(L_{10})+O_{01}(L_{01})+O_{00}(L_{00})+O_{10}(L_{10})O_{00}(L_{00})},
$$
which agrees with Theorem~\ref{thm::Codds}.
The probability of observing patterns $10$ and $01$ are $P(R=10|L) = \kappa(11\rightarrow 10|L) = \pi(L)O_{10}(L_{10})$ and 
$P(R=01|L) = \kappa(11\rightarrow01|L)= \pi(L)O_{01}(L_{01})$,
respectively.
Pattern $00$ occurs with a probability of 
\begin{align*}P(R=00|L) & = \kappa(11\rightarrow 00|L)+\kappa(11\rightarrow 10\rightarrow 00|L)\\
&= \pi(L)O_{00}(L_{00})+\pi(L)O_{10}(L_{10})O_{00}(L_{00}) 
\end{align*} 
The first component $\pi(L)O_{00}(L_{00})$
represents scenario $11\rightarrow00$, i.e., the individual directly drops both variables. 
The other component $\pi(L)O_{10}(L_{10})O_{00}(L_{00}) $
corresponds to scenario $11\rightarrow10\rightarrow00$, i.e.,
variable $L_2$ is missing first, and then variable $L_1$ is missing.
Therefore, the paths in the pattern graph represent
possible hidden scenarios that
generate a response pattern.
\end{example}

\begin{remark}
\cite{robins1997non}
proposed a randomized monotone missing (RMM) process to construct a class of  MAR assumptions 
for the nonmonotone missing data problems that also admits a graph representation 
on how the missingness of one variable is associated with others. 
This method may look similar to ours; however, the two ideas (RMM and pattern graphs)
are very different.
First, RMM constructs a MAR assumption, whereas pattern graphs are generally MNAR
(generalizations of RMM to MNAR can be found in \citealt{Robins97} and \citealt{robins2000sensitivity}). 
Second, each node in the RMM graph is a variable,
whereas each node in a pattern graph is a response pattern. 
Third, in the next section, we demonstrate that the selection odds model in a pattern graph has an equivalent pattern mixture model
representation; however,t it is unclear whether the RMM process has a desirable pattern mixture model representation or not. 
\end{remark}



\subsection{Pattern graph and pattern mixture models}


Another common strategy for handling missing data is  pattern mixture models \citep{Little93},
which factorize
$$
p(\ell, r) = p(\ell|R=r) P(R=r) = p(\ell_{\bar r}|\ell_r,R=r) p(\ell_r|R=r) P(R=r).
$$
The above factorization provides a clear separation between
observed and unobserved quantities. 
The first part, $p(\ell_{\bar r}|\ell_r,R=r)$, is called the extrapolation density \citep{Little93}, which
corresponds to the distribution of unobserved entries given the observed entries. 
This part cannot be inferred from the data without making additional assumptions.
The latter part, $p(\ell_r|R=r) P(R=r)$, is called the observed-data distribution,
which characterizes the distribution of the observed entries and can
be estimated from the data without any identifying assumptions.

An interesting insight is that different response patterns provide
information on different variables.
Thus, we can associate an extrapolation density
to the observed parts of
another pattern. 
This motivates us to consider a graphical approach to factorize the distribution
using pattern mixture models. 

Formally, 
{\bf the pattern mixture model of $(L, R)$ factorizes} with respect to a pattern graph $G$
if 
\begin{equation}
p(x_{\bar r}|x_r,R=r) = p(x_{\bar r}|x_r,R\in {\sf PA}_r).
\label{eq::PMM}
\end{equation}
Equation \eqref{eq::PMM} states that
the extrapolation density of pattern $r$
can be identified by its parent(s).
Namely, we  model the unobserved part of pattern $r$
using the information from its parents. 
This is a reasonable choice because
condition (G2) implies that a parent pattern is more informative than its child pattern.
Pattern mixture model factorization leads to the following identifiability property.



\begin{theorem}
Assume that the pattern mixture model of $(L, R)$ factorizes with respect to a regular pattern graph $G$,
then $p(\ell,r)$ is nonparametrically identifiable/saturated.
\label{thm::PMM1}
\end{theorem}

Theorem~\ref{thm::PMM1} states 
that graph factorization using pattern mixture models implies
a nonparametrically identifiable full-data distribution. 
Namely, the implied observed distribution of $F(\ell,r)$
coincides with the observed-data distribution that generates our data for patterns $r$ such that $P(R=r)>0$.
Thus, the identifying restriction derived from the graph  never contradicts the
observed data \citep{robins2000sensitivity}.
Nonparametric identification is also known as nonparametric saturation or just-identification in \cite{Robins97,Vansteelandtetal06, DanielsHogan08, HoonhoutRidder18}.


Thus far, we have discussed two different methods of associating a pattern graph to a full-data distributions. 
The following theorem states that they are equivalent under the positivity condition 
($p(\ell_r,r)>0$ for all $\ell_r\in \mathbb{S}_r$ and $r\in \mathcal{R}$).

\begin{theorem}
If $G$ is a regular pattern graph and $p(\ell_r,r)>0$ for all $\ell_r\in \mathbb{S}_r$ and $r\in \mathcal{R}$,
then 
the following two statements are equivalent:
\begin{itemize}
\item The selection odds model of $(L, R)$ factorizes with respect to $G$.
\item The pattern mixture model of $(L, R)$ factorizes with respect to $G$.
\end{itemize}

\label{thm::PMM2}
\end{theorem}


With Theorem~\ref{thm::PMM2}, we can interpret the graph factorization using
either the selection odds model or the pattern mixture model,
both of which lead to
the same full-data distribution.
Because of Theorem~\ref{thm::PMM2},
when we say $(L,R)$ factorizes with respect to $G$,
this factorization may be interpreted using the selection odds model or pattern mixture model.
Note that this equivalence is not surprising, as
\cite{robins2000sensitivity} demonstrated that 
certain classes of selection odds models and pattern mixture models
are equivalent.
Theorem~\ref{thm::PMM2} shows that the identifying restrictions
from pattern graphs form another class of  restrictions
with this elegant property.


\begin{figure}
\center
\includegraphics[width=2in]{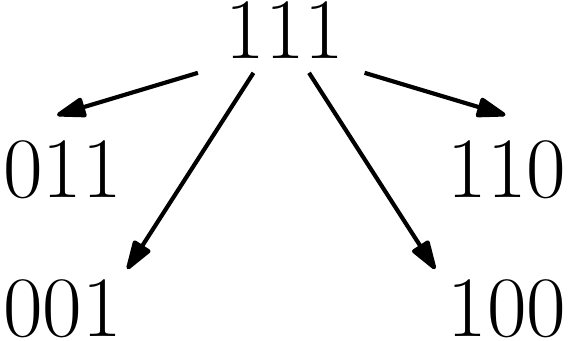}\qquad\qquad
\includegraphics[width=2in]{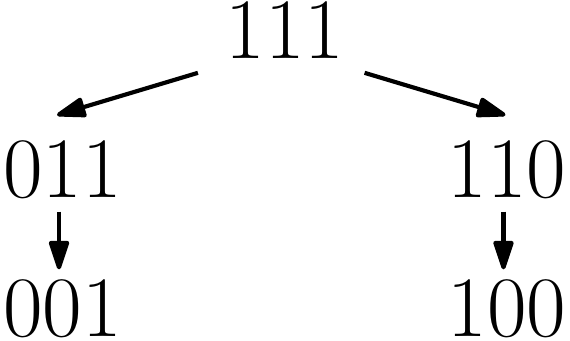}
\caption{Examples of regular pattern graphs of three variables
with only 5 possible patterns $\mathcal{R} = \{111,110,100,011,001\}$.
{\bf Left:} The left panel shows the pattern graph that CCMV restriction corresponds.
{\bf Right:}
The right panel shows a pattern graph that is related to the transform-observed-data restriction in \cite{Linero17}.}
\label{fig::ex03}
\end{figure}

\begin{example}[Complete-case missing value restriction]
The CCMV restriction \citep{little1993pattern}
is an assumption in pattern mixture models. 
It requires that 
\begin{equation}
p(\ell_{\bar r}|\ell_r,R=r) = p(\ell_{\bar r}|\ell_r,R=1_d)
\label{eq::CCMV1}
\end{equation}
for all pattern $r\in \mathcal{R}$. 
The corresponding pattern graph is a graph where every node (except the node of $1_d$)
has only one parent: the completely-observed case; namely, ${\sf PA}_r = 1_d$ for all $r\neq 1_d$.
The left panel in Figure~\ref{fig::ex03} presents an example of the pattern graph of CCMV.
Using Theorem~\ref{thm::PMM2} and the selection odds model, 
equation \eqref{eq::CCMV1} is equivalent to
\begin{equation}
\frac{P(R=r|L=\ell)}{P(R=1_d|L=\ell)} = \frac{P(R=r|L=\ell_r)}{P(R=1_d|L=\ell_r)},
\label{eq::CCMV2}
\end{equation}
which is the key formulation in \cite{tchetgen2018discrete} that 
establishes a multiply-robust estimator.
\end{example}

\begin{remark}[Transform-observed-data restriction]
\cite{Linero17} proposed a transform-observed-data restriction
that is related to a particular pattern graph under a special case.
Consider a three-variable scenario in which only five patterns are available $111,110,100, 011,001$, and
there are two paths of arrows: $111\rightarrow110\rightarrow 100$
and  $111\rightarrow011\rightarrow 001$.
The right panel of Figure~\ref{fig::ex03} displays this graph.
The first path implies  $p(x_3|x_1,x_2,110) = p(x_3|x_1,x_2,111)$
and $p(x_2,x_3|x_1,100) = p(x_2,x_3|x_1,110)$,
which further implies $p(x_2|x_1,100) = p(x_2|x_1,110)$, which is 
a requirement of the transform-observed-data restriction in this case.
Similarly, the other path implies
$
p(x_2|x_3,001) = p(x_2|x_3,011),
$
which is another requirement of the transform-observed-data restriction.

\end{remark}

\begin{remark}[Monotone missing data problem]
Suppose that the missingness is monotone; 
then, the pattern graph reduces to special cases of the interior family \citep{Thijs} 
and donor-based identifying restriction \citep{chen2019nonparametric}.
In particular, the parent set ${\sf PA_r}$ is the donor set of the dropout time $t = |r|.$
The available-case missing value restriction \citep{ACMV} corresponds to the pattern
graph with ${\sf PA}_r = \{s: |s|>|r|\}$, i.e., the graph with all possible arrows/edges.
The neighboring-case missing value restriction \citep{Thijs}
is the pattern graph with ${\sf PA}_r = \{s: |s| = |r|+1\}$.


\end{remark}



\section{Estimation with pattern graphs}	\label{sec::estimator}

In this section,
we present several strategies for estimating
the parameter of interest using the pattern graph. 
Here, we consider the parameter of interest that can be written in the form 
$\theta_0 = \E(\theta(L))$, where $\theta(L)$ is a known function.
Note that all analyses can be applied to the case of estimating equations.

With a slight abuse of notation, the observed data are written as  IID random elements
$$
(L_{1,R_1},R_1),\cdots, (L_{n,R_n},R_n),
$$
where $R_1,\cdots,R_n\in\mathcal{R}$
denote the response pattern of each observation and $L_{i,R_i}$ denotes the observed variables
of the $i$-th individual
and $L_i\in\mathbb{R}^d$ denotes the vector of study variables of the $i$-th individual.
Note that not every entry of $L_i$ is observed; we only observe $L_{i,R_i}$, while
$L_{i,\bar R_i}$ is missing.

\subsection{Inverse probability weighting}

The parameter of interest can be written as
$$
\theta_0 = \E(\theta(L)) = \E\left(\frac{\theta(L)I(R=1_d)}{P(R=1_d|L)}\right)=\E\left(\frac{\theta(L)I(R=1_d)}{\pi(L)}\right).
$$
This formulation implies that as long as we can estimate $\pi(\ell)$,
we can construct a consistent estimator of $\theta$ via the concept of IPW. 


From Theorem~\ref{thm::Codds}, the propensity score can be expressed as 
$$
\pi(\ell) = \frac{1}{\sum_{r}Q_r(\ell)},\quad Q_{1_d}(\ell) = 1,\quad Q_r(\ell)= O_r(\ell_r) \sum_{s\in {\sf PA}_r} Q_s(\ell).
$$
By the above recursive property, an estimator of $O_r(\ell_r)$ leads to
an estimator of $Q_r(\ell)$ and $\pi(\ell)$.
The odds 
$$
O_r(\ell_r) = \frac{P(R=r|\ell_r)}{P(R\in{\sf PA}_r|\ell_r)} 
$$
can be estimated by comparing the distribution of patterns $R=r$ with patterns $R\in {\sf PA}_r$.
This can be achieved by 
constructing a generative binary classifier \citep{friedman2001elements} such that label $1$ refers to $R=r$ and label $0$ refers to $R\in{\sf PA}_r$
or by a regression function with the same binary outcome and the feature/covariate is $\ell_r$.
In Example~\ref{ex::logistic} of Appendix \ref{sec::assumptions},
we describe a logistic regression approach to estimate $O_r(\ell_r)$.


Suppose that we have an estimator $\hat \pi(\ell)$ of the propensity score.
%
Then, we can estimate $\theta$ using the IPW approach as follows:
$$
\hat \theta_{\sf IPW}  = \frac{1}{n}\sum_{i=1}^n \frac{\theta(L_i) I(R_i=1_d)}{\hat \pi(L_i)}.
$$
As an example, suppose that we estimate $\pi(\ell)$ by placing parametric models over the odds, i.e., 
$$
\hat O_r(\ell_r) = O_r(\ell_r;\hat \eta_r),
$$
where $\hat \eta_r\in\Theta_r$ is the estimated parameter of the selection odds $\frac{P(R=r|\ell_r)}{P(R\in {\sf PA}_r|\ell_r)}$. 
We can estimate the selection odds using a maximum likelihood approach or moment-based approach.
With the estimated selection odds, we estimate the propensity score $\hat \pi(\ell) = \pi(\ell;\hat \eta)$
using the recursive relation. 
Let $\hat \eta = (\hat\eta_r: r\in\mathcal{R})$
be the set of the estimated parameters.

\begin{theorem}
Assume (L1-4)  in Appendix \ref{sec::assumptions} and that 
the selection odds model of $(L,R)$ factorizes
with respect to a regular pattern graph $G$.
Then $\hat \theta_{\sf IPW}$ is a consistent estimator and satisfies
$$
\sqrt{n}(\hat \theta_{\sf IPW}  - \theta_0) \overset{D}{\rightarrow}N(0,\sigma_{IPW}^2),
$$
for some $\sigma^2_{IPW}>0$.
\label{thm::IPW}
\end{theorem}

%

Theorem~\ref{thm::IPW} shows the asymptotic normality
of the IPW estimator and can be used to construct a confidence interval.
A traditional approach is to obtain a sandwich estimator of $\sigma^2_{IPW}$
and use it with the normal score to construct a confidence interval. 
However, the actual form of $\sigma^2_{IPW}$ is complex because patterns are correlated
based on the graph structure
and there is no simple way to disentangle them. 
Thus, we recommend using the bootstrap approach \citep{efron1979,efron1994introduction} to construct a confidence interval.
This can be acheived without knowing the form of $\sigma^2_{IPW}$.
Note that the bootstrap method often requires a third moment condition of the score \citep{hall2013bootstrap};
for smooth parametric models such as logistic regression with a bounded covariates,
this condition holds.

%

We can rewrite the IPW estimator as 
$$
\hat\theta_{\sf IPW} = \frac{1}{n}\sum_{i=1}^n \theta(L_i) I(R_i=1_d) \sum_{r} Q_r(L_i;\hat\eta).
$$
So the quantity $Q_r(L_i;\hat\eta)$
behaves like a score from pattern $r$ on observation $L_i$.

\subsubsection{Recursive computation}

Although the IPW estimator has desirable properties,
the propensity score does not have a simple closed form; therefore, the computation
of
Equation \eqref{eq::naive} is not easy.
To resolve this problem, we provide a computationally friendly approach 
to evaluate $\pi(\ell)$ (or its estimator $\hat \pi(\ell)$) using the recursive relation in Theorem~\ref{thm::Codds}.

From Theorem~\ref{thm::Codds}, $\pi(\ell) = \frac{1}{\sum_{r}Q_r(\ell)}$;
thus, it is only necessary to
compute $Q_r(\ell)$. 
The recursive form in Theorem~\ref{thm::Codds},
$$
Q_{1_d}(\ell) = 1,\quad Q_r(\ell)= O_r(\ell_r) \sum_{s\in {\sf PA}_r} Q_s(\ell),
$$
demonstrates that
we can compute $Q_r(L)$ recursively.

Algorithm \ref{alg::IPW}  summarizes the procedure 
for computing $\hat \pi(L)$.
We first compute cases where $|r| = d-1$. 
Having computed $\{Q_r(L): |r|=d-1\}$,
we can easily compute $\{Q_r(L): |r|=d-2\}$
because $\{Q_r(L): |r|=d-2\}$ only depend on $\{Q_r(L): |r|=d, d-1\}$ and each $O_r(L)$. 
Thus, 
by  sequentially computing  (noting that $Q_{1_d}(L) = 1$)
$$
\{Q_r(L): |r|=d-1\},\quad \{Q_r(L): |r|=d-2\},\quad\cdots, \{Q_r(L): |r|=1\},
$$
we obtain every $Q_r(L)$, which 
then leads to $\pi(L) = \frac{1}{\sum_r Q_r(L)}$.

\begin{algorithm}[tb]
\caption{Recursive computation of the propensity score} 
\label{alg::IPW}
\begin{algorithmic}
\State 1. Input: $\hat Q_{1_d}(\ell) =1$ and a given fully-observed vector $L$ and estimators $\hat O_r(\ell_r)$ for each $r\in \mathcal{R}$.
\State 2. Starting from $j=1,\cdots, d-1$, do the following:
	\State 2-1. For each $r \in \{s\in\mathcal{R}: |s| = d-j\}$, do the following:
		\State 2-1-1. Compute $\hat O_r(L_r)$. In the case of logistic regression, 
		$\hat O_r(L_r) = \exp(\hat \beta_r^T \tilde{L}_r)$.
		\State 2-1-2. Compute $\hat Q_r(L)  = \hat O_r(L_r) \sum_{s\in {\sf PA}_r} \hat Q_s(L)$.
\State 3. Return: $\hat \pi(L) = \frac{1}{\sum_{r} \hat Q_r(L)}$. 
\end{algorithmic}
\end{algorithm}


Suppose that evaluating $O_r(L_r)$ takes  $\Omega(1)$ units of operations;
then, total cost of evaluating $\pi(L)$ using Algorithm \ref{alg::IPW} is $\Omega(\sum_{r}|{\sf PA}_r|)$ units, where $|{\sf PA}_r|$
is the number of parents of node $r$.
However, if we use equation \eqref{eq::naive}, 
the total cost is $\Omega(\sum_r\sum_{\Xi\in \Pi_r}|\Xi|)$,
where $|\Xi|$ is the number of vertices in the path.
It can be seen that $|{\sf PA}_r|\leq \sum_{\Xi\in \Pi_r}|\Xi|$
and the number of parents can be much smaller than the total number of paths.
Therefore,  Algorithm \ref{alg::IPW} is much more efficient than directly using equation \eqref{eq::naive}.

\subsection{Regression adjustments}	\label{sec::RA}

We can rewrite the parameter of interest as
$$
\theta_0= \E(\theta(L)) = \int m(\ell_r,r) P(d\ell_r, dr),\quad m(\ell_r,r) = \E(\theta(L)|L_r=\ell_r, R=r).
$$
Thus,
if we have an estimator $\hat m(\ell_r,r)$ for every $r$,
we can  estimate $\E(\theta(L))$ using
the regression adjustment approach
$$
\hat \theta_{\sf RA} = \frac{1}{n}\sum_{i=1}^n \hat m(L_{i,R_i}, R_i).
$$
In Appendix \ref{sec::imputation::alg}, we demonstrate that a Monte Carlo approximation 
of this estimator is the imputation-based estimator \citep{LittleRubin02,rubin2004multiple, tsiatis2007semiparametric}.

Regression adjustment is feasible because the regression function $m(\ell_r,r) = \E(\theta(L)|L_r=\ell_r, R=r)$
is identifiable. 
To see this, using the PMM factorization in equation \eqref{eq::PMM}, 
\begin{align*}
m(\ell_r,r) &= \E(\theta(L)|L_r=\ell_r,R=r) \\
&=\int \theta(\ell_{\bar r}, \ell_r)p(\ell_{\bar r}|\ell_r,R=r) d\ell_{\bar r}\\
&=\int \theta(\ell_{\bar r}, \ell_r)p(\ell_{\bar r}|\ell_r,R\in {\sf PA}_r) d\ell_{\bar r}\\
&= \E(\theta(L)|L_r=\ell_r,R\in {\sf PA}_r),
\end{align*}
and $p(\ell_{\bar r}|\ell_r,R\in {\sf PA}_r) $ is identifiable due to Theorem~\ref{thm::PMM1}.

In practice, we first estimate  $\hat p(\ell_r|R=r)$ using a parametric model
for every $r$.
With this, we then estimate $p(\ell_{\bar r}|\ell_r,R\in {\sf PA}_r)$.
Note that we can use  a nonparametric density estimator as well,
but it often suffers from
the curse of dimensionality. 

For pattern $r$,
let $\lambda_r\in\Lambda_r$ be the parameter of the model $L_r|R=r$.
Namely,
$$
p(\ell_r|R=r) = p(\ell_r|R=r; \lambda_r).
$$
We can  estimate $\lambda_r$ via the maximum likelihood estimator (MLE). Let $\hat \lambda_r$ be the MLE. 
We model it in this way to avoid model conflicts; see
Appendix \ref{sec::MC} in the supplementary material \citep{chen2020supp} for more details.
Let $\lambda = (\lambda_r: r\in \mathcal{R})$ be the collection of all parameters
in the model, let $\Lambda$ be the corresponding parameter space, and let $\hat\lambda$ be the MLE. 
The regression function is then estimated by
\begin{align*}
\hat m(\ell_r,r)& =  m(\ell_r,r; \hat \lambda)\\
& = \int \theta(\ell_{\bar r}, \ell_r)p(\ell_{\bar r}|\ell_r,R\in {\sf PA}_r; \hat \lambda) d\ell_{\bar r}.
\end{align*}
Note that in the above expression, the expression of the estimator 
depends on the entire set of  parameters $\hat \lambda =(\hat \lambda_r: r\in\mathcal{R})$,
but $\hat m(\ell_r,r)$ actually only depends on the parameter belonging to its ancestor.
We express it using $\hat \lambda$ to simplify the notation.
 

%
%
%
%


\begin{theorem}
Assume (R1-3) in Appendix \ref{sec::assumptions} and that the pattern mixture model of $(L,R)$ factorizes with respect to a regular pattern graph $G$.
Then $\hat \theta_{\sf RA}$ is a consistent estimator and satisfies
$$
\sqrt{n}(\hat \theta_{\sf RA}  - \theta_0) \overset{D}{\rightarrow}N(0,\sigma_{RA}^2)
$$
for some $\sigma_{RA}^2>0$.
\label{thm::RA}
\end{theorem}

Theorem~\ref{thm::RA} shows that if the density estimators are consistent,
the resulting  regression adjustment estimator is asymptotically normal.
Similar to the IPW estimator, this provides a way to construct
a confidence interval using the bootstrap.
In Appendix \ref{sec::imputation::alg}, we describe a Monte Carlo approach to compute $\hat \theta_{\sf RA}$.
In addition, we show that when the pattern graph is a tree graph, there may be a closed form of
the regression adjustment estimator; thus,o we do not need a numerical procedure (Appendix~\ref{sec::SG}).

\subsection{Semi-parametric estimators}	\label{sec::semi}



We now study the semi-parametric theory of the pattern graph and propose an efficient estimator.
We start with a derivation of the efficient influence function (EIF) of $\E(\theta(L))$.
For any pattern $r\in G$, recall that $\Pi_r$ denotes all paths from $1_d$ to $r$
and $\Pi = \cup_r \Pi_r$ is the collection of all paths.

By Theorem~\ref{thm::Codds} and equation \eqref{eq::path::factor}, 
the inverse of the propensity score can be written as
\begin{align*}
\frac{1}{\pi(L)} = \sum_r Q_r(L) =1+\sum_{r\neq 1_d}\sum_{\Xi\in \Pi_r} \prod_{s\in\Xi}O_s(L_s).
\end{align*}
Thus, the IPW formulation can be decomposed as
\begin{equation}
\begin{aligned}
\theta &=\E(\theta(L)) \\
& = \E\left(\frac{\theta(L)I(R=1_d)}{\pi(L)}\right)\\
& = \E\left(\theta(L)I(R=1_d)\right) + \sum_{r\neq 1_d}\sum_{\Xi\in \Pi_r} \E\left(\theta(L)I(R=1_d)\prod_{s\in \Xi} O_s(L_s)\right) \\
& = \theta_{1_d} + \sum_{r\neq 1_d}\sum_{\Xi \in \Pi_r} \theta_{\Xi}.
\end{aligned}
\label{eq::path::odds}
\end{equation}

For a path $\Xi \in \Pi$ and an element $s\in \Xi$, 
we define
\begin{equation}
\begin{aligned}
\EIF_{\Xi,s}&(L_s,R) \\
&= \mu_{\Xi,s}(L_s)\left(I(R=s) - O_s(L_s) I(R\in {\sf PA}_s)\right) \prod_{w\in\Xi, w<s} O_w(L_w),
\end{aligned}
\label{eq::EIF::pi_s}
\end{equation}
where 
\begin{align}
 \mu_{\Xi,s}(L_s) &= \frac{m_{\Xi,s}(L_s)}{P(R\in {\sf PA}_s|L_s)}, \label{eq::EIF::mu}\\
 m_{\Xi,s}(L_s) &=\E\left(\theta(L)I(R=1_d) \prod_{\tau\in\Xi, \tau>s} O_\tau(L_\tau)\bigg|L_s \right).
 \label{eq::EIF::m}
\end{align}
The following proposition demonstrates that $\sum_{s\in \Xi}\EIF_{\Xi,s}(L_s,R)$
is the EIF of $\theta_\Xi$;
therefore, we obtain a closed form of the EIF of $\theta$.

\begin{theorem}[Efficient influence function]
Suppose that the selection odds model of $(L,R)$
factorizes with respect to a regular pattern graph $G$ and $p(\ell_r,r)>0$.
The EIF of $\theta_\Xi$
is 
$$
\EIF_{\Xi}(L, R)   =   \sum_{s\in \Xi} \EIF_{\Xi,s}(L_s,R).
$$
Thus, the EIF of $\theta$
is
$$
\EIF(L,R) = 
\sum_{r\neq 1_d}\sum_{\Xi\in \Pi_r} \EIF_{\Xi}(L, R).
$$

\label{thm::EIF}
\end{theorem}

Theorem~\ref{thm::EIF}
provides an analytical form of the EIF of both $\theta$
and a pathwise version of it. 
Theorem~\ref{thm::EIF}  also illustrates how a pattern graph informs the construction of the EIF. 
In Appendix \ref{sec::example}, we derive the expression of the EIF of Example~\ref{ex::ex01}.
A key element in the EIF is the function $ \mu_{\Xi,s}(L_s)$ defined in equation \eqref{eq::EIF::mu}. 
In what follows, we describe how $ \mu_{\Xi,s}(L_s)$ is associated with the regression adjustment estimator
in Section~\ref{sec::RA}.

%


\begin{proposition}[Relation to regression adjustment]
Let ${\sf Ans}_r$ denote the ancestors of $r$ 
including $r$ itself.
For $s\in{\sf Ans}_r$, let $\Upsilon_{s,r}$ be the collection of all paths from $s$ to $r$. 
Then 
\begin{enumerate}
\item Function $\mu_{\Xi,s}(L_s)$
is identifiable from $
\{p(\ell_r|R=r): r\in {\sf Ans}_s\};
$
\item 
$\sum_{\Xi\in\Pi_s} \mu_{\Xi,s}(\ell_s) = m(\ell_s, s)$, where $m(\ell_s,s)$ is the regression function defined in Section \ref{sec::RA};
\item 
The EIF of pattern $r$, $\EIF_{r}  = \sum_{\Xi\in\Pi_r}\EIF$, can be written as 
$$
\EIF_{r}(L,R) = \sum_{s\in {\sf Ans}_r} \underbrace{m(L_s,s)(I(R=s)-O_s(L_s)I(R\in {\sf PA}_s)) \sum_{\zeta\in\Upsilon_{s,r}}\prod_{w\in\zeta, w<s} O_w(L_w)}_{=\EIF_{s,r}(L,R)}.
$$
\end{enumerate}

\label{prop::mu_RA}
\end{proposition}

%
Suppose that we have a collection of models $\{p(\ell_\tau|R=\tau; \lambda_\tau): \tau\in\mathcal{R}\}$,
where $\lambda_\tau$ is the underlying parameters.
By Proposition \ref{prop::mu_RA}, we can identify $\mu_{\Xi,r}(L_r)$ using these models, leading to $\mu_{\Xi,r}(L_r; \lambda)$
without any knowledge of the selection odds. 
This insight leads to the construction of a semi-parametric estimator in the next section.

In addition, Theorem~\ref{thm::EIF} and Proposition \ref{prop::mu_RA} provide two equivalent expressions of the EIF.
The first one is a \emph{path expression}:
$$
\EIF(L,R)= \sum_{r\neq 1_d}\sum_{\Xi\in\Pi_r}\sum_{s\in \Xi}\EIF_{\Xi,s}(L,R),
$$
while the second is an \emph{ancestor expression}:
$$
\EIF(L,R) = \sum_{r\neq 1_d} \sum_{s\in {\sf Ans}_r} \EIF_{s,r}(L,R),
$$
where $\EIF_{s,r}(L,R)$ is defined in Proposition \ref{prop::mu_RA}.
The path expression provides insight into how each path's information
contributes to the efficiency of a node,
whereas the ancestor expression demonstrates how an ancestor improves the efficiency of its descendent. 
Moreover, the path expression provides a clear picture of the multiple robustness property (Section~\ref{sec::MR})
while the ancestor expression leads to a simpler numerical procedure (Algorithm \ref{alg::semi}),
which is a mild modification of the regression adjustment.

%

\subsubsection{Construction of semi-parametric estimators}

With the EIF, we can derive a
semi-parametric estimator. 
Since our derivation of EIF is based on the IPW approach,
the linear form of the semi-parametric estimator
is the IPW added to the augmentation from the EIF, i.e.,
\begin{align*}
\mathcal{L}_{\sf semi}(L,R) &= \frac{\theta(L)I(R=1_d)}{\pi(L)} + \EIF(L,R)\\
&= \frac{\theta(L)I(R=1_d)}{\pi(L)} + \sum_{r\neq 1_d}\sum_{\Xi\in\Pi_r}\sum_{s\in \Xi}\EIF_{\Xi,s}(L,R)\\
&= \frac{\theta(L)I(R=1_d)}{\pi(L)} + \sum_{r\neq 1_d}\sum_{s\in {\sf Ans}_r} \EIF_{s,r}(L,R).
\end{align*}
It can be seen that $\E[\mathcal{L}_{\sf semi}(L,R)] = \theta$.
We use the path expression in the following derivation, as it leads to an elegant multiple robustness property (see next section).


Let $ O_r (L_r;\hat \eta_r)$ be the estimated selection odds
and let $ p(\ell_r|R=r; \hat \lambda_r)$ be the estimated density used
in the regression adjustment method. 
By Proposition \ref{prop::mu_RA}, the collection $\{ p(\ell_r|R=r; \hat\lambda_r): r\in\mathcal{R}\}$
implies the collection $\{ \mu_{\Xi,s}(\ell_s, r; \hat \lambda): s\in\Xi, \Xi \in \Pi_r, r\neq 1_d\}$, where $\hat \lambda = (\hat \lambda_r:r\in\mathcal{R})$.
In addition, let $O_r(L_r; \hat \eta_s)$ be the estimated selection odds of pattern $r$.

With these estimators, we estimate the EIF by
\begin{equation*}
\begin{aligned}
\EIF_{\Xi,s}&(L_s,R; \hat \lambda,\hat \eta) \\
&=  \mu_{\Xi,s}(L_s;\hat \lambda)\left[I(R=s) -  O_s(L_s; \hat \eta_s) I(R\in {\sf PA}_s)\right] \prod_{w\in\Xi, w<s}  O_w(L_w; \hat \eta_w)
\end{aligned}
\end{equation*}
and construct the semi-parametric estimator 
\begin{equation}
\begin{aligned}
\hat \theta_{\sf semi}  &=  \frac{1}{n}\sum_{i=1}^n \mathcal{L}_{\sf semi}(L_i,R_i; \hat \lambda, \hat \eta)\\
&= 
\frac{1}{n}\sum_{i=1}^n  \underbrace{\frac{\theta(L_i)I(R_i=1_d)}{\pi(L_i; \hat \eta)}}_\text{IPW} + \underbrace{\sum_{r\neq 1_d}\sum_{\Xi\in\Pi_r}\overbrace{\sum_{s\in \Xi}\EIF_{\Xi,s}(L_i,R_i; \hat \lambda,\hat \eta)}^{=\EIF_{\Xi}(L_i,R_i; \hat \lambda,\hat \eta)}}_\text{augmentation}.
\end{aligned}
\label{eq::semi}
\end{equation}
The semi-parametric estimator contains an IPW component and an augmentation component,
so it is an augmented IPW estimator (see Appendix \ref{sec::eff} for more details). 
Semi-parametric theory ensures that this estimator is the most efficient estimator
when both the selection odds $\{O_r(L_r; \eta_r):r\in\mathcal{R}\}$ and the regression functions  $\{\mu_{\Xi,s}(L_r;\lambda_r):r\in\mathcal{R}\}$
are correctly specified.
Algorithm \ref{alg::semi} provides a Monte Carlo procedure to compute the semi-parametric estimator,
which is a combination of the recursive algorithm in Algorithm \ref{alg::IPW}
and the multiple imputation in Algorithm \ref{alg::impute2} in the supplementary materials. 
The key is to use the ancestor expression, which leads to a simpler form
of the semi-parametric estimator.
Note that similar to the regression adjustment estimator,
if the pattern graph is a tree graph, we can avoid using Algorithm~\ref{alg::semi}
to compute the estimator; see Appendix \ref{sec::SG}.


\begin{algorithm}[tb]
\caption{Monte Carlo approximation of the semi-parametric estimator} 
\label{alg::semi}
\begin{algorithmic}
\State Input models: $\{p(\ell_r|R=r; \hat \lambda_r), O_r(L_r;\hat \eta_r):r\in\mathcal{R}\}$.
\State 1. Apply the multiple imputation method (Algorithm \ref{alg::impute2} in the appendix) to obtain an approximation $\tilde m(\ell_r,r;\hat \lambda)$
for each $r$.
\State 2. For each $r$ and an ancestor $s\in {\sf Ans}_r$, compute
$$
\tilde\EIF_{s,r}(L,R)  =\tilde m(L_s,s; \hat \lambda)(I(R=s)-O_s(L_s; \hat \eta_s)I(R\in {\sf PA}_s)) \sum_{\zeta\in\Upsilon_{s,r}}\prod_{w\in\zeta, w< s} O_w(L_w; \hat \eta_w)
$$

\State 3. Compute the EIF as $\tilde\EIF(L,R) = \sum_{r\neq 1_d} \sum_{s\in{\sf Ans}_r} \tilde\EIF_{s,r}(L,R)$. 
\State 4. Compute the propensity score $\pi(L; \hat \eta)$ by Algorithm~\ref{alg::IPW}.
\State 5. Return: $\tilde \theta_{\sf semi}$  as 
$$
\tilde \theta_{\sf semi} = \frac{1}{n}\sum_{i=1}^n \frac{\theta(L_i)I(R_i=1_d)}{\pi(L_i; \hat \eta)} + \tilde\EIF(L_i,R_i).
$$
\end{algorithmic}
\end{algorithm}

\begin{remark}
In the pattern graph of the
CCMV restriction, arrows are in the form $1_d\rightarrow r$ for each $r\neq 1_d$.
In this case, $\Pi_r = \{r\}$ and $\Xi = r$, so 
$$
\mu_{\Xi,r} (\ell_r) = \frac{\E(\theta(L)I(R=1_d)|L_r = \ell_r)}{P(R=1_d|\ell_r)} =  \E(\theta(L)|R=1_d, L_r=\ell_r).
$$
Thus, the semi-parametric estimator in equation \eqref{eq::semi} 
is the same as the semi-parametric estimator in \cite{tchetgen2018discrete}.

\end{remark}

\subsubsection{Multiple robustness}	\label{sec::MR}



In many scenarios, a semi-parametric estimator
often exhibits a double robustness or multiple robustness property \citep{robins2000sensitivity,tsiatis2007semiparametric, seaman2018introduction}. 
We  demonstrate that our semi-parametric estimator in equation \eqref{eq::semi}
also enjoys a multiple robustness property.
Here, we assume that the parameters $\hat \lambda\overset{P}{\rightarrow}\lambda^*$
and $\hat \eta\overset{P}{\rightarrow}\eta^*.$
Note that  equation \eqref{eq::semi} can be factorized as
\begin{align*}
\mathcal{L}_{\sf semi}(L,R;\lambda^*,\eta^*)  &=\theta(L)I(R=1_d)+ \sum_{r\neq 1_d}\sum_{\Xi\in\Pi_r}\mathcal{L}_{\sf semi, \Xi}(L,R;\lambda^*,\eta^*),\\
\mathcal{L}_{\sf semi, \Xi}(L,R;\lambda^*,\eta^*)& = \theta(L)I(R=1_d)\prod_{s\in \Xi} O_s(L_s;\eta^*) + \EIF_\Xi(L,R; \lambda^*, \eta^*).
\end{align*}
We demonstrate the multiple robustness properties of each component $\mathcal{L}_{\sf semi, \Xi}(L,R;\lambda^*,\eta^*)$.
Note that we let $O_s(L_s)$ and $\mu_{\Xi,s}(L_s)$
denote the correct selection odds and regression function for each $s\in\Xi$
and each path $\Xi$, respectively.




\begin{theorem}[Multiple robustness]
Suppose that the selection odds model of $(L,R)$
factorizes with respect to a regular pattern graph $G$ and $p(\ell_r,r)>0$.
Let $r\in \mathcal{R}$ be a response pattern. 
For a path $\Xi \in \Pi_r$,
if either $ O_s(L_s;\eta^*) = O_s(L_s)$ or $\mu_{\Xi,s}(L_s;\lambda^*) = \mu_{\Xi,s}(L_s)$
for each $s\in \Xi$, then
$$
\E\left(\mathcal{L}_{\sf semi, \Xi}(L,R;\lambda^*,\eta^*)\right) = \theta_{\Xi}.
$$

\label{thm::MR}
\end{theorem}

Using the fact that $\theta = \theta_{1_d}  +  \sum_{r\neq 1_d}\sum_{\Xi\in\Pi_r}\theta_\Xi$, 
it is evident that if we can consistently estimate $\theta_\Xi$ for each $\Xi$,
we can estimate $\theta$ consistently.

Let $\mcM^O_s = \{O_s(\cdot;\eta^*) = O_s(\cdot)\}$ 
be the case where the selection odds of pattern $s$ is correctly specified. 
For $\Xi\in\Pi_r, r\neq 1_d$ and $s\in \Xi$,
let $\mcM^\mu_{\Xi,s} = \{\mu_{\Xi,s}(\cdot; \lambda^*) = \mu_{\Xi,s}(\cdot)\}$ be the case where $\mu_{\Xi,s}$ is correctly specified.
Theorem~\ref{thm::MR} shows that 
under the intersection of models
$$
\mcM_\Xi = \bigcap_{s\in \Xi} (\mcM^O_s\cup \mcM^{\mu}_{\Xi,s}),
$$
the quantity $\mathcal{L}_{\sf semi, \Xi}(L,R;\lambda^*,\eta^*)$
leads to a consistent estimator of $\theta_\Xi$, i.e.,
$$
\hat \theta_\Xi = \frac{1}{n}\sum_{i=1}^n\mathcal{L}_{\sf semi, \Xi}(L_i,R_i;\hat \lambda,\hat \eta)\overset{P}{\rightarrow} \theta_\Xi.
$$
Thus, to estimate $\theta = \sum_r \theta_r$,
we must select a model in
\begin{equation}
\mcM =\bigcap_{r\neq 1_d}  \bigcap_{\Xi\in\Pi_r} \bigcap_{s\in \Xi} (\mcM^O_s\cup \mcM^{\mu}_{\Xi,s}).
\label{eq::MR::full}
\end{equation}
If our model falls within $\mcM$, we have 
$
\hat \theta_{\sf semi}\overset{P}{\rightarrow} \theta.
$
This describes the multiple robustness property of the semi-parametric estimator in equation \eqref{eq::semi}.

Similar to a conventional multiply robust estimator \citep{tchetgen2018discrete}, $\hat \theta_{\sf semi}$ is a $\sqrt{n}$-rate efficient normal estimator of $\theta$
if for any path $\Xi$, 
$$
\sum_{s\in \Xi} \|\mu_{\Xi,s}(\cdot; \hat \lambda) - \mu_{\Xi,s}(\cdot)\|_{L_2(P)}\|O_s(\cdot; \hat \eta_s) - O_s(\cdot)\|_{L_2(P)}=  o_P\left(\frac{1}{\sqrt{n}}\right),
$$
where $\|f\|_{L_2(P)}  = (\int |f(\ell)|^2dP(\ell))^{1/2}$ is the $L_2(P)$ norm of a function $f$.
This occurs when all (L1-L4) and (R1-R3) conditions in Appendix \ref{sec::assumptions} hold.



\section{Discussion}	\label{sec::discussion}

In this paper, we,
introduce the concept of pattern graphs and use it to represent an identifying restriction for missing data problems.
Pattern graphs provide a new way to construct identifying restrictions.
We demonstrate that  pattern graphs  can be interpreted 
using a selection odds model or pattern mixture model. 
In addition, we propose various estimators using different modeling strategies and study statistical and computational properties with a pattern graph.
The theories developed in Section~\ref{sec::semi}
demonstrate the elegant association between
the semi-parametric theory and pattern graphs.
We believe that the pattern graph approach 
can provide
a new direction in  missing data research.
Below, we discuss  possible future directions that  arevworth pursuing.

\begin{itemize}

\item {\bf Choice of pattern graph.}
In this paper, we mainly focus on the theoretical analysis of pattern graphs and assume that
a pattern graph is given.
In practice, determining how to select a pattern graph  is an open problem. 
Since a pattern graph leads to an identifying restriction, 
it should be chosen based on background knowledge of how  missingness 
occurs.
In Appendix \ref{sec::data}, we provide a data analysis example
and attempt to choose a pattern graph based on  prior knowledge of the data generating process.
In this particular example, we use
the path selection interpretation of pattern graphs (Proposition \ref{lem::pathG} and related discussion)
to select a plausible pattern graph. 
Although this approach is reasonable for this particular data, it may not apply to other problems.
We plan to develop a general principle for selecting a pattern graph in future work.

\item {\bf Inference with multiple restrictions.}
Although a pattern graph may be derived from  scientific knowledge, sometimes
there may be uncertainties regarding the graph to be used.
As a result, there may be a set of possible graphs $\{G_1,\cdots,G_k\}$ that are reasonable.
In this scenario, determining how to perform statistical inference is an open question. 
One possible solution is to derive a nonparametric bound \citep{manski1990nonparametric, horowitz2000nonparametric}
or an uncertainty interval \citep{Vansteelandtetal06}
in which
we compute an estimator of each graph
and use the range of these estimators as an interval estimate.
Alternatively, one can consider a Bayesian approach that 
assigns a prior distribution over possible graphs and derives the posterior distribution
of the parameter of interest. 
The posterior mean behaves like a Bayesian model averaging estimator \citep{hoeting1999bayesian},
and the posterior distribution includes uncertainties from both estimation and graphs. 

\item {\bf MAR and conditional independence.}
The MAR restriction can be written as a pattern graph with ${\sf PA}_r = \mathcal{R}\backslash \{r\}$.
It is not a regular pattern graph; however, it still leads to a uniquely identified full-data distribution \citep{Gilletal97}.
This implies that pattern graphs that are not DAGs may still lead to an identifying restriction. 
Pattern graph factorization implies the following conditional independence: 
\begin{equation}
I(R=r) \perp L_{\bar r}|L_{ r}, R\in E_r,\quad E_r = \{r\}\cup {\sf PA}_r
\label{eq::Cind}
\end{equation}
for each $r$.
When $E_r = \mathcal{R}$, this is equivalent to the MAR restriction. 
The choice of $E_r$ is equivalent to the choice of the parents, which may provide
a way to study identifying restrictions beyond acyclic pattern graphs. 
Thus, studying the conditions on $E_r$ that lead to an identifiable full-data distribution
is a future direction that is worth pursuing.


\end{itemize}

%
%
%

\section*{Acknowledgement}
We thank Adrian Dobra, Mathias Drton, Mauricio Sadinle, Daniel Suen, Thomas Richardson for very helpful comments on the paper. 
This work is partially supported by 
NSF grant DMS 1810960 and DMS - 195278 and NIH grant U01 AG016976.


\appendix

\section{Sensitivity analysis}	\label{sec::sensitivity}

Sensitivity analysis is a common task in handling missing data \citep{little2012prevention}.
It aims to analyze the effect
of perturbing an identifying restriction 
on the final estimate, and also serves as a means of incorporating the
uncertainties of the identifying restriction into the inference.
Here, we introduce three approaches for sensitivity analysis based on pattern graphs.


\subsection{Perturbing selection odds}

The first approach involves perturbing the selection odds model. 
Using the concept of exponential tilting \citep{kim2011semiparametric, shao2016semiparametric,zhao2017semiparametric}, 
the selection odds model in equation \eqref{eq::Codds}
can be perturbed as 
$$
\frac{P(R=r|\ell)}{P(R\in {\sf PA}_r|\ell)} = \frac{P(R=r|\ell_r)}{P(R\in {\sf PA}_r|\ell_r)}e^{\delta_{\bar r}^T \ell_{\bar r}},
$$
where $\delta_{\bar r} \in \R^{|\bar r|}$  is a given vector
that controls the amount of perturbation. 
If we set $\delta_{\bar r}= 0$, this reduces to the usual graph factorization. 

When we use a logistic regression model, 
the exponential tilting approach 
leads to an elegant form of the selection odds:
\begin{equation}
\log \left(\frac{P(R=r|L)}{P(R\in {\sf PA}_r|L)}\right) = \log O_r(L_r) + \delta_{\bar r}^T L_{\bar r}= \gamma_r^T \tilde L,
\label{eq::SA}
\end{equation}
where $\gamma_r = (\beta_r, \delta_{\bar r})$ and $\tilde{L} = (1, L_r, L_{\bar r})$. 
Thus, computing the estimator of the propensity score $\hat \pi(L)$
is simple: we  modify Algorithm~\ref{alg::IPW}
by replacing $\hat O_r(L_r)$ by $\hat\gamma_r^T \tilde L$,
where $\hat \gamma_r = (\hat \beta_r, \delta_{\bar r})$. 
The recursive computation approach in Algorithm~\ref{alg::IPW} can be easily adapted
to this case.
Appendix \ref{sec::SA::PISA} 
provides a data example of  this concept.


\subsection{Perturbing pattern mixture models}


Alternatively, we can perturb the PMMs. 
From equation \eqref{eq::PMM}, the graph factorization of a PMM implies that 
$$
p(\ell_{\bar r}|\ell_r, R=r) = p(\ell_{\bar r}|\ell_r, R\in {\sf PA}_r),
$$
and we use the exponential tilting again to perturb it as
\begin{equation}
p(\ell_{\bar r}|\ell_r, R=r) = p(\ell_{\bar r}|\ell_r, R\in {\sf PA}_r) e^{\omega_{\bar r}^T \ell_{\bar r}}.
\label{eq::pert::PMM1}
\end{equation}
Again, $\omega_{\bar r} = 0$ implies that there is no perturbation,
which is the case where graph factorization is assumed to be correct.

Interestingly, perturbations on selection odds and on PMMs
are the same, as illustrated by the following theorem. 

\begin{theorem}
Let $r$ be a response pattern and $g(\ell_{\bar r})$ be any function of the unobserved entries 
and $p(\ell_r,r)>0$ for all $\ell_r\in \mathbb{S}_r$ and $r\in\mathcal{R}$.
Then the assumption
$$
\frac{P(R=r|\ell)}{P(R\in {\sf PA}_r|\ell)} = \frac{P(R=r|\ell_r)}{P(R\in {\sf PA}_r|\ell_r)} \cdot g(\ell_{\bar r})
$$
is equivalent to the assumption 
$$
p(\ell_{\bar r}|\ell_r, R=r) = p(\ell_{\bar r}|\ell_r, R\in {\sf PA}_r)\cdot  g(\ell_{\bar r}).
$$
\label{thm::perturb}
\end{theorem}

Theorem~\ref{thm::perturb} demonstrates that a perturbation
on the selection odds is the same as a perturbation on the PMMs.
This result is not limited to the exponential tilting approach: any other
perturbation, as long as the perturbation is only on unobserved variables, 
will lead to the same result. 

As  mentioned before, we generally use the multiple imputation procedure
to compute an estimator 
when a PMM factorization is used. 
This procedure must be modified when using the sensitivity analysis of equation \eqref{eq::pert::PMM1}. 
If $L$ is bounded and with a known upper bound $U$ such that $L_j\leq U_j$, 
we can then modify Algorithm~\ref{alg::impute2} by combining it with rejection sampling. 
We change steps 2-4 in Algorithm~\ref{alg::impute2} to the following two steps:
\begin{quote}
2.4' If $R_{\sf now}\neq 1_d$, return to 2-2; otherwise 
draw $V \sim {\sf Uni}[0,1]$.\\
2.5' If $V\leq \frac{e^{\omega^T_{\bar r}L_{\sf now, \bar r}}}{e^{\omega^T_{\bar r}U_{\bar r}}}$,
then update $L_i = L_{\sf now}$; otherwise return to 2-1.
\end{quote}
Step 2.5' states that with a probability of 
$\frac{e^{\omega^T_{\bar r}L_{\sf now, \bar r}}}{e^{\omega^T_{\bar r}U_{\bar r}}} = e^{\omega^T_{\bar r}(L_{\sf now, \bar r}-U_{\bar r})}$,
we accept this proposal. 
This additional rejection-acceptance step rescales the density
so that we are indeed sampling from \eqref{eq::pert::PMM1}. 
Note that it is possible to modify the algorithm using  Markov chain Monte Carlo (MCMC; \citealt{liu2008monte});
however, the computational cost of MCMC would be enormous, as
we would have to perform it for every observation. 



\subsection{Perturbing the graph}


In addition to performing sensitivity analysis on the selection odds and pattern mixture models,
we can consider perturbing the graph.
Before we proceed, we provide description of the number of identifying restrictions
that can be represented by regular pattern graphs.
Let
$M_d$ be the total number of distinct graphs 
that satisfy (G1-2) when there are $d$ variables subject to missingness.

\begin{proposition}
If all study variables in $L\in\mathbb{R}^d$ are subject to missingness, 
then there are 
$$
M_d = \prod_{k=0}^{d-1}  (2^{2^{d-k}-1}-1)^{{d\choose k} }
$$
distinct graphs satisfying conditions (G1-2) .
\label{lem::Lnumber}
\end{proposition}

The first few values of $M=M_d$ are as follows:
$$
M_1 =1,\quad M_2 = 7,\quad M_3 = 43561, \quad M_4 >10^{18}.
$$
Proposition~\ref{lem::Lnumber} demonstrates 
that the collection of regular pattern graphs is a rich class.
It contains an astronomical number of identifying restrictions
when only four variables are subject to missingness. 
Given the richness of this class,
we can examine the effect of perturbing the graph
on the final estimate.

Here, we formally describe our perturbation of a graph.
Suppose that $G$ is the graph used in our original analysis
that leads to an estimate $\hat \theta_G$.
We wish to know how $\hat\theta_G$ changes
if we slightly perturb $G$.
A simple perturbation is by
using graph $G'$ such that $G$
and $G'$ differ by only one edge. 


Let $G$ be a graph satisfying (G1-2). 
We define $\Delta_1 G$ to be the collection of graphs
such that 
$$
\Delta_1 G = \{G': |G'-G| = 1, \mbox{condition (G1-2) holds for $G'$}\},
$$
where $|G'-G| = 1$ represents the case in which the two graphs only differ by one edge (arrow). 
Namely, $\Delta_1 G$ is the collection of graphs satisfying (G1-2) 
and only differ from $G$ by one edge (arrow). 
The class $\Delta_1 G$ can be decomposed into
$$
\Delta_1 G = \Delta_{+1} G\cup \Delta_{-1} G,
$$
where 
\begin{align*}
\Delta_{+1} G &= \{G':G\subset G',  |G'-G| = 1, \mbox{condition (G1-2) holds for $G'$}\},\\
\Delta_{-1} G &= \{G':G'\subset G, |G'-G| = 1, \mbox{condition (G1-2) holds for $G'$}\}.
\end{align*}
Namely, $\Delta_{+1}G$ is the collection of graphs with one more edge than $G$,
whereas $\Delta_{-1}G$ is the collection of graphs with one less edge than $G$.

The following proposition provides an explicit characterization of $\Delta_{+1}G$ and $\Delta_{-1}G$. 
\begin{proposition}
Assume that $G$ is a regular pattern graph.
Let $s,r$ be vertices of $G$.
We define $G\oplus e_{s\rightarrow r}$ to be the graph where edge $s\rightarrow r$ is added
and $G\ominus e_{s\rightarrow r}$ to be the graph where edge ${s\rightarrow r}$ is removed.
Then 
\begin{align*}
\Delta_{+1}G &= \{G\oplus e_{s\rightarrow r}: s>r, s\notin {\sf PA}_r\},\\
\Delta_{-1}G &=  \{G\ominus e_{s\rightarrow r}: s\in {\sf PA}_r, |{\sf PA}_r|>1\}.
\end{align*}
\label{lem::Gpm}
\end{proposition}

Proposition~\ref{lem::Gpm} provides a simple description of the possible
perturbed graphs from $G$. 
$\Delta_{+1}G$ is the collection of graphs
in which we add an arrow from a potential parent (the set $\{s: s>r, s\notin {\sf PA}_r\}$ is the potential parent of $r$). 
In other words, the constraint of $\Delta_{+1}$ is that the added edge must preserve the partial order among patterns. 
The set $\Delta_{-1}G$ is the collection of graphs in which we drop one parent if there are at  least two possible parents.
Namely, the constraint of $\Delta_{-1}$ is that we can only remove an arrow if it is not the only arrow
pointing toward a pattern.
Given graph $G$, finding these two sets is straightforward: the first one can be obtained by
enumerating all possible edges that are not yet presented in $G$. 
To find all graphs in $\Delta_{-1}G$, we  identify all arrows pointing to a node with multiple parents;
each arrow represent a graph in $\Delta_{-1}G$.

\section{Acyclic pattern graphs and equivalence classes}	\label{sec::GPG}

In this section, we investigate the scenario of relaxing the regular pattern graph conditions (G1-2).
A pattern graph is called an {\bf acyclic pattern graph} if it satisfies (G1) and (DAG).
An acyclic pattern graph also leads to an identifying restriction.

%

\begin{theorem}
For a pattern graph $G$ that satisfies (G1) and (DAG) and $p(\ell_r,r)>0$ for all $\ell_r\in \mathbb{S}_r$ and $r\in \mathcal{R}$,
the following holds:
\begin{enumerate}
\item The selection odds model and pattern mixture model factorizations are equivalent. 
\item Graph factorization leads to an identifiable full-data distribution.
\end{enumerate}

\label{thm::GPG}
\end{theorem}

Namely, Theorem~\ref{thm::GPG} states that if we replace the descending property (G2; $s\rightarrow r$ implies $s>r$)
by the DAG condition (DAG), 
graph factorization still defines an identifying restriction. 
This is not a surprising result because using PMM factorization,
as long as the source is identifiable (i.e., $p(\ell|R=1_d)$ is estimatable), 
its children and all descendants are identifiable.
Although an  acyclic pattern graph defines an identifying restriction,
it may be difficult to interpret the implied
restriction.



\begin{figure}
\includegraphics[height=1.2in]{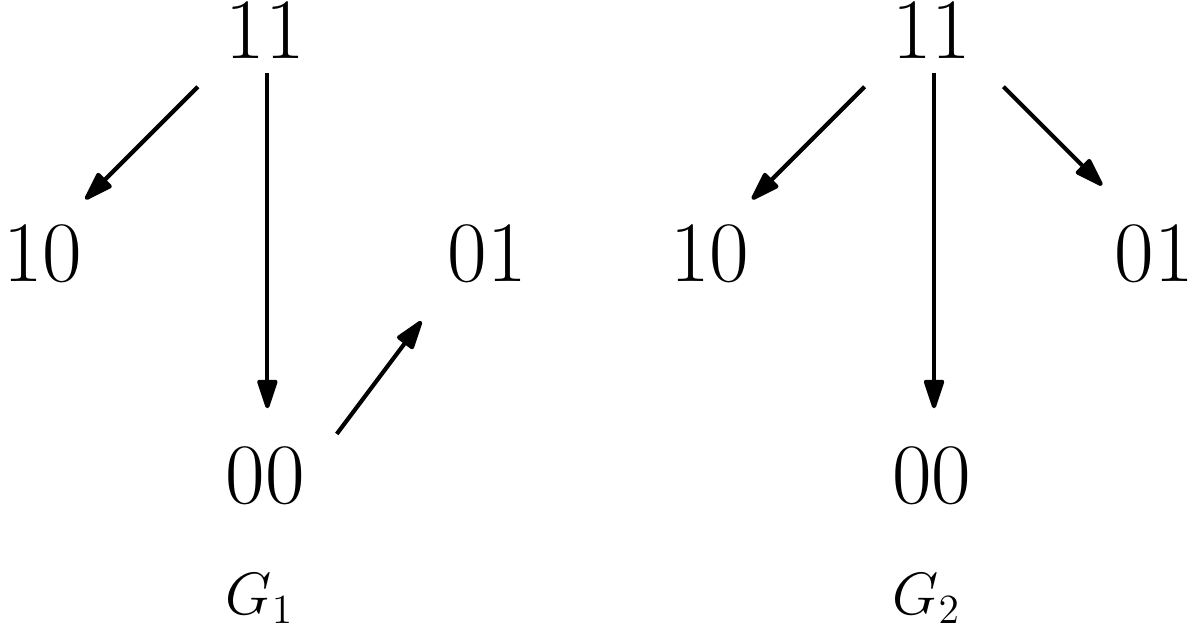}
\qquad
\includegraphics[height=1.2in]{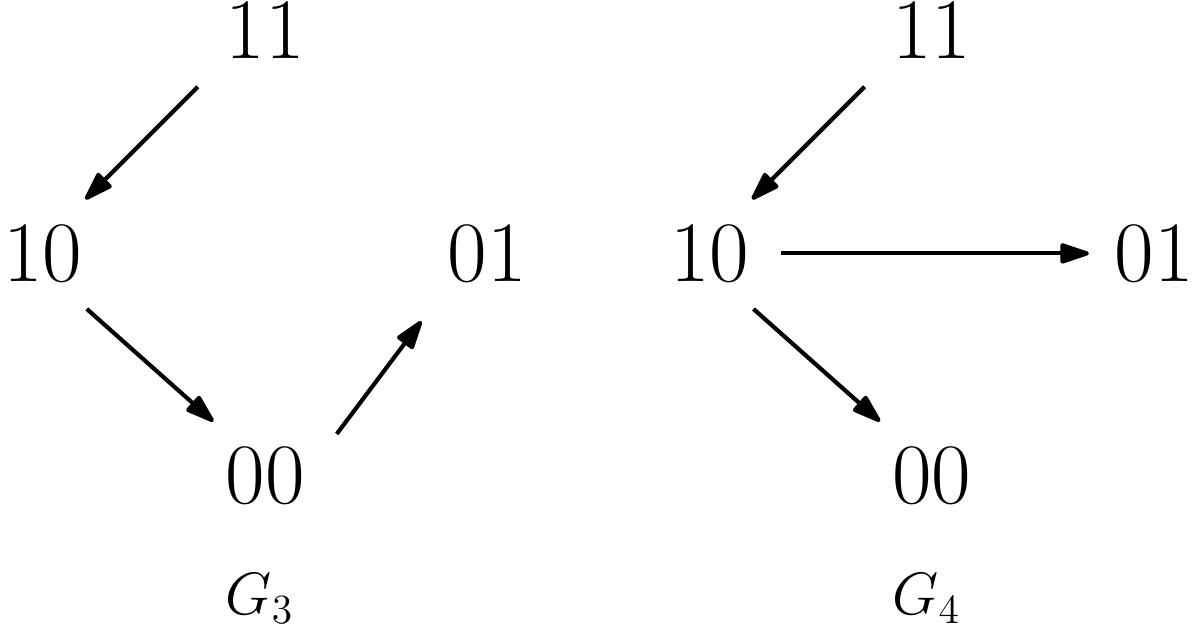}
\caption{Four acyclic pattern graphs.
Only $G_2$ is a regular pattern graph; the other three graphs are not regular.
All full-data distributions of $G_1$ and $G_2$
are the same, so they belong to the same equivalence class. 
Graphs $G_3$ and $G_4$ belong to another equivalence class,
and there is no regular pattern graph that belongs to the same equivalence class
containing $G_3$ and $G_4$.}
\label{fig::equiv1}
\end{figure}

Two graphs are equivalent if the implied full-data distributions are the same. 
An equivalence class is a collection of graphs that are all equivalent. 
This idea is similar to the Markov equivalence class in graphical model literature 
\citep{andersson1997characterization, gillispie2002size, ali2009markov}.
Figure~\ref{fig::equiv1} presents four examples of acyclic pattern graphs (note that $G_2$ is also a regular pattern graph),
and
they form two equivalence classes:
$G_1,G_2$ are equivalent and $G_3, G_4$ are equivalent. 
Although $G_1$ is not a regular pattern graph, it represents the same full-data distribution
as a regular pattern graph $G_2$. Thus, some acyclic pattern graphs
are equivalent to regular pattern graphs.
However, there are cases in which acyclic pattern graphs 
are different from regular pattern graphs.
Graphs $G_3,G_4$ form another equivalent class, but there is no
regular pattern graph in the same class.

The example in Figure~\ref{fig::equiv1} motivates us to investigate graphical criteria
leading to the equivalence of two acyclic pattern graphs.
The following theorem provides a graphical criterion for this purpose.

%




\begin{theorem}
Let $G$ be an acyclic pattern graph
and $r,s$ be two patterns
such that $s\notin{\sf PA}_r$.
Graph $G$ is equivalent to graph
$
G' = G\oplus e_{s\rightarrow r}\ominus\{e_{\tau\rightarrow r}: \tau \in {\sf PA}_r\}
$
if 
the following conditions hold:
\begin{enumerate}
\item {\bf (blocking)} All paths from $1_d$ to $ r$ intersect $s$. 
\item {\bf (uninformative)} For any pattern $q$ that is on a path from $s$ to $r$,  $q<r$.
\end{enumerate}

\label{thm::equiv}
\end{theorem}


Theorem~\ref{thm::equiv} provides a graphical criterion for how to construct
an equivalent graph. 
It also provides a sufficient condition for the equivalence of two graphs.
The two equivalence classes in Figure~\ref{fig::equiv1} can be obtained by applying Theorem~\ref{thm::equiv}. 
This theorem states that if we can identify a pattern $s$
such that 
$s$ blocks all paths from the source to $r$ (blocking condition) and all descendants on a path from $s$ to $r$
do not provide any information on the missing variables of $r$ (uninformative condition),
then we can remove all arrows to $r$ and replace them with
an arrow from $s$ to $r$.

\begin{figure}
\center
\includegraphics[height=1.2in]{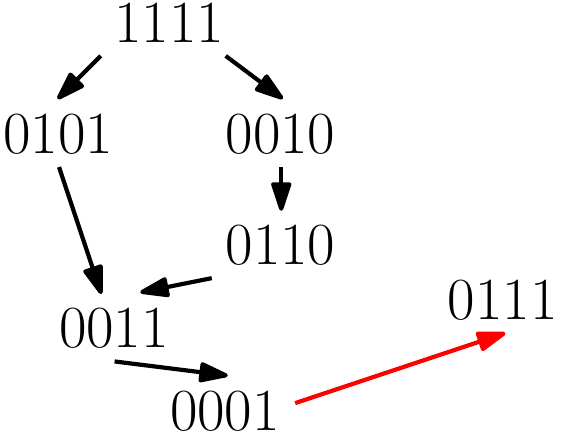}
\includegraphics[height=1.2in]{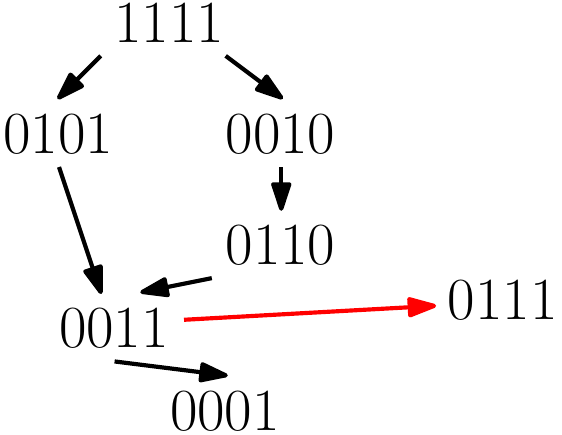}
\includegraphics[height=1.2in]{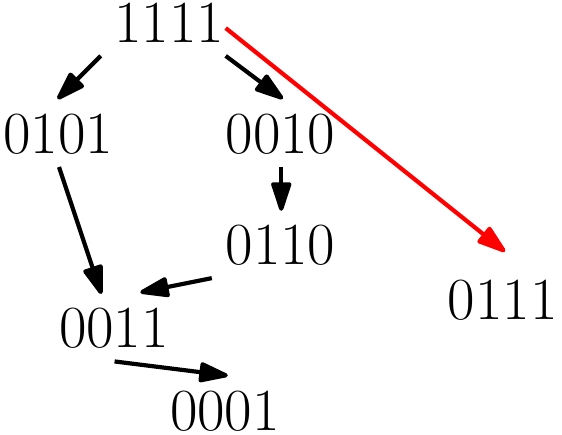}
\caption{Three acyclic pattern graphs that are all equivalent.
This is an example of four variable and  seven possible patterns.
Three equivalent graphs are displayed.
Note that the only difference is the location of the red arrow. 
Equivalence is implied by Theorem~\ref{thm::equiv}: patterns $0011$ and $1111$
both satisfy all conditions in Theorem~\ref{thm::equiv}.
}
\label{fig::equiv2}
\end{figure}

Figure~\ref{fig::equiv2} presents an example of applying Theorem~\ref{thm::equiv}
to obtain equivalent graphs. 
Starting from the left panel, we can see that patterns $0011$ and $1111$
both satisfy all conditions in Theorem~\ref{thm::equiv},
which leads to the graphs in the middle and right panels.

Note that Theorem~\ref{thm::equiv} does not provide necessary conditions for
the equivalence between two graphs. 
There may be other examples in which two acyclic pattern graphs are equivalent,
but do not satisfy Theorem~\ref{thm::equiv}.
We leave this for future work.






%
%

%
%
%

\section{Data Analysis}	\label{sec::data}

To demonstrate the applicability of pattern graphs, we use the Programme for International Student Assessment (PISA)
data from the year 2009\footnote{The data can be obtained from \url{http://www.oecd.org/pisa/data/pisa2009database-downloadabledata.htm}}.
We focus on Germany (country code: 276) as there is a higher proportion of missing entries for Germany's students. 
There are a total of 4,979 students in Germany's dataset. 
We consider the following three study variables: \texttt{MATH}: the plausible score of mathematics,
\texttt{FA}:  whether the father has higher education or not, \texttt{MA}, whether the mother has higher education or not. 
There are five plausible scores of mathematics (due to five different item response models to balance the fact that 
students may be taking different exams), and we use their average.
Variables \texttt{FA} and \texttt{MA} are binary variables (we use H/L to avoid confusion with the response indicator)
such that H represents yes (with a college degree or higher) and L represents no. 
Missingness occurs in the variables \texttt{FA} and \texttt{MA} while the mathematics scores are always observed.
Note that the original data contains finer categories for educational level of the father/mother; if any of them are missing, we treat variable as missing. 
The distribution of missingness is presented in Table~\ref{tab::missing}.

\begin{table}
\center
\begin{tabular}{rrrrr}
$(R_{\texttt{FA}}, R_{\texttt{MA}})=$&11 &10  & 01 & 00   \\
\hline
$n=$& 3282 &230  & 340 & 1126   \\
Proportion$=$&$ 65.9\%$ &$ 4.6\%$  &$6.8\%$  & $22.6\%$  
\end{tabular}
\caption{The distribution of missingness in the PISA data.}
\label{tab::missing}
\end{table}



\begin{figure}
\center
\includegraphics[width=1.2in]{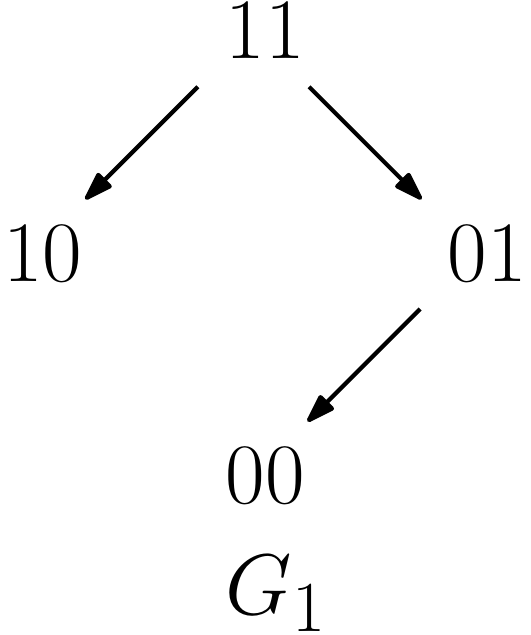}\qquad
\includegraphics[width=1.2in]{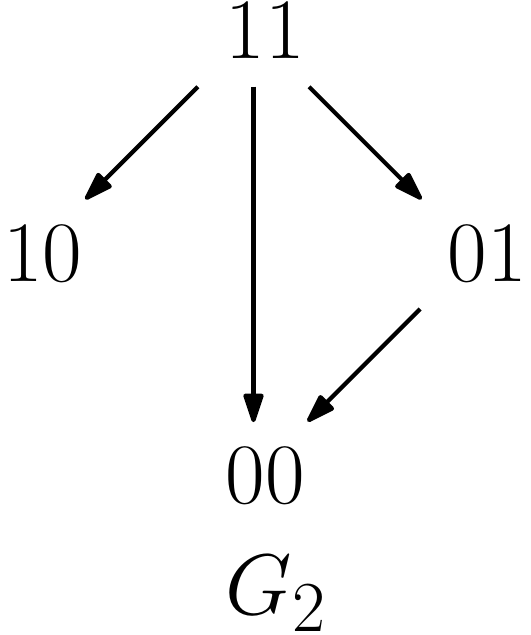}
\includegraphics[height=2in]{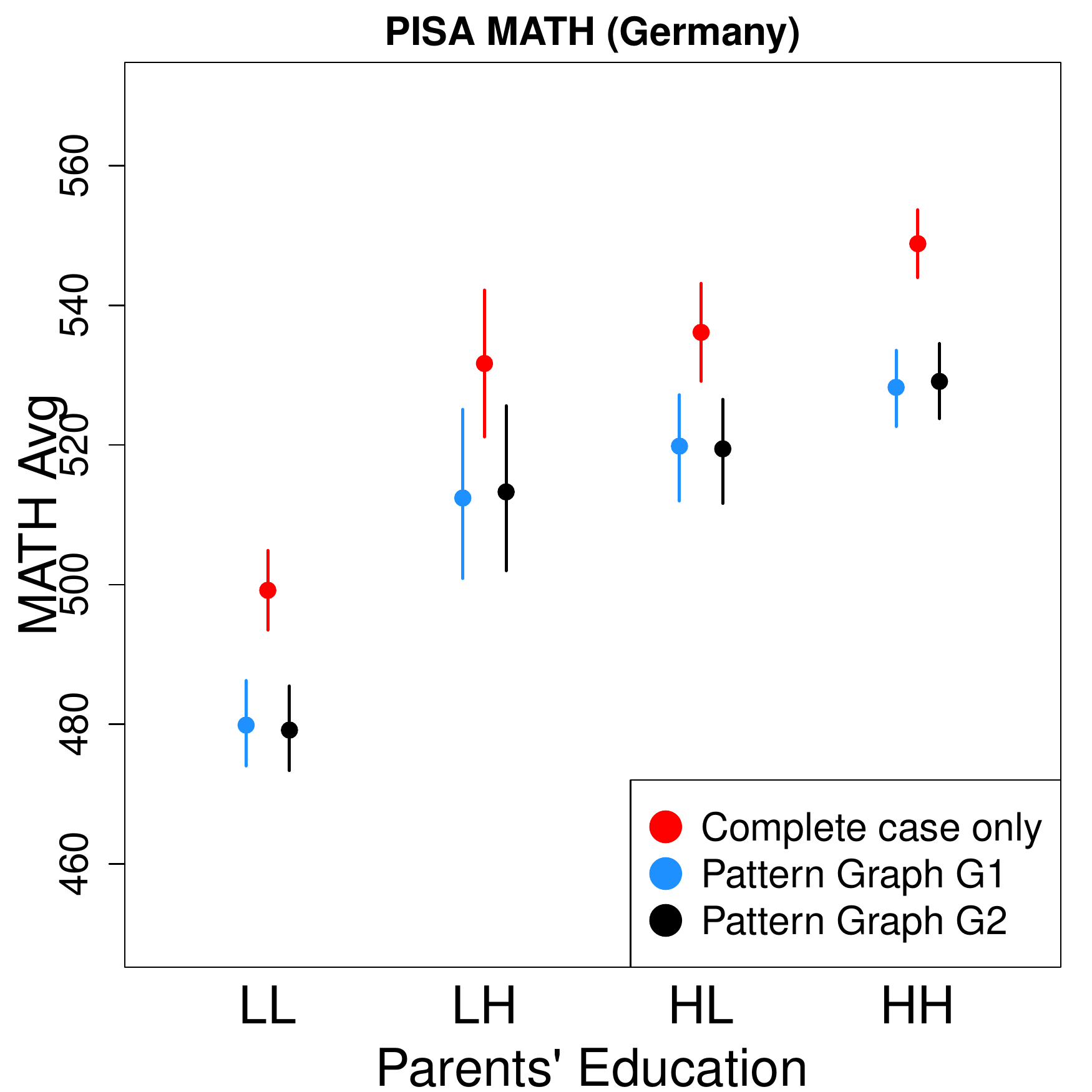}
\caption{{\bf Left and middle:} Two pattern graphs corresponding to the response pattern of \texttt{FA} and \texttt{MA}
in the PISA data.
The choice of a pattern graph reflects our prior knowledge of the generation process of the missing pattern.
{\bf Right:} Average math score of students with different parent education levels under complete-case analysis (red),
patten graph $G_1$ (blue), and pattern graph $G_2$ (black).
}
\label{fig::PISA01}
\end{figure}

Here, we present  a possible approach of choosing a pattern graph using prior knowledge of the data.
Variables \texttt{FA} and \texttt{MA}  are collected by a questionnaire before a student takes the  exam. 
Suppose that the question asking about the father's education precedes asking about the mother's education.
In addition, suppose that if a student chooses to report \texttt{FA} and moves to the question about the mother's education, he or she will not
change his/her mind to remove the value of \texttt{FA}.


Before we ask a student a question,  
there is an answer to that question.
Thus, every individual starts with a response pattern $(1,1)$ in the beginning.
%
When we ask the first question (the father's education level), the student may answer it or not.
If the student answers it, the pattern remains $(1,1)$ and the student moves to the second question.
If the student does not answer it, then the pattern becomes $(0,1)$ and the student moves to the second question.
Before asking the second question, the response pattern is $(R_\texttt{FA}, 1)$. 
If the student  answers the second question, then the pattern remains $(R_\texttt{FA}, 1)$;
however, if the student does not answer it, the pattern becomes $(R_\texttt{FA}, 0)$.
To sum up, there are four possible scenarios and each  can be represented by a particular path:
\begin{align*}
\mbox{Answer \texttt{FA} and then answer \texttt{MA}}&\Rightarrow11\triangleright11\triangleright11\\
&\Rightarrow \mbox{path = }11\rightarrow 11\\
\mbox{Answer \texttt{FA} and then not answer \texttt{MA}}&\Rightarrow11\triangleright11\triangleright10\\
&\Rightarrow \mbox{path = }11\rightarrow 10\\
\mbox{Not answer \texttt{FA} but then answer \texttt{MA}}&\Rightarrow11\triangleright01\triangleright01\\
&\Rightarrow \mbox{path = }11\rightarrow 01\\
\mbox{Not answer \texttt{FA} and then not answer \texttt{MA}}&\Rightarrow11\triangleright01\triangleright00\\
&\Rightarrow \mbox{path = }11\rightarrow 01\rightarrow 00.
\end{align*}
The notation $\triangleright$ denotes the decision of whether to answer one question;
$r_1 \triangleright r_2$ becomes an arrow in a DAG when $r_1\neq r_2$.
The only exception is the scenario in which
$1_d\triangleright1_d\triangleright\cdots\triangleright1_d$; in this case, we denote it as $1_d\rightarrow 1_d$.
We do not have the arrow $10\rightarrow 00$ because the decision to report \texttt{FA} precedes the decision to report \texttt{MA}.
Using the path selection interpretation, the graph $G_1$ in Figure~\ref{fig::PISA01}
is a reasonable pattern graph that contains all these scenarios.
Now, if we include a new scenario in which the individual can skip any questions about the parents' education at the same time,
this corresponds to the path $11\rightarrow 00$, so the graph $G_2$ in Figure~\ref{fig::PISA01}
is a plausible pattern graph in this case.
Although the above procedure provides a simple and perhaps interpretable way to select a pattern graph,
it should be emphasized that this procedure is merely a tool for selecting a plausible pattern graph and
is not a model of the  mechanism of how an individual responds to the questions.

With a given pattern graph,
we study the students' average math scores
under different parents' education levels (\texttt{FA}, \texttt{MA}).
Figure~\ref{fig::PISA01} presents the results using both $G_1$ (blue) and $G_2$ (black),
and the result using a complete-case only (red) as a reference.
We use the IPW estimator with a logistic regression model for the selection odds and compute the uncertainty using the (empirical) bootstrap. 
The intervals are 95\% confidence intervals. 
We observe that both $G_1$ and $G_2$ produce very similar results,
and the complete-case analysis indicate a higher average score across all groups. 
Note that using both $G_1$ and $G_2$ in the analysis can be viewed as a sensitivity analysis
in which we perturb the underlying mechanism (graphs) to investigate the effect on the final estimate.

\subsection{Sensitivity analysis on  PISA data}	\label{sec::SA::PISA}

\begin{figure}
\center
\includegraphics[height=2in]{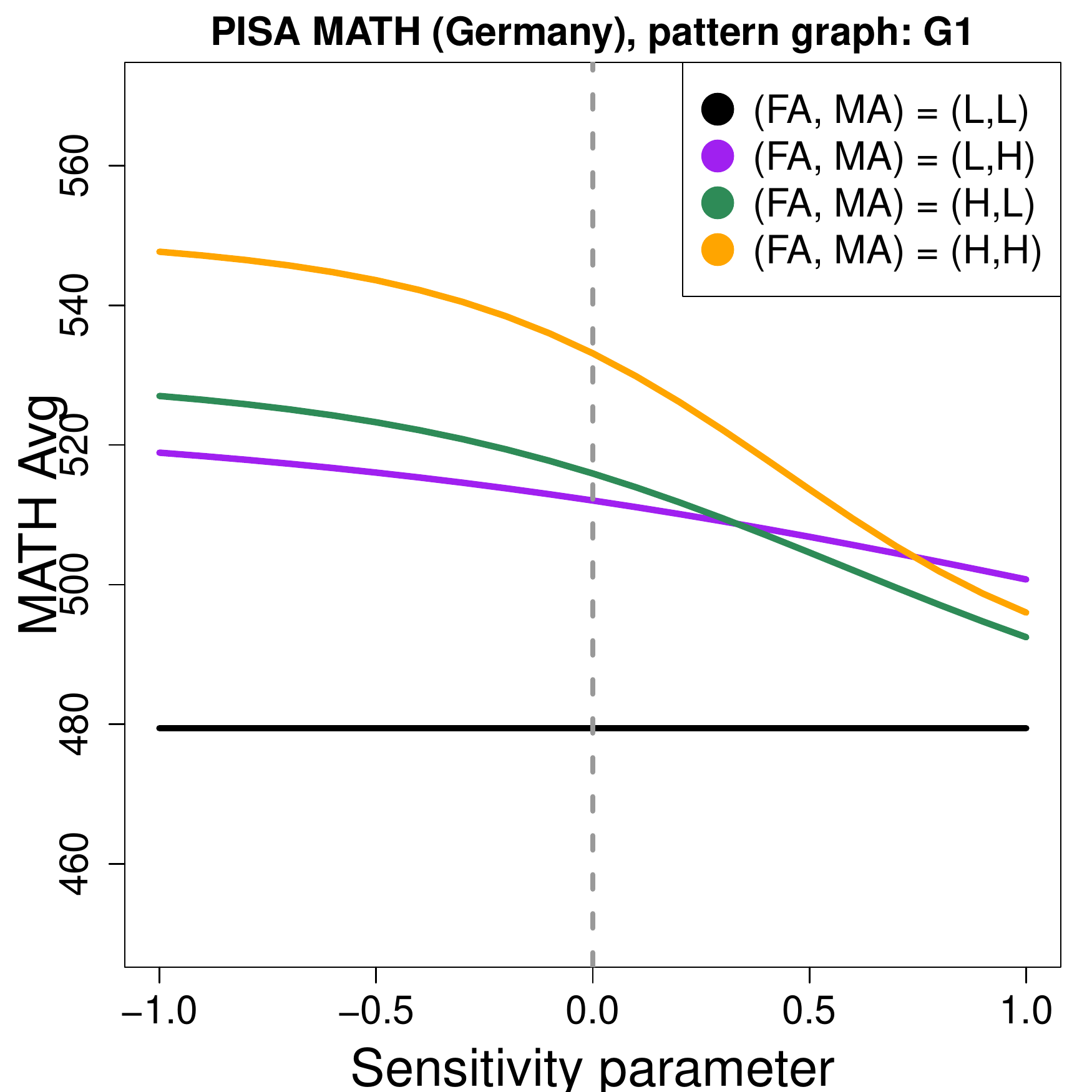}
\includegraphics[height=2in]{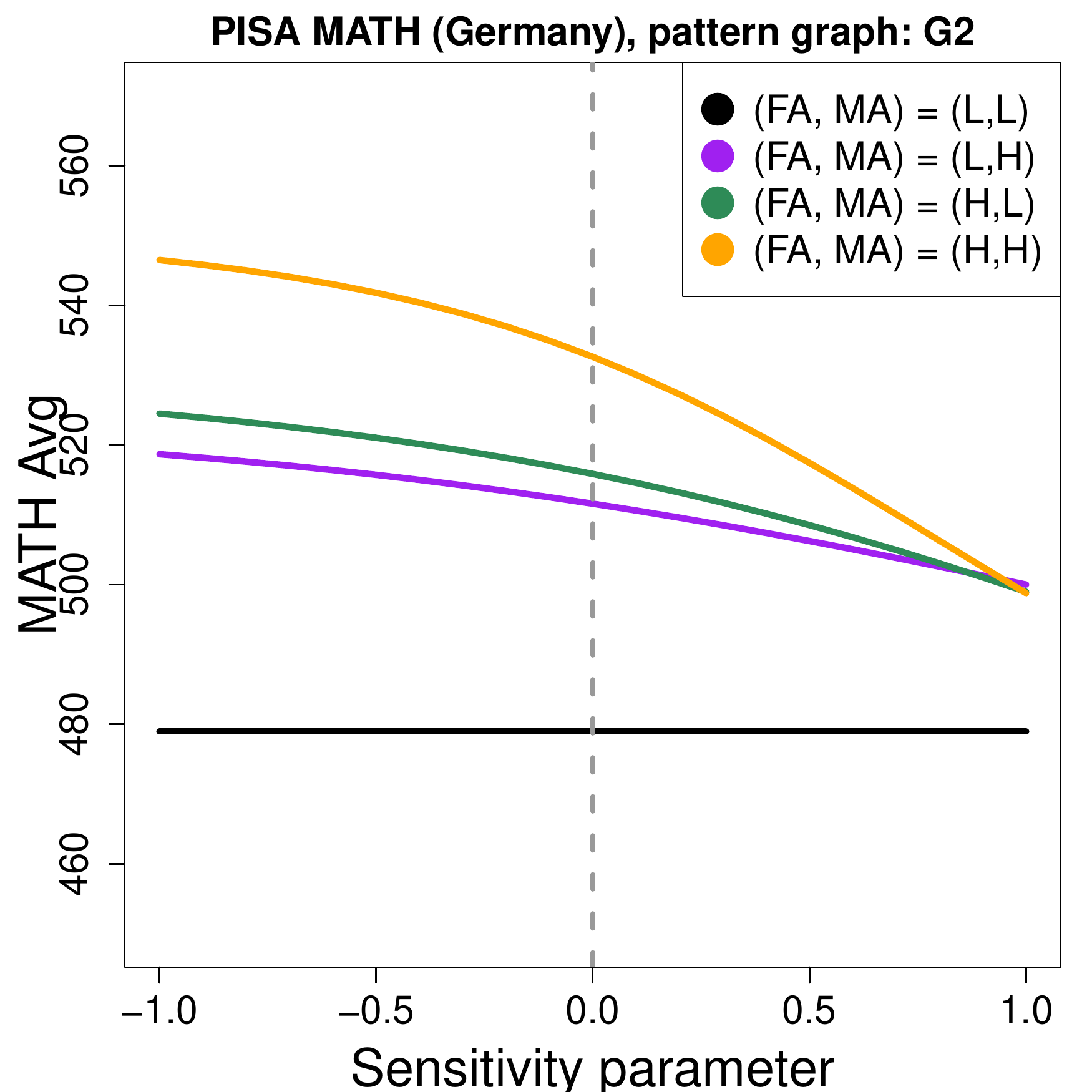}
\caption{
Sensitivity analysis of the pattern graph by exponential tilting. 
We use the exponential tilting idea in equation \eqref{eq::SA}
and examine how the result changes with respect to different values of
the sensitivity parameter. 
}
\label{fig::PISA02}
\end{figure}

For  completeness of analysis, we perform a simple sensitivity analysis on the PISA data
by the exponential tilting approach introduced in Section~\ref{sec::sensitivity}. 
We use the same sensitivity parameter for all patterns and all values, i.e., 
every element of $\delta_{\bar r} $ in equation \eqref{eq::SA} is identical.
Note that because only \texttt{FA} and \texttt{MA} are subject to missingness,
the sensitivity parameter only applies to these two variables.

Figure~\ref{fig::PISA02} presents the average math score when we vary the sensitivity parameter
in both graphs $G_1$ and $G_2$.
In both panels, we observe that group $(L,L)$ is unaffected by the sensitivity parameter.
This is because when both \texttt{FA} and \texttt{MA} are L (the binary representation of L is $0$ and H is $1$), 
the sensitivity parameter does not affect any odds ($L_{\bar r}^T\delta_{\bar r} = 0$ when $L_{\bar r}= 0$). 
Group $(H,H)$ is strongly influenced by the sensitivity parameter
because both variables are non-zero; thus, the effect is strongest. 
In most cases (except case $(L,L)$), we see a decreasing trend.
This can be understood by comparing it to the complete-case analysis (red dots in Figure~\ref{fig::PISA01}). 
When we only use complete data, all values are higher than pattern graphs. 
A small value (negatively large) of the sensitivity parameter provides low selection odds in the graph,
leading to a result that is similar to the complete-case analysis. 
This is why a decreasing trend is observed.

\section{Derivation of Example~1 (conditional MAR)}	\label{sec::ex02}

Recall that in Example~\ref{ex::ex02}, $L = (Z,Y_1,Y_2,Y_3)$ and $R_1$ is the response indicator of $Z$
and $R_{2},R_3,R_4$ are the response indicators of $Y_1,Y_2,Y_3$
and $T = R_2+R_3+R_4$ is the dropout time. 
Let $R_z= R_1$ be the response indicator of $Z$.
For two patterns $r_1,r_2$
we use the notation $r_1\lor r_2$ to denote $r_1$ or $r_2$. 

We present the result for the case in which $Z$ is observed.
The case in which $Z$ is unobserved can be derived in a similar manner.

{\bf Case $T=1$.}
In the case of observing $Z$, the selection odds model implies
\begin{align*}
P(R=1100|L) &= P(R_z=1,T=1|L) \\
&= P(R=1110\lor 1111|L)\frac{P(R=1100|Z,Y_1)}{P(R=1110\lor 1111|Z,Y_1)}\\
&= P(R_z=1,T=2\lor 3|L)\frac{P(R_z=1,T=1|Z,Y_1)}{P(R_z=1, T=2\lor3|Z,Y_1)}\\
&= P(R_z=1,T=2\lor 3|L)\frac{P(T=1|R_z=1,Z,Y_1)}{P(T=2\lor3|R_z=1,Z,Y_1)}.
\end{align*}
Dividing both sides by $P(R_z=1|L)$, we obtain
$$
P(T=1|R_z=1,L) = P(T=2\lor 3|R_z=1,L)\frac{P(T=1|R_z=1,Z,Y_1)}{P(T=2\lor3|R_z=1,Z,Y_1)}.
$$

Using the fact that $1 = P(T=1|R_z=1,L) + P(T=2\lor 3|R_z=1,L)$, we have
\begin{align*}
1& = P(T=2\lor 3|R_z=1,L)\left(1+\frac{P(T=1|R_z=1,Z,Y_1)}{P(T=2\lor3|R_z=1,Z,Y_1)}\right)\\
& =  P(T=2\lor 3|R_z=1,L) \cdot \frac{1}{P(T=2\lor3|R_z=1,Z,Y_1)}
\end{align*}
so $P(T=2\lor 3|R_z=1,L) = P(T=2\lor 3|R_z=1,Z,Y_1)$,
which further implies 
$$
P(T=1|R_z=1,L) = P(T=1|R_z=1,Z,Y_1),
$$
the conditional MAR of $T=1$ given $Z=1$.

{\bf Case $T=2$.}
The selection odds model implies that 
\begin{align*}
P(R=1110|L) &= P(R_z=1,T=2|L) \\
&= P(R=1111|L)\frac{P(R=1110|Z,Y_1,Y_2)}{P(R= 1111|Z,Y_1,Y_2)}\\
&= P(R_z=1,T= 3|L)\frac{P(R_z=1,T=2|Z,Y_1,Y_2)}{P(R_z=1, T=3|Z,Y_1,Y_2)}\\
&= P(R_z=1,T=3|L)\frac{P(T=2|R_z=1,Z,Y_1,Y_2)}{P(T=3|R_z=1,Z,Y_1,Y_2)}.
\end{align*}
Dividing both sides by $P(R_z=1|L)$, we obtain
$$
P(T=2|R_z=1,L) = P(T=3|R_z=1,L)\frac{P(T=2|R_z=1,Z,Y_1)}{P(T=3|R_z=1,Z,Y_1)}.
$$

The case of $T=1$ also implies that 
$$
P(T=1|R_z=1,L) = P(T=1|R_z=1,Z,Y_1) = P(T=1|R_z=1,Z,Y_1,Y_2).
$$ 
Thus, 
using the equality $1 = P(T=1|R_z=1,L) + P(T=2|R_z=1,L)+P(T=3|R_z=1,L)$ again, we have
\begin{align*}
1& = P(T=1|R_z=1,Z,Y_1,Y_2) \\
&\quad+ P(T=3|R_z=1,L)\left(1+\frac{P(T=2|R_z=1,Z,Y_1,Y_2)}{P(T=3|R_z=1,Z,Y_1,Y_2)}\right)
\end{align*}
Using $1- P(T=1|R_z=1,Z,Y_1,Y_2) = P(T=2\lor 3|R_z=1,Z,Y_1,Y_2)$, 
the above equality becomes
$$
P(T=2\lor 3|R_z=1,Z,Y_1,Y_2) = P(T=2\lor 3|R_z=1,Z,Y_1,Y_2) \frac{P(T=3|R_z=1,L)}{P(T=3|R_z=1,Z,Y_1,Y_2)},
$$
which implies $P(T=3|R_z=1,L) = P(T=3|R_z=1,Z,Y_1,Y_2)$. 
Using the fact that 
\begin{align*}
1&= P(T=1|R_z=1,L) + P(T=2|R_z=1,L) + P(T=3|R_z=1,L) \\
&= P(T=1|R_z=1,Z,Y_1,Y_2) + P(T=2|R_z=1,Z,Y_1,Y_2) \\
&\quad+ P(T=3|R_z=1,Z,Y_1,Y_2) ,
\end{align*}
we conclude that 
$P(T=2|R_z=1,L) = P(T=2|R_z=1,Z,Y_1,Y_2)$, which proves the case of $T=2$.

Note that $T=3$ is a trivial case and is thus omitted.
Therefore, the above analysis demonstrates that the graph in Example~\ref{ex::ex02}
implies 
$$
P(T=t|R_z=1,L) = P(T=t|R_z=1,Z,Y_1,\cdots,Y_t).
$$
The case of unobserved $Z$ can be derived in a similar manner by replacing $R_z=1$ by $R_z=0$
and removing all conditioning on $Z$.
Thus, we also have 
$$
P(T=t|R_z=0,L) = P(T=t|R_z=0,Y_1,\cdots,Y_t).
$$

Note that the graph in Example~\ref{ex::ex02} can be generalized to cases in which there are more time points.
The pattern graph will correspond to similar conditional MAR assumptions.

\section{Computation: logistic regression}	\label{sec::logistic}
In Theorem~\ref{thm::Codds},
a key quantity for the IPW estimator 
is $Q_r(L) = \frac{P(R=r|L)}{P(R=1_d|L)}$ and Proposition \ref{lem::pathG}
presents a simple form for $Q_r(L)$.
With logistic regression,
we can further express
$Q_r(L)$ in an elegant way.

\begin{proposition}
Assume that $(L,R)$ factorizes with respect to graph $G$,
and let $Q_r(L)$ be defined as in Theorem~\ref{thm::Codds}.
We assume a logistic regression model for the selection odds as equation \eqref{eq::logistic}
and denote $\beta_{[r]} \in \mathbb{R}^{1+d}$ as $\beta_{[r], r} = \beta_r$ and $\beta_{[r], \bar r} = 0$.
Namely, $\beta_{[r]}$ is the vector $\beta_{r}$ augmented with $0$'s on the coordinates of unobserved patterns.
Then
$$
Q_r(L) 
= \sum_{\Xi\in \Pi_r} \exp\left(\tilde L^T\sum_{s\in\Xi}\beta_{[s]}\right).
$$
Thus, equation \eqref{eq::naive}  becomes
$$
\pi(L) = \frac{1}{\sum_r \sum_{\Xi\in \Pi_r} \exp\left(\tilde L^T\sum_{s\in\Xi}\beta_{[s]}\right)}.
$$

\label{lem::path}
\end{proposition}

Using the path selection interpretation in Section~\ref{sec::Codds},
each path contributes the amount of $\exp\left(\tilde L^T\sum_{s\in\Xi}\beta_{[s]}\right)$ to $Q_r(L)$,
thus, the quantity $\sum_{s\in\Xi}\beta_{[s]}$ can be interpreted as a path-specific 
parameter in the logistic regression model.
The intuition behind this is that the augmented parameter has the following useful property:
$$
\tilde L^T \beta_{[r]} = \tilde L_r ^T \beta_r.
$$
As a result, using the form of $ \beta_{[r]} $ simplifies the representation.

%


\begin{example}
Consider an example in which we have three study variables and a total of four
possible patterns: $001,101,011,111$ (the third variable is always observed). 
Suppose that there are four arrows: $111\rightarrow 101$, $111\rightarrow 001$, $111\rightarrow011\rightarrow 001$. 
In this case, 
$$
\Pi_{001} = \{(001,111), (001,011,111)\},\quad\Pi_{101} = \{(101,111)\},\quad\Pi_{011} = \{(011,111)\}.
$$
For a vector of study variables $L\in\mathbb{R}^3$, and the corresponding $\tilde L= (1,L_1,L_2,L_3)^T$, and the parameters $\beta_{[r]}$ are
$$
\beta_{[001]} = \begin{pmatrix}
\beta_{001,1}\\
0\\
0\\
\beta_{001,2}
\end{pmatrix},
\beta_{[011]} = \begin{pmatrix}
\beta_{011,1}\\
0\\
\beta_{011,2}\\
\beta_{011,3}
\end{pmatrix},
\beta_{[101]} = \begin{pmatrix}
\beta_{101,1}\\
\beta_{101,2}\\
0\\
\beta_{101,3}
\end{pmatrix},
\beta_{[111]} = \begin{pmatrix}
\beta_{111,1}\\
\beta_{111,2}\\
\beta_{111,3}\\
\beta_{111,4}
\end{pmatrix}.
$$
Thus, $O_{001}(L) = O_{001}(L_{001})$ only depends on the last variable, which implies
\begin{align*}
Q_{001}(L) = \exp\left(
\tilde L^T \beta_{[001]}
\right) + 
\exp\left(\tilde L^T (\beta_{[001]}+\beta_{[011]})\right).
\end{align*}
The other two cases are very simple: $Q_{011}(L) = \exp(\tilde L^T\beta_{[011]}), Q_{101}(L) = \exp(\tilde L^T\beta_{[101]})$. 
With these quantities, we can compute 
\begin{align*}
\pi(L) &= \frac{1}{1+Q_{011}(L)+Q_{101}(L)+Q_{001}(L)}\\
&=\frac{1}{1+e^{\beta_{001}^T \tilde L_{001}}+e^{\beta_{001}^T\tilde{L}_{001} + \beta_{011}^T\tilde{L}_{011}}+e^{\beta_{011}^T\tilde{L}_{011}}+e^{\beta_{101}^T\tilde{L}_{101}}}.
\end{align*}
If we have estimators $\hat\beta_{r}$ for each $r$,
the estimated propensity score is 
$$
\hat \pi(L) = \frac{1}{1+e^{\tilde L^T\hat\beta_{[001]}}+e^{\tilde L^T(\hat\beta_{[001]}+ \hat\beta_{[011]})}+e^{\tilde L^T\hat\beta_{[011]}}+e^{\tilde L^T\hat\beta_{[101]}}}.
$$ 


\end{example}

\begin{proof}[ of Proposition~\ref{lem::path}]

With the logistic regression model, the selection odds can be written as
$$
O_r(L_r) = e^{\tilde L_r^T \beta_r}.
$$
Using the fact that 
$$
\tilde L^T \beta_{[r]} = \tilde L_r^T \beta_r,
$$
we can rewrite the odds as 
$$
O_r(L_r) = \exp(\tilde L^T \beta_{[r]}).
$$
Using Proposition~\ref{lem::pathG},
equation \eqref{eq::path1} can then be rewritten as 
\begin{align*}
Q_r(L) &= \sum_{\Xi\in \Pi_r} \prod_{s\in \Xi} \exp(\tilde L^T \beta_{[s]})\\
& = \sum_{\Xi\in \Pi_r} \exp\left(\tilde L^T \left(\sum_{s\in \Xi} \beta_{[s]}\right)\right),
\end{align*}
which implies the final assertion
$$
\pi(L) = \frac{1}{\sum_r Q_r(L)} = \frac{1}{\sum_r \sum_{\Xi\in \Pi_r} \exp(\tilde L^T (\sum_{s\in \Xi} \beta_{[s]}))}.
$$

\end{proof}

\section{More about regression adjustment}

\subsection{Model congeniality}	\label{sec::MC}

The  parametric model in regression adjustment is constructed from modeling the observed-data distribution first
and then deriving the model on the conditional expectation (regression function).
One may wonder whether we can directly start with a model on the conditional expectation, i.e.,
make a parametric model $m(\ell_r,r;\eta_r)$ for each $r$.
We do not recommend this approach because these models may not be variationally independent. 
For instance, suppose that pattern $s$ is a parent of pattern $r$. 
Then, the regression model $m(\ell_r,r)$ and  regression model $m(\ell_s,s)$
are linked via the following equality:
\begin{align*}
m(\ell_r,r) &= \E(\theta(L)|\ell_r,R=r)\\
& = \E(\theta(L)|\ell_r,R\in{\sf PA}_r) \\
& = \E(\theta(L)|\ell_r,R=s) P(R=s|R\in{\sf PA}_r, \ell_r)\\ &\quad+  \sum_{\tau \in {\sf PA}_r\backslash\{s\}}\E(\theta(L)|\ell_r,R=\tau)P(R=\tau|R\in{\sf PA}_r, \ell_r).
\end{align*}
The quantity $\E(\theta(L)|\ell_r,R=s) $ can be further written as 
\begin{align*}
\E(\theta(L)|\ell_r,R=s) &=\int \E(\theta(L)|\ell_s,R=s)p(\ell_{s-r}|\ell_r,R=s)d\ell_r\\
&=\int m(\ell_s,s)p(\ell_{s-r}|\ell_r,R=s)d\ell_r.
\end{align*}
Thus, $m(\ell_r,r)$ and $m(\ell_s,s)$ are associated. If
we do not specify them properly, the two models may conflict with each other.

\subsection{Imputation algorithm of PMMs}	\label{sec::imputation::alg}

Despite the power of Theorem~\ref{thm::RA},
the regression adjustment estimator is generally not easy to compute.
A major challenge is that the conditional expectation $\E(\theta(L)|L_r=\ell_r,R\in {\sf PA}_r)$
often does not have a simple form. 
Here we propose to compute the expectation using a Monte Carlo approach.
For a given $L_R=\ell_r,R=r$, we generate many values of $L_{\bar r}$ as follows:
$$
L_{\bar r,1},\cdots, L_{\bar r,N}\sim \hat p(\ell_{\bar r}|\ell_r,R\in {\sf PA}_r)
$$
and then approximate the expectation using 
$$
\hat \E_N(\theta(L)|L_r=\ell_r,R\in {\sf PA}_r) = \frac{1}{N} \sum_{k=1}^N\theta(L_{\bar r,k}, \ell_r).
$$
We perform this approximation for every observation, and then obtain our final estimator as 
\begin{equation}
\begin{aligned}
\hat \theta_{\sf RA, N} &= \frac{1}{n}\sum_{i=1}^n \hat \E_N(\theta(L)|L_r=L_{i,R_i},R\in {\sf PA}_{R_i})\\
&= \frac{1}{n}\sum_{i=1}^n \frac{1}{N} \sum_{k=1}^N\theta(L_{\bar R_i,k}, L_{i,R_i})\\
&= \frac{1}{nN}\sum_{k=1}^N\sum_{i=1}^n \theta(L_{\bar R_i,k}, L_{i,R_i})\\
&= \frac{1}{N}\sum_{k=1}^N\hat \theta^*_{\sf RA, k},
\end{aligned}
\label{eq::MI1}
\end{equation}
where 
$$
\hat \theta^*_{\sf RA, k} = \frac{1}{n}\sum_{i=1}^n \theta(L_{\bar R_i,k}, L_{i,R_i})
$$
is an estimator using a completely imputed dataset. 
Namely, the Monte Carlo approximated estimator
combines several individually imputed estimators;
thus, it
is essentially a \emph{multiple imputation} estimator \citep{LittleRubin02,rubin2004multiple, tsiatis2007semiparametric}. 

To compute the regression adjustment estimator, 
we must be able to sample from $\hat p(\ell_{\bar r}|\ell_r,R=r)= \hat p(\ell_{\bar r}|\ell_r,R\in {\sf PA}_r)$. 
However,
sampling from $\hat p(\ell_{\bar r}|\ell_r,R\in {\sf PA}_r)$
may be difficult.
Here, we provide a simple procedure to sample from the estimated extrapolation density
with access to only (i) sampling from $\hat p(\ell_r|R=r)$ and (ii) evaluating the function $\hat p(\ell_r|R=r)$.

\begin{algorithm}[tb]
\caption{Imputation algorithm} 
\label{alg::impute}
\begin{algorithmic}
\State 1. Input: variables $\ell_r$; the pattern $r$ is determined by the input $\ell_r$.
\State 2. Generate a random pattern $S$ from the parent set ${\sf PA}_r$ with probability
$$
P(S=s) = \frac{\hat p(\ell_r|R=s)n_s}{\sum_{\tau \in {\sf PA}_r} \hat p(\ell_r|R=\tau)n_\tau},
$$
where $n_s = \sum_{i=1}^n I(R_i=s)$. 
\State 3. Impute the variables $S-r$ by sampling from the conditional density:
$$
L^\dagger_{S-r} \sim \hat p(\ell_{S-r}|\ell_r,R=S).
$$
\State 4. Impute the missing entries $L_{S-r} =L^\dagger_{S-r}$.
\end{algorithmic}
\end{algorithm}

Algorithm~\ref{alg::impute}
is a simple approach that imputes missing entries of an observation.
The output is an observation with a smaller number of missing entries 
(there may still be missing entries after executing Algorithm~\ref{alg::impute} once).
Suppose that the input response pattern is $r_1$ and Algorithm \ref{alg::impute}
imputes some missing entries, making it a new response pattern $r_2\neq 1_d$.
Then, we can treat this observation as if it was an observation with response pattern $r_2$
and apply Algorithm~\ref{alg::impute} again to impute more missing entries. 
By repeatedly executing Algorithm~\ref{alg::impute} until no missingness remains,
we impute all missing entries of this observation.
Note that this algorithm is based on equation \eqref{eq::PMM::IM} in the proof of Theorem~\ref{thm::PMM1}.

\begin{algorithm}[tb]
\caption{Imputing the entire data} 
\label{alg::impute2}
\begin{algorithmic}
\State 1. Input: estimators $\hat p_(\ell_r|R=r)$; a graph $G$ 
satisfying assumption (G1-2).
\State 2. For $i=1,\cdots, n$, do the following
	\State 2-1. Set $L_{\sf now} = L_{i,R_i}$ and $R_{\sf now} = R_i$. 
	\State 2-2. Execute Algorithm \ref{alg::impute} with input $L_{\sf now} $ and $R_{\sf now}$. 
	\State 2-3. Update $L_{\sf now}, R_{\sf now}$ to be the return of the algorithm.
	\State 2-4. If $R_{\sf now}\neq 1_d$, return to 2-2; otherwise update $L_i = L_{\sf now}$.
\end{algorithmic}
\end{algorithm}

Algorithm~\ref{alg::impute2} summarizes the procedure of obtaining one imputed dataset. 
Each estimator $\hat \theta^*_{\sf RA, k} $ in equation \eqref{eq::MI1} is the estimator computed 
from one imputed dataset. 
By applying Algorithm~\ref{alg::impute2} $N$ times,
we obtain $N$ estimators, and their average is the final estimator in equation \eqref{eq::MI1}.


\section{Improving efficiency by augmentation}	\label{sec::eff}

It is known from semi-parametric theory that the IPW estimator may not be efficient,
and it is possible to
improve the efficiency by augmenting it with
additional quantities \citep{tsiatis2007semiparametric}. 
We propose to augment it by the form
\begin{equation}
\sum_{r\neq 1_d}\left(I(R=r) - I(R\in {\sf PA}_r) O_r(L_r)\right) \Psi_r(L_r),
\label{eq::aug1}
\end{equation}
where $\Psi_r(L_r)$ is a pattern $r$-specific function of variable  $L_r$. 
This augmentation is inspired by the following equality
\begin{align*}
\E((I(R=r) - &I(R\in {\sf PA}_r) O_r(L_r)) \Psi_r(L_r)) \\
&= \E(\E(\left(I(R=r) - I(R\in {\sf PA}_r) O_r(L_r)\right)|L_r) \Psi_r(L_r)) = 0.
\end{align*}
Therefore, the augmented inverse probability weighting (AIPW) estimator 
\begin{equation}
\begin{aligned}
\hat \theta_{\sf AIPW} &= \frac{1}{n}\sum_{i=1}^n \frac{\theta(L_i) I(R_i=1_d)}{\hat \pi(L_i)} \\
&\qquad+\sum_{r\neq 1_d} \left(I(R_i=r) - I(R_i\in {\sf PA}_r) O_r(L_{i,r})\right) \Psi_r(L_{i,r})
\end{aligned}
\label{eq::AIPW}
\end{equation}
is  an unbiased estimator of $\theta$.
The semi-parametric estimator in Section~\ref{sec::semi} is an AIPW estimator. 


To investigate the augmentation of equation \eqref{eq::AIPW},
let 
\begin{align*}
\mathcal{G} = \bigg\{\frac{\theta(L)I(R=1_d)}{\pi(L)}&+ g(L_R,R):\\
&\qquad\E(g(L_R,R)) = 0, \E(g^2(L_R,R))<\infty\bigg\}
\end{align*}
be the collection of all possible augmentations leading to an unbiased estimator
and
\begin{equation}
\begin{aligned}
\mathcal{F} = &\Bigg\{\theta_{\sf AIPW,r}(L,R) = \frac{\theta(L) I(R=1_d)}{ \pi(L)} 
+ \sum_{r\neq 1_d}\big[I(R=r) \\
&\qquad\qquad- I(R\in {\sf PA}_r) O_r(L_{r})\big] \Psi_r(L_{r}):
\E(\Psi^2_r(L_r))< \infty\Bigg\}
\end{aligned}
\label{eq::AIPW2}
\end{equation}
be the augmentation via equation \eqref{eq::AIPW}.

\begin{proposition}
Assume that $(L,R)$ factorizes with respect to a regular pattern graph $G$ and $p(x_r,r)>0$ for all $r\in\mathcal{R}$.
Then 
$
\mathcal{G} = \mathcal{F}.
$
\label{prop::aug}
\end{proposition}

Proposition \ref{prop::aug} presents a powerful result--the augmentation in the form of $\hat \theta_{\sf AIPW}$
spans the entire augmentation space.
Therefore, 
any augmented IPW estimator can be written in the form of equation \eqref{eq::AIPW}.
Alternatively, one can interpret this proposition as stating that
a typical element orthogonal to the observed data tangent space 
can be expressed via the augmentation in equation \eqref{eq::AIPW2}.

\begin{remark}[Another representation of augmentations]
A common augmentation \citep{tsiatis2007semiparametric, tchetgen2018discrete} is 
in the form of 
$$
\sum_{r\neq 1_d}\left(I(R=r) - I(R\in 1_d)\frac{P(R=r|L)}{P(R=1_d|L)}\right) \Psi_r(L_r).
$$
A notable fact is that this augmentation is the same as equation \eqref{eq::AIPW}
if the pattern graph is constructed using ${\sf PA}_r = \{1_d\}$ for all $r\neq 1_d$.  
Namely, this is the augmentation from the CCMV restriction \citep{tchetgen2018discrete}. 
Although this is a valid augmentation (see Lemma~\ref{lem::augG}), the optimal $\Psi^*_r(L_r)$ may not have a simple form
due to the fact that
the odds $\frac{P(R=r|L)}{P(R=1_d|L)}$ depend on every variable in $L$. 
As a result, we commend to construct the augmentation using equation \eqref{eq::AIPW2}.
\end{remark}

\section{An example of efficient influence function}	\label{sec::example}

Here, we provide a closed form expression of the EIF of
Example \ref{ex::ex01} (based on the pattern graph of right panel of Figure~\ref{fig::ex01}) in the main document. 
There are a total of four patterns: $11,10,01,00$ and four edges $11\rightarrow10, 11\rightarrow01, 11\rightarrow00, 10\rightarrow00$. 

For each $r\neq 1_d$, the collection of paths $\Pi_r$ is
\begin{align*}
\Pi_{10} &= \{11\rightarrow 10\}\\
\Pi_{01} &= \{11\rightarrow 01\}\\
\Pi_{00} &= \{11\rightarrow00, 11\rightarrow 10\rightarrow00\}.
\end{align*}
The corresponding regression function $\mu_{\pi,s}$ is
\begin{align*}
\mu_{11\rightarrow10, 10}(L_{10}) & = \frac{\E(\theta(L)I(R=1_d)|L_{10})}{P(R=10|L_{10})}\\
\mu_{11\rightarrow01, 01}(L_{01}) & = \frac{\E(\theta(L)I(R=1_d)|L_{01})}{P(R=01|L_{01})}\\
\mu_{11\rightarrow00, 00}(L_{00}) & = \frac{\E(\theta(L)I(R=1_d)|L_{00})}{P(R=01|L_{00})}\\
\mu_{11\rightarrow10\rightarrow00, 10}(L_{10}) & = \frac{\E(\theta(L)I(R=1_d)|L_{10})}{P(R=10|L_{10})}\\
\mu_{11\rightarrow10\rightarrow00, 00}(L_{00}) & = \frac{\E(\theta(L)I(R=1_d)O_{00}(L_{00})|L_{00})}{P(R=10|L_{00})}.
\end{align*}
One can observe that 
$
\mu_{11\rightarrow10, 10}(L_{10})= \mu_{11\rightarrow10\rightarrow00, 10}(L_{10}),
$
which is
a property that used in the part 3 of the proof of Proposition \ref{prop::mu_RA}. 
Note that $L_{00} = \emptyset$ so $O_{00}(L_{00}) = O_{00}$ is merely a constant.

With this, the EIF of each path is
\begin{align*}
\EIF_{11\rightarrow10,10}(L,R) &= \mu_{11\rightarrow10, 10}(L_{10})[I(R=10) - O_{10}(L_{10})I(R=11)], \\
\EIF_{11\rightarrow01,01}(L,R) &= \mu_{11\rightarrow01, 01}(L_{01})[I(R=01) - O_{01}(L_{01})I(R=11)], \\
\EIF_{11\rightarrow00,00}(L,R) &= \mu_{11\rightarrow00, 00}(L_{10})[I(R=00) - O_{00}(L_{00})I(R=11\vee 10)], \\
\EIF_{11\rightarrow10\rightarrow00,10}(L,R) &= \mu_{11\rightarrow10\rightarrow00, 10}(L_{10})[I(R=10) - O_{10}(L_{10})I(R=11)] O_{00}(L_{00}), \\
\EIF_{11\rightarrow10\rightarrow00,00}(L,R) &= \mu_{11\rightarrow10\rightarrow00, 00}(L_{00})[I(R=10) - O_{00}(L_{00})I(R=11\vee 10)],
\end{align*}
where $11\vee 10$ signifies $11$ or $10$.
The function $\EIF(L,R)$ is the summation of all these terms.

\section{Tree graph} \label{sec::SG}

In this section, we discuss a particularly interesting family of pattern graphs
called \emph{tree graphs}. 
We  demonstrate that 
for this family, if the observed-data  distribution  $p(\ell_r|R=r)$ is Gaussian for all $r$
and the parameter of interest $\theta(L)$
is a simple function,
then we can avoid the use of Monte Carlo approach
in the regression adjustment estimator
and the semi-parametric estimator.

A tree graph is a pattern graph such that for all patterns $r\neq 1_d$, 
$|{\sf PA}_r|=1$.
Namely, 
every node has only one parent,
so it looks like a tree with the unique source $1_d$.
We use $\mathcal{TG}$ to denote the collection of all tree graphs. 
One can easily see that the pattern graph corresponding to the CCMV restriction  is a tree graph.

Since every node in a tree graph has only one parent, the selection
odds have an elegant expression. 
Specifically, supposing that $s$ is the parent of $r$,
we have 
\begin{equation}
O_r(\ell_r)\equiv \frac{P(R=r|L_r= \ell_r)}{P(R\in {\sf PA}_r|L_r=\ell_r)} = \frac{P(R=r|L_r=\ell_r)}{P(R=s|L_r=\ell_r)}  = \frac{p(\ell_r,r)}{p(\ell_r,s)}.
\label{eq::odds::prod}
\end{equation}
With this, we now demonstrate that if the pattern graph $G\in\mathcal{TG}$
and $p(\ell_r|R=r)$ is a (multivariate) normal density, 
the  function $\mu_{\pi,s}$ may have  a closed  form,
so by property 2 of Proposition \ref{prop::mu_RA}, function $m(\ell_s,s)$
has a closed form.

Without loss of generality, consider a path 
$$
\pi = \{1_d = s_0\rightarrow s_1\rightarrow s_2\rightarrow \cdots s_{m-1}\rightarrow s_m=r\}.
$$
By equation \eqref{eq::odds::prod} and $1_d=s_0, r=s_m$, we have
\begin{align*}
\mu_{\pi,r}(\ell_r) &= \frac{\E(\theta(L)I(R=1_d)\prod_{j=1}^{m-1} O_{s_j}(L_{s_j})|L_{r} = \ell_r)}{P(R=s_{m-1}|L_r=\ell_r)}\\
& = \int \theta(\ell)  \left[\prod_{j=1}^{m-1} O_{s_j}(\ell_{s_j})\right] \frac{p(\ell,1_d)}{p(\ell_r,s_{m-1})}d\ell_{\bar r}\\
& = \int \theta(\ell) p(\ell_{s_0},s_0)\frac{p(\ell_{s_{1}}, s_{1})}{p(\ell_{s_{1}}, s_{0})}\cdots \frac{p(\ell_{s_{m-1}}, s_{m-1})}{p(\ell_{s_{m-1}}, s_{m-2})} \frac{1}{p(\ell_{s_m},s_{m-1})}d\ell_{\bar s_{m}}\\
& = \int \theta(\ell) {p(\ell_{s_0}|\ell_{s_1},s_0)p(\ell_{s_1-s_{2}}|\ell_{s_{2}}, s_{1}) \cdots p(\ell_{s_{m-1}-s_{m}}|\ell_{s_{m}} ,s_{m-1})}d\ell_{\bar s_m}\\
& = \int \theta(\ell) p(\ell_{s_0}|\ell_{s_1},s_0)d\ell_{s_0-s_1}p(\ell_{s_1-s_{2}}|\ell_{s_{2}}, s_{1})d\ell_{s_1-s_2}\times\\ 
&\qquad\cdots\times p(\ell_{s_{m-1}-s_{m}}|\ell_{s_{m}} ,s_{m-1})d\ell_{s_{m-1}-s_m}\\
& = \E[\E[\cdots \E[\E[\theta(L)|L_{s_1},s_0]|L_{s_2},s_1]\cdots|L_{s_{m-1}}, s_{m-2}]|L_{s_m}, s_{m-1}].
\end{align*}
Thus, if functional $\theta(L)$ is a simple function such as $\theta(L) = \sum_{j} a_j L_j$ for some fixed $\{a_j:j=1,\cdots, d\}$),
the quantity $\E[\theta(L)|L_{s_1},s_0]$ is a linear function of $L_{s_1}$ and parameters of this function are determined by the mean
and covariance of $p(\ell_{\bar s_0}|\ell_{s_0},s_0)$ because of Gaussian assumption. 
By iteratively applying this fact, we conclude that $\mu_{\pi,r}(\ell_r)$
is a linear function of $\ell_r$ and the parameters of this function
are determined by the coefficients of Gaussians on path $\pi$.
Moreover, using property 2 of Proposition \ref{prop::mu_RA}, we also obtain a closed form of the function $m(\ell_s,s)$.
In this case, we do not need to use any numerical methods to compute $m(\ell_s,s)$.

\section{Technical assumptions}	\label{sec::assumptions}

\subsection{Assumptions of IPW estimators}

Let $\eta = (\eta_r:r\in\mathcal{R})\in \Theta$ be any parameter value, where $\Theta$ is the total parameter space. 
To obtain the asymptotic normality of the IPW estimator, we assume the following conditions:
\begin{itemize}
\item[(L1)] there exists $\underline{O},\overline{O}$ such that
$$
0<\underline{O}\leq O_r(\ell_r;\eta) \leq \overline{O}<\infty
$$
for all $\ell_r\in \mathbb{S}_r$ and $r\in\mathcal{R}$ and $\eta \in \Theta$.
\item[(L2)] there exists $\eta^* = (\eta^*_r:r\in\mathcal{R})$ in the interior of $\Theta$ such that $O_r(\ell_r;\eta^*) = \frac{P(R=r|\ell_r)}{P(R\in{\sf PA}_r|\ell_r)}$
and 
$$
\sqrt{n}(\hat \eta_r - \eta_r^*)\rightarrow N(0,\sigma_r^2),\quad \int\theta^2(\ell) (O_r(\ell_r;\hat\eta) - O_r(\ell_r;\eta^*))^2 F(d\ell) = o_P(1),
$$
for some $\sigma_r^2>0$ for all $r$.
\item[(L3)] for every $r$,
the class $\{f_{\eta_r}(\ell_r) = O_r(\ell_r;\eta_r): \eta_r \in\Theta_r\}$ is a Donsker class.

\item[(L4)] for every $r$, the differentiation of $O_r(\ell_r;\eta_r)$ with respect to $\eta_r$, $O_r'(\ell_r;\eta_r) = \nabla_{\eta_r} O_r(\ell_r;\eta_r)$, exists and 
$\int \|O_r'(\ell_r;\eta_r) \| F(d\ell_r) <\infty$
for a ball $B(\eta^*,\tau_0)$ for some $\tau_0>0$.

\end{itemize}

Assumption (L1) avoids the scenario that the selection odds diverge. 
The second assumption (L2) requires that the model is correctly specified
and
the estimator $\hat \eta_r$ 
is asymptotic normal and the implied estimated odds converges in $L_2(P)$ norm. 
The Donsker condition (L3) is a common condition that many parametric models satisfy;
see Example 19.7 in \cite{van1998asymptotic} for a sufficient condition on the parametric family. 
The bounded integral condition (L4) is relatively weak after assuming (L2) and (L3).
The quantity $O_r'(\ell_r;\eta_r) $ is essentially a score equation so this assumption
requires that the score equation exists and has a finite norm.

\begin{example}[Logistic regression]	\label{ex::logistic}
Here, we discuss a special case of modeling the selection odds $O_r(L_r)$ via logistic regression. 
For each $r$, the logistic regression models 
the selection odds as 
\begin{equation}
\log O_r(L_r) =\log \left(\frac{P(R=r|L_r)}{P(R\in{\sf PA}_r|L_r)} \right)= \beta_{r}^T \tilde L_r ,
\label{eq::logistic}
\end{equation}
where $\beta_r  \in \R^{1+|r| }$ is the coefficient vector 
and $\tilde L_r  = (1, L_r)$ is the vector including $1$ as the first variable.
We include $1$ in $\tilde L_r$ so the intercept is the first element of $\beta_r.$
$\beta_r$ can be estimated by applying a logistic regression for the pattern $R=r$ against the pattern $R\in {\sf PA}_r$. 
Now we discuss
conditions (L1-4) in Theorem~\ref{thm::Codds}
under the logistic regression model.
(L1) holds if the support of study variable $L$ is (elementwise) bounded.
The second condition holds if the logistic regression model
correctly describes the selection odds. 
The asymptotic normality follows from
the regular conditions of a logistic regression model. 
The Donsker class condition (L3) holds for the logistic regression model with a bounded study variable. 
Condition (L4) also holds when $L$ is bounded and the true parameter $\eta^*_r$ is away
from boundary because of $O'_r(L_r;\eta_r) = L_r \cdot e^{\eta_r^TL_r}$. 
Note that the logistic regression has a special computational benefit, as described in Appendix \ref{sec::logistic}.
\end{example}

\subsection{Assumptions of RA estimators}

The regression adjustment estimator has asymptotic normality 
under the following conditions:
\begin{itemize}
\item[(R1)] There exists $\lambda^*_r \in\Lambda_r$ such that the true conditional density $p(\ell_r|R=r) = p(\ell_r|R=r;\lambda_r^*)$
for every $r$.

\item[(R2)] For every $r$,  the class
$$
\{f_{\lambda}(\ell_r) =  m(\ell_r,r;  \lambda): \lambda\in\Lambda\}
$$
is a Donsker class.
\item[(R3)] For every $r$,  $q_r(\lambda) = \E(m(L_r,r;\lambda)I(R=r))$ is bounded twice-differentiable and
\begin{align*}
\int (m(\ell_r,r; \hat \lambda)- m(\ell_r,r;  \lambda))^2 F(d\ell_r,r) &= o_P(1)\\
\sqrt{n}(\hat \lambda_r -\lambda^*_r) &\rightarrow N(0, \sigma^2_{r}).
\end{align*}

\end{itemize}

(R1) requires that the parametric model is correct, which is common for establishing
the asymptotic normality centering at the true parameter. 
The Donsker class condition in (R2) is a common assumption
to establish a uniform central limit theorem of a likelihood estimator (see Chapter 19 of \citealt{van1998asymptotic}). 
In general, if the parametric model is sufficiently smooth and the statistical functional $\theta(L)$
is smooth such as being a linear functional \citep{van1998asymptotic}, 
we have this condition. 
Condition (R3) is a consistency condition--we need $\hat\lambda$ to  be a consistent
estimator of $\lambda$ in the sense that the implied regression function converges in $L_2(P)$ norm
and has asymptotic normality.

\section{Proofs}		\label{sec::proof}

\begin{proof}[ of Theorem~\ref{thm::Codds}]

We first prove 
the closed form of $\pi(L)$ and 
the recursive form of $Q_r(L)$.
Because $Q_{r}(L)  = \frac{P(R=r|L)}{P(R=1_d|L)}$, 
it is easy to see that 
$$
\frac{1}{\pi(L)} = \frac{1}{P(R=1_d|L)} = \frac{\sum_{r} P(R=r|L)}{P(R=1_d|L)} = \sum_{r} Q_r(L),
$$
which is the closed form of $\pi(L)$.
For the recursive form, 
a direct computation shows that
\begin{align*}
Q_r(L)& = \frac{P(R=r|L)}{P(R=1_d|L)}\\ 
&\overset{\eqref{eq::Codds}}{=} \frac{P(R\in {\sf PA}_r|L)O_r(L_r)}{P(R=1_d|L)}\\
& = O_r(L_r)\frac{\sum_{s\in {\sf PA}_r}P(R=s|L)}{P(R=1_d|L)} \\
&= O_r(L_r)\sum_{s\in{\sf PA}_r} Q_s(L).
\end{align*}

For the identifiability, we use the proof by induction.
We will show that each $Q_r(L)$ is identifiable so $\pi(L)$ is identifiable. 
Since $Q_r(L) = 1$, it is immediately identifiable. 
For $Q_r(L)$ with $|r| = d-1$, they only have one parent: ${\sf PA}_r = 1_d$
so $Q_r(L) = O_r(L_r)$ is identifiable. 

Now we assume that $Q_\tau(L)$ is identifiable for all $|\tau|> k$,
and we consider a pattern $r$ such that $|r| = k$. 
We will show that $Q_r(L)$ is also identifiable. 
By the recursive form, 
$$
Q_r(L) = O_r(L_r)\sum_{s\in{\sf PA}_r} Q_s(L).
$$
Assumption (G2) implies that any $s\in{\sf PA}_r$ must satisfies $s>r$
so $|s|>k$. The assumption of induction implies that $Q_s(L)$ is identifiable. 
Thus, both $O_r(L_r)$ and $\sum_{s\in{\sf PA}_r} Q_s(L)$ are identifiable, which implies that $Q_r(L)$
is identifiable. 
By induction, $Q_r(L)$ is identifiable for all $r$ and $\pi(L)$ is identifiable.

\end{proof}

\begin{proof}[ of Proposition \ref{lem::pathG}]

We  first prove that 
\begin{equation}
\pi(L) = \frac{1}{\sum_{\Xi\in \Pi} \prod_{s\in\Xi}O_s(L_s)}
\label{eq::piL}
\end{equation}
and then equation \eqref{eq::naive} follows immediately.

Recall from Theorem~\ref{thm::Codds} that 
$$
\pi(L) = \frac{1}{\sum_r Q_r(L)}
$$
so
proving equation \eqref{eq::piL} is equivalent to proving 
\begin{equation}
Q_r(L) = \sum_{\Xi\in \Pi_r} \prod_{s\in \Xi} O_s(L_s).
\label{eq::path1}
\end{equation}
We prove this by induction from $|r| = d, d-1,d-2,\cdots, 0$. 
It is easy to see that when $|r|=d$ and $d-1$,
this result holds.

We now assume that the statement holds for any pattern $\tau$  with $|\tau|=d,d-1,\cdots, k+1$.
Consider the pattern $r$ such that $|r| = k$.
By induction and assumption (G2), any parent pattern of $r$ must satisfy equation \eqref{eq::path1}.

Due to the construction of a path (from $1_d$ to $r$), 
any path $\Xi \in \Pi_r$ can be written as 
$$
\Xi = (\Xi', r),
$$
where $\Xi'\in \Pi_q$ for some $q\in {\sf PA}_r$
except for the case where $q = 1_d$.
Suppose that $1_d\in {\sf PA}_r$, then there is only one path that 
corresponds to the parent pattern being $1_d$,
and this path contributes in the right-hand-sided of equation \eqref{eq::path1} the amount of $O_r(L_r)$.
With the above insight, we can rewrite the right-hand-side of equation \eqref{eq::path1} as
\begin{align*}
 \sum_{\Xi\in \Pi_r} \prod_{s\in \Xi} O_s(L_s)
& = 
O_r(L_r)I(1_d\in {\sf PA_r})+
\sum_{q\in {\sf PA}_r,q\neq1_d}\sum_{\Xi\in \Pi_q} \prod_{s\in (\Xi, r)} O_s(L_s)\\
&=O_r(L_r)I(1_d\in {\sf PA_r})+\sum_{q\in {\sf PA}_r, q\neq1_d}\sum_{\Xi\in \Pi_q} \left(\prod_{s\in \Xi} O_s(L_s)\right)O_r(L_r)\\
&=O_r(L_r)I(1_d\in {\sf PA_r})+O_r(L_r)\sum_{q\in {\sf PA}_r, q\neq1_d}\underbrace{\sum_{\Xi\in \Pi_q} \prod_{s\in \Xi} O_s(L_s)}_{=Q_r(L)\mbox{ by induction}}\\
&=O_r(L_r)\sum_{q\in {\sf PA}_r}Q_r(L)\\
&=Q_r(L) \quad \mbox{(by Theorem~\ref{thm::Codds})}.
\end{align*}
Thus, equation \eqref{eq::path1} holds and thus, equation \eqref{eq::naive} is true.

\end{proof}

\begin{proof}[ of Theorem~\ref{thm::PMM1}]

We prove this by induction from patterns with $|r|=d, d-1,\cdots, 0$. 
We first prove that it identifies the joint distribution $p(\ell, r)$.
The case of $|r|=d$ is trivially true since everything is identifiable under this case. 

For $|r|=d-1$, they only have one parent $1_d$ and
recall from equation \eqref{eq::PMM}, 
$$
p(\ell_{\bar r}|\ell_r,R=r) = p(\ell_{\bar r}|\ell_r,R\in{\sf PA}_r).
$$
This implies 
$$
p(\ell_{\bar r}|\ell_r,R=r) = p(\ell_{\bar r}|\ell_r,R=1_d).
$$
Clearly, $p(\ell_{\bar r}|\ell_r,R=1_d)$ is identifiable so $p(\ell_{\bar r}|\ell_r,R=r)$ is identifiable.

Now we assume that $p(\ell_{\bar \tau}|\ell_\tau,R=\tau)$ is identifiable for all $|\tau|>k$
and consider a pattern $r$ with $|r| = k$.
Equation \eqref{eq::PMM} implies that the extrapolation density
\begin{equation}
\begin{aligned}
p(\ell_{\bar r}|\ell_r,R=r) &= p(\ell_{\bar r}|\ell_r,R\in{\sf PA}_r)\\
&= \frac{p(\ell_{\bar r},R\in{\sf PA}_r|\ell_r)}{P(R\in{\sf PA}_r|\ell_r)}\\
& = \frac{\sum_{s\in {\sf PA}_r} p(\ell_{\bar r},R=s|\ell_r)}{P(R\in{\sf PA}_r|\ell_r)}\\
& = \sum_{s\in{\sf PA}_r}p(\ell_{\bar r}|\ell_r,R=s) P(R=s|R\in{\sf PA}_r, \ell_r).
\end{aligned}
\label{eq::PMM::IM}
\end{equation}
Since $s$ is a parent of $r$,
condition (G2) implies that $s>r$ so $P(R=s|R\in{\sf PA}_r, \ell_r)$ is identifiable. 
Also, by the assumption of induction, $p(\ell_{\bar r}|\ell_r,R=s)$ is identifiable for $s\in {\sf PA}_r.$
Thus, $p(\ell_{\bar r}|\ell_r,R=r) $ is identifiable, which proves the result.

Since equation \eqref{eq::PMM} only places conditions on the extrapolation densities, 
it is easy to see that the resulting full-data distribution $F(\ell,r)$
is nonparametrically identifiable.

\end{proof}

\begin{proof}[ of Theorem~\ref{thm::PMM2}]
This proof consists of a sequence of ``if and only if" statements.
We start with the selection odds model:
$$
\frac{P(R=r|L=\ell)}{P(R\in{\sf PA}_r|L=\ell)}= \frac{P(R=r|L_r=\ell_r)}{P(R\in{\sf PA}_r|L_r=\ell_r)}.
$$
The left-hand-side equals $\frac{p(R=r,L=\ell)}{p(R\in{\sf PA}_r,L=\ell)}$
whereas the right-hand-side equals $\frac{p(R=r,L_r=\ell_r)}{p(R\in{\sf PA}_r,L_r=\ell_r)}$.
So the selection odds model is equivalent to 
\begin{align*}
&\frac{p(R=r,L=\ell)}{p(R\in{\sf PA}_r,L=\ell)}  = \frac{p(R=r,L_r=\ell_r)}{p(R\in{\sf PA}_r,L_r=\ell_r)}\\
\Longleftrightarrow&\frac{p(R=r,L=\ell)}{p(R=r,L_r=\ell_r)}  = \frac{p(R\in{\sf PA}_r,L=\ell)}{p(R\in{\sf PA}_r,L_r=\ell_r)}\\
\Longleftrightarrow&\frac{p(L=\ell|R=r)}{p(L_r=\ell_r|R=r)}  = \frac{p(L=\ell|R\in {\sf PA}_r)}{p(L_r=\ell_r|R\in{\sf PA}_r)}\\
\Longleftrightarrow&p(L_{\bar r}=\ell_{\bar r}|L_r = \ell_r, R=r)=p(L_{\bar r}=\ell_{\bar r}|L_r = \ell_r, R\in{\sf PA}_r),
\end{align*}
which is what the pattern mixture model factorization refers to.

\end{proof}

Before proceeding to the proof of Theorem~\ref{thm::IPW},
we introduce some notations from the empirical process theory. 
For a function $f(\ell,r)$,
we write
$$
\int f(\ell,r) F(d\ell, dr) = \E(f(L,R))
$$
and the empirical version of it
$$
\int f(\ell,r) \hat F(d\ell, dr) = \frac{1}{n}\sum_{i=1}^n f(L_i,R_i).
$$
Although $L_i$ may not be fully observed when $R_i\neq 1_d$,
the indicator function $I(R=1_d)$ has an appealing feature that 
$$
\int f(\ell,r) I(r=1_d) \hat F(d\ell,dr) = \frac{1}{n}\sum_{i=1}^n f(L_i,R_i)I(R_i=1_d)
$$
so the IPW estimator can be written as 
\begin{align*}
\hat \theta_{IPW} = \frac{1}{n}\sum_{i=1}^n \frac{\theta(L_i)I(R_i=1_d)}{\pi(L_i;\hat\eta)} &= \int \frac{\theta(\ell)I(r=1_d)}{\pi(\ell;\hat\eta)} \hat F(d\ell, dr)\\
& = \int \xi(\ell,r;\hat\eta) \hat F(d\ell, dr),
\end{align*}
where $\xi(\ell,r;\hat\eta) = \frac{\theta(\ell)I(r=1_d)}{\pi(\ell;\hat\eta)} $.
Note that when the model is correct, the parameter of interest
\begin{align*}
\theta_0 = \E(\theta(L)) = \E\left(\frac{\theta(L)I(R=1_d)}{\pi(L;\eta^*)}\right) = \int \xi(\ell,r;\eta^*) F(d\ell, dr),
\end{align*}
where $\eta^*$ is true parameter value.

%

\begin{proof}[ of Theorem~\ref{thm::IPW}]


Using the notation of the empirical process, we can rewrite the difference
$\hat\theta_{IPW} - \theta_0$ as
\begin{align*}
\hat\theta_{IPW} - \theta_0& =\int \xi(\ell,r;\hat\eta) \hat F(d\ell, dr) -\int \xi(\ell,r;\eta^*) F(d\ell, dr)\\
& = \underbrace{\int \xi(\ell,r;\hat\eta) (\hat F(d\ell, dr) - F(d\ell,dr))}_{(I)} +\underbrace{\int (\xi(\ell,r;\hat\eta) - \xi(\ell,r;\eta^*)) F(d\ell, dr)}_{(II)}.
\end{align*}
Thus, we only need to show that both (I) and (II) have asymptotic normality. 
Note that formally we need to show that the asymptotic correlation between (I) and (II) is not -1, but this is clearly the case
so we ignore this step.
We analyze (I) and (II) separately.

{\bf Part (I):}
The asymptotic normality is based on Theorem~19.24 of \cite{van1998asymptotic}
that this quantity is asymptotically the same as the case if we replace $\hat\eta$ by $\eta^*$
when we have the following:
\begin{itemize}
\item[(C1)] $\int (\xi(\ell,r;\hat\eta)  - \xi(\ell,r;\eta^*) )^2 F(d\ell,dr) = o_P(1)$ and
\item[(C2)] the class $\{\xi(\ell,r;\eta) :\eta\in\Theta\}$ is a Donsker class. 
\end{itemize}
Thus, we will show both conditions in this proof. 

{\bf Condition (C1).}
A direct computation shows that 
\begin{align*}
\xi(\ell,r;\hat\eta) - \xi(\ell,r;\eta^*)
&=\frac{\theta(\ell)I(r=1_d)}{ \pi(\ell;\hat\eta)} - \frac{\theta(\ell)I(r=1_d)}{ \pi(\ell;\eta^*)}\\
& =\theta(\ell)I(r=1_d)\sum_r Q_r(\ell;\hat\eta)- \theta(\ell)I(r=1_d) \sum_r Q_r(\ell;\eta^*)\\
& =\theta(\ell)I(r=1_d)\sum_r\left( Q_r(\ell;\hat\eta)-Q_r(\ell;\eta^*) \right).
\end{align*}
Thus,
\begin{align*}
\int (\xi(\ell,r;\hat\eta) &- \xi(\ell,r;\eta^*))^2 F(d\ell,dr)\\
& = \int  \theta^2(\ell)I(r=1_d) (\sum_r Q_r(\ell;\hat\eta)-Q_r(\ell;\eta^*) )^2 F(d\ell,dr)\\
& \leq \int  \theta^2(\ell)I(r=1_d)\|\mathcal{R}\| \sum_r( Q_r(\ell;\hat\eta)-Q_r(\ell;\eta^*) )^2 F(d\ell,dr)\\
& \leq \|\mathcal{R}\|\sum_r\int  \theta^2(\ell) ( Q_r(\ell;\hat\eta)-Q_r(\ell;\eta^*) )^2 F(d\ell,dr),
\end{align*}
where $\|\mathcal{R}\|$  is the number of elements in $\mathcal{R}$ and $C_0>0$ is some constant.
So a sufficient condition to (C1) is 
\begin{equation}
\int  \theta^2(\ell) ( Q_r(\ell;\hat\eta)-Q_r(\ell;\eta^*) )^2 F(d\ell,dr) = o_P(1)
\label{eq::pf::Qr1}
\end{equation}
for each $r$.


From the likelihood condition (L2), we have $\int \theta^2(\ell) (O_r(\ell_r;\hat\eta) - O_r(\ell_r;\eta^*))^2 F(d\ell) = o_P(1)$. 
Namely, we have the desired weighted $L_2$ convergence result  of $O_r(\ell_r;\hat\eta)$.
To convert this into $Q_r$, we use the proof by induction.
%
%
It is easy to see that when $r=1_d$, this holds trivially because $Q_{1_d}=1$. 
Suppose for a pattern $r$, the $L_2$ convergence holds for all its parents ${\sf PA}_r$, i.e., 
$$
\int \theta^2(\ell) (Q_s(\ell;\hat\eta) - Q_s(\ell;\eta^*))^2 F(d\ell) = o_P(1)
$$
for all $s\in {\sf PA}_r$.
By Theorem~\ref{thm::Codds},
\begin{align*}
Q_r(\ell;\hat\eta) - Q_r(\ell;\eta^*) &= O_r(\ell_r;\hat\eta_r) \sum_{s\in {\sf PA}_r} Q_s(\ell;\hat\eta) - O_r(\ell_r;\eta^*_r) \sum_{s\in {\sf PA}_r} Q_s(\ell;\eta^*)\\
&= \underbrace{(O_r(\ell_r;\hat\eta_r) - O_r(\ell_r;\eta^*_r))\sum_{s\in {\sf PA}_r} Q_s(\ell;\hat\eta)}_{A(\ell)} \\
&\quad+ \underbrace{O_r(\ell_r;\eta^*_r)  \sum_{s\in {\sf PA}_r} (Q_s(\ell;\hat\eta) - Q_s(\ell;\eta^*) )}_{B(\ell)}.
\end{align*}

{\bf Quantity $A(\ell)$:}
The  boundedness assumption of $O_r$ in (L1) implies that $Q_s(\ell;\hat\eta)\leq \bar Q$ for some constant $\bar Q$
so $\sum_{s\in {\sf PA}_r} Q_s(\ell;\hat\eta) \leq \bar Q\|\mathcal{R}\| =  C_1$ is uniformly bounded. 
As a result, 
\begin{align*}
\int \theta^2(\ell)A^2(\ell)F(d\ell)&=
\int\theta^2(\ell) \Big[(O_r(\ell_r;\hat\eta_r) - O_r(\ell_r;\eta^*_r))\sum_{s\in {\sf PA}_r} Q_s(\ell;\hat\eta)\Big]^2F(d\ell) \\
&\leq C_1^2 \int\theta^2(\ell) (O_r(\ell_r;\hat\eta_r) - O_r(\ell_r;\eta^*_r))^2F(d\ell) \\
& = o_P(1)
\end{align*}
by assumption (L2).

{\bf Quantity $B(\ell)$:}
Assumption (L1) implies that $O_r$ is uniformly bounded so 
\begin{align*}
\int \theta^2(\ell)B^2(\ell)F(d\ell)&=\int \theta^2(\ell) \Big[O_r(\ell_r;\eta^*_r)  \sum_{s\in {\sf PA}_r} (Q_s(\ell;\hat\eta) - Q_s(\ell;\eta^*) )\Big]^2 F(d\ell)\\
&\leq C_2\int \theta^2(\ell) \Big[\sum_{s\in {\sf PA}_r} (Q_s(\ell;\hat\eta) - Q_s(\ell;\eta^*) )\Big]^2 F(d\ell)\\
&\leq C_3\int \theta^2(\ell) \sum_{s\in {\sf PA}_r} (Q_s(\ell;\hat\eta) - Q_s(\ell;\eta^*) )^2 F(d\ell)\\
& = o_P(1)
\end{align*}
by the induction assumption and $C_2,C_3>0$ are some constant.

Therefore, both $A(\ell)$ and $B(\ell)$ converges in the weighted $L_2$ sense, which implies that 
\begin{align*}
\int \theta^2(\ell)(Q_r(\ell;\hat\eta) - Q_r(\ell;\eta^*))^2 F(d\ell) 
& = \int \theta^2(\ell)(A(\ell)+B(\ell))^2 F(d\ell)\\
& \leq 2 \int \theta^2(\ell)(A^2(\ell)+B^2(\ell)) F(d\ell)\\
&= o_P(1)
\end{align*}
so condition (C1) holds.

{\bf Condition (C2).}
The derivation of
this property follows from a similar idea as condition (C1)
that we start with $O_r$ and then $Q_r$ and finally $\xi$. 
The Donsker class follows because $\{f_{\eta_r}(\ell_r) = O_r(\ell_r;\eta_r): \eta_r \in\Theta_r\}$
is a uniformly bounded Donsker class. 
The multiplication of uniformly bounded Donsker class is still a Donsker class (see, e.g., Example 2.10.8 of \citealt{van1996weak}). 
Thus, the class $\{f_{\eta}(\ell)  = Q_r(\ell;\eta): \eta \in \Theta\}$  is a uniformly bounded Donsker class. 

By Theorem~\ref{thm::Codds}, $\pi(\ell;\eta) = \frac{1}{\sum_r Q_r(\ell;\eta)}$
so $\xi(\ell,r;\eta) = \theta(\ell)I(r=1_d) \sum_r Q_r(\ell;\eta)$, which
implies that 
$\{f_\eta(\ell,r) = \xi(\ell,r;\eta): \eta\in\Theta\}$ is a Donsker class.
So condition (C2) holds.

With condition (C1) and (C2), applying Theorem~19.24 of \cite{van1998asymptotic}
shows that the quantity (I) has asymptotic normality.

{\bf Part (II):}
Using the fact that $\pi(\ell;\eta) = \frac{1}{\sum_r Q_r(\ell;\eta)}$,
we can rewrite $\xi$ as
$$
\xi(\ell,r;\eta) = \frac{\theta(\ell)I(r=1_d)}{\pi(\ell;\eta)} = \theta(\ell)I(r=1_d)\sum_r Q_r(\ell;\eta).
$$
Thus, quantity (II) becomes
\begin{align*}
(II)&=\int (\xi(\ell,r;\hat\eta) - \xi(\ell,r;\eta^*)) F(d\ell, dr)\\
& = \int \theta(\ell)I(r=1_d)\sum_s[Q_s(\ell;\hat \eta)-Q_s(\ell;\eta^*)]F(d\ell,dr)\\
& = \sum_s\int \theta(\ell)I(r=1_d)[Q_s(\ell;\hat \eta)-Q_s(\ell;\eta^*)]F(d\ell,dr).
\end{align*}
Thus, we only need to show that this quantity is either $0$ or has an asymptotic normality for each pattern $s$
(and at least one of them is non-zero).

Clearly, when $s=1_d$, this quantity is 0 so we move onto the next case. 
For $s$ being a pattern with only one variable missing, we have $Q_s(\ell;\eta) = O_s(\ell_s;\eta)$,
which leads to
\begin{align*}
\int \theta(\ell)I(r=1_d)&[Q_s(\ell;\hat \eta)-Q_s(\ell;\eta^*)]F(d\ell,dr) \\
&= \int \theta(\ell)I(r=1_d)[O_s(\ell_s;\hat \eta_s)-O_s(\ell_s;\eta^*_s)]F(d\ell,dr).
\end{align*}
Applying the Taylor expansion of $O_s$ with respect to $\eta_s$,
assumption (L4) implies that 
\begin{align*}
\int \theta(\ell)I(r=1_d)&[Q_s(\ell;\hat \eta)-Q_s(\ell;\eta^*)]F(d\ell,dr) \\
&= \int \theta(\ell)I(r=1_d)(\hat\eta_s -\eta^*_s)^TO'_s(\ell_s;\eta^*_s)F(d\ell,dr)\\
& = (\hat\eta_s -\eta^*_s)^T \int \theta(\ell)I(r=1_d)O'_s(\ell_s;\eta^*_s)F(d\ell,dr),
\end{align*}
where $O'_s(\ell_s;\eta_s) = \nabla_{\eta_s}O_s(\ell_s;\eta_s)$  is the derivative with respect to $\eta_s$.
It has asymptotic normality due to assumption (L2). 
Note that the variance is finite because of the boundedness assumption (L1). 
Using the induction, one can show that for a pattern $s$, if its parents have either asymptotic normality or equals to $0$ (but not all $0$), 
we have asymptotic normality of $\int \theta(\ell)I(r=1_d)[Q_s(\ell;\hat \eta)-Q_s(\ell;\eta^*)]F(d\ell,dr)$.
Thus, the quantity in (II) converges to a normal distribution after rescaling.

Since both (I) and (II) both have asymptotic normality, 
$\hat\theta_{IPW} - \theta_0 = (I)+(II)$ also has asymptotic normality by the continuous mapping theorem, which completes the proof.





\end{proof}


\begin{proof}[ of Theorem~\ref{thm::RA}]

This proof utilizes tools from empirical process theory that are similar to the proof of Theorem~\ref{thm::IPW}.
Let $\hat F(\ell_r,r)$ and $ F(\ell_r,r)$ be the empirical and probability measures of variable $L_r$ and pattern $R=r$, respectively.

The regression adjustment estimator can be written as 
\begin{align*}
\hat \theta_{\sf RA} &= \frac{1}{n}\sum_{i=1}^n m(L_{i,R_i}, R_i;\hat\lambda)\\
&= \frac{1}{n}\sum_{i=1}^n \sum_{r}m(L_{i,r}, r;\hat\lambda)I(R_i=r)\\
&= \sum_{r}\frac{1}{n}\sum_{i=1}^n m(L_{i,r}, r;\hat\lambda)I(R_i=r)\\
&= \sum_{r}\hat\theta_{\sf RA, r},
\end{align*}
where 
\begin{equation*}
\hat\theta_{\sf RA, r} = \frac{1}{n}\sum_{i=1}^n m(L_{i,r}, r;\hat \lambda)I(R_i=r) = \int m(\ell_r, r;\hat \lambda) \hat F(d\ell_r,r).
\end{equation*}
A population version of the above quantity is 
$$
\theta_{\sf RA, r} = \E(m(L_{r}, r;\lambda^*)I(R=r)) = \int m(\ell_r, r;\lambda^*)  F(d\ell_r,r).
$$
It is easy to see that the parameter of interest $\theta_0 = \sum_r \theta_{\sf RA, r}$. 
Thus, if we can show that 
\begin{equation}
\sqrt{n}(\hat \theta_{\sf RA,r} -  \theta_{\sf RA,r}) \overset{D}{\rightarrow} N(0,\sigma^2_{\sf RA,r})
\label{eq::RAr}
\end{equation}
for each $r$, we have completed the proof (by the continuous mapping theorem).

To start with, we decompose the difference 
\begin{align*}
\sqrt{n}(\hat \theta_{\sf RA,r} -  \theta_{\sf RA,r})& = \int m(\ell_r, r;\hat\lambda) \hat F(d\ell_r,r)- \int m(\ell_r, r;\lambda^*)  F(d\ell_r,r)\\
& = \underbrace{\sqrt{n}\int m(\ell_r, r;\hat\lambda) (\hat F(d\ell_r,r) - F(d\ell_r,r))}_{=(I)} \\
&\quad+ \underbrace{\sqrt{n}\int (m(\ell_r, r;\hat\lambda) - m(\ell_r, r;\lambda^*) ) F(d\ell_r,r)}_{=(II)}.
\end{align*}

{\bf Analysis of (I).}
By Theorem~19.24 of \cite{van1998asymptotic} and condition (R2) and the first equality of (R3), 
\begin{align*}
(I)&= \sqrt{n}\int m(\ell_r, r;\lambda^*) (\hat F(d\ell_r,r) - F(d\ell_r,r))+o_P(1)\\
& = \frac{1}{\sqrt{n}}\sum_{i=1}^n \left[m(L_{i,r}, r;\lambda^*)I(R_i=r) - \E\left(m(L_{i,r}, r;\lambda^*)I(R_i=r) \right)\right] + o_P(1),
\end{align*}
which has asymptotic normality. 

{\bf Analysis of (II).}
Recall that $q_r (\lambda) = \E(m(L_r,r;\lambda)I(R=r))$.
Using Tayloy expansion of $q_r$, we can rewrite (II) as
\begin{align*}
(II) &=\sqrt{n}( q_r(\hat \lambda) - q_r(\lambda^*))\\
& = \sqrt{n} \nabla q_r(\lambda^*)^T (\hat \lambda-\lambda^*) + o_P(1)
\end{align*}
due to the assumption on the boundedness of derivatives of $q_r$ and the rate of $\hat \lambda$ in (R3). 
The asymptotic normality assumption of $\sqrt{n}(\hat \lambda-\lambda^*) $ implies the asymptotic normality of (II).

Thus, both (I) and (II) are asymptotically normal
so we have the asymptotic normality of $\sqrt{n}(\hat \theta_{\sf RA,r} -  \theta_{\sf RA,r})$ via the continuous mapping theorem, i.e., equation \eqref{eq::RAr},
which implies the desired result.

\end{proof}


\begin{proof}[ of Theorem~\ref{thm::EIF}]

Let $\Xi\in\Pi_q$ be a path of pattern $q$.
For a pathwise effect $\theta_\Xi$, it equals to
\begin{align*}
\theta_\Xi &= \E(\theta(L) I(R=1_d) \prod_{s\in\Xi} O_s(L_s))\\
&=\int \theta(\ell) I(r=1_d) \prod_{s\in\Xi} O_s(\ell_s)p(\ell,r)d\ell d_r
\end{align*}

Let $p_0(\ell,r)$ be the correct model,
and we consider a pathwise perturbation
$p_\epsilon(\ell,r) = p_0(\ell,r) (1+\epsilon \cdot g(\ell,r))$
such that $g$ satisfies
$\int p_0(\ell,r)g(\ell,r) d\ell dr = 0$. 

Under the correct model $p_0$, the effect is $\theta_{\Xi,0}$.
Under the model $p_{\epsilon}$, the effect is $\theta_{\Xi,\epsilon}$.

By the semi-parametric theory (see, e.g., Section 25.3 of \citealt{van1998asymptotic}), 
the EIF is a function ${\sf EIF}_\Xi(\ell,r)$ such that  $\E({\sf EIF}_\Xi(L,R))=0$ and
\begin{equation}
\lim_{\epsilon\rightarrow0} \frac{\theta_{\Xi,\epsilon} - \theta_{\Xi,0}}{\epsilon}  = \int {\sf EIF}_\Xi(\ell,r) p_0(\ell,r) g(\ell,r)d\ell dr .
\label{eq::EIF::pf1}
\end{equation}
So we just need to find the proper expression of ${\sf EIF}_\Xi(\ell,r)$. 

Our strategy is very simple.
We  compute $\theta_{\Xi,\epsilon}$
and keep those terms involving the first order of $\epsilon$
and ignore anything involving $\epsilon^2$ since the higher-order terms
varnish in the above limit. 

Under the model $p_\epsilon(\ell,r)$, we have perturbed quantities $p_{\epsilon}(\ell_r ,r)$
and $O_{r,\epsilon} (\ell_r) = \frac{p_\epsilon(r, \ell_r)}{p_\epsilon({\sf PA}_r, \ell_r)}$.
We denote 
$$
\Delta O_r(\ell) = O_{r,\epsilon}(\ell_r) - O_{r,0}(\ell_r).
$$
A direct computation shows that 
\begin{equation}
\begin{aligned}
\theta_{\Xi,\epsilon} & = \int \theta(\ell) I(r=1_d) \prod_{s\in\Xi} O_{s,\epsilon}(\ell_s)p_{\epsilon}(\ell,r)d\ell dr\\
& =  \int \theta(\ell) I(r=1_d) \prod_{s\in\Xi} O_{s,0}(\ell_s)p_0(\ell,r)(1+\epsilon g(\ell,r))d\ell dr\\
&\quad+  \int \theta(\ell) I(r=1_d) \sum_{s\in \Xi}\left[\prod_{\tau\in\Xi,\tau\neq s} O_{\tau,0}(\ell_\tau)\right]\Delta O_s(\ell_s)p_{0}(\ell,r)d\ell dr+ O(\epsilon^2)\\
& = \theta_{\Xi,0} + \underbrace{\epsilon \int \theta(\ell) I(r=1_d) \prod_{s\in\Xi} O_{s,0}(\ell_s)p_0(\ell,r) g(\ell,r)d\ell dr}_{\mathbf{A}}\\
&\quad+  \underbrace{\int \theta(\ell) I(r=1_d) \sum_{s\in \Xi}\left[\prod_{\tau\in\Xi,\tau\neq s} O_{\tau,0}(\ell_\tau) \right]\Delta O_s(\ell_s)p_{0}(\ell,r)d\ell dr}_{\mathbf{B}}+ O(\epsilon^2).
\end{aligned}
\label{eq::EIF::pf2}
\end{equation}
Clearly, part $\mathbf{A}$ is already in the form of an EIF so we  focus on derivations of part $\mathbf{B}$.

Part $\mathbf{B}$ has several components, and we can write it as 
\begin{align*}
\mathbf{B} & = \sum_{s\in \Xi} \mathbf{B}_s\\
\mathbf{B}_s & = \int \theta(\ell) I(r=1_d) \left[\prod_{\tau\in\Xi,\tau\neq s} O_{\tau,0}(\ell_\tau)\right]\Delta O_s(\ell_s)p_{0}(\ell,r)d\ell dr
\end{align*}
We expand the difference $\Delta O_s(\ell_s)$:
\begin{align*}
\Delta O_r(\ell_r) &= O_{r,\epsilon}(\ell_r) - O_{r,0}(\ell_r)\\
& = \frac{p_\epsilon(\ell_r,r)}{p_\epsilon(\ell_r, {\sf {PA}_r)}} - \frac{p_0(\ell_r,r)}{p_0(\ell_r, {\sf {PA}_r})}\\
& = \frac{1}{p_0(\ell_r, {\sf PA}_r)} \left(\Delta p(\ell_r,r) - O_r(\ell_r)\Delta p(\ell_r,{\sf PA_r})\right) + O(\epsilon^2),\\
\Delta p(\ell_r,r)& =  p_\epsilon(\ell_r,r) - p_0(\ell_r,r) = \epsilon \int I(w=r) p_0(\ell, w) g(\ell, w)d\ell_{\bar r} dw,\\
\Delta p(\ell_r,{\sf PA_r})& =  p_\epsilon(\ell_r,{\sf PA_r}) - p_0(\ell_r,{\sf PA}_r) = \epsilon \int I(r'\in {\sf PA}_r) p_0(\ell, w) g(\ell,w)d\ell_{\bar r} dw.
\end{align*}
Thus, we can further write $\Delta O_r(\ell_r)$ as 
\begin{equation}
\begin{aligned}
\Delta O_r(\ell_r)  &=  \frac{\epsilon}{p_0(\ell_r, {\sf PA}_r)}\int [I(w=r) -O_{r,0}(\ell_r) I(w\in{\sf PA_r})] \\
&\qquad \times p_0(\ell, w) g(\ell,w)d\ell_{\bar r} dw +O(\epsilon^2).
\end{aligned}
\label{eq::EIF::pf3}
\end{equation}

Now going back to $\mathbf{B}_s$, 
note that we can decompose 
$$
\prod_{\tau\in\Xi,\tau\neq s} O_{\tau,0}(\ell_\tau) = \prod_{\tau\in\Xi,\tau> s} O_{\tau,0}(\ell_\tau)\times \prod_{\tau\in\Xi,\tau< s} O_{\tau,0}(\ell_\tau).
$$
The first part involving terms in $\ell_{\bar s}$ while the second part is fixed when $L_s = \ell_s$. 
Thus, we can rewrite $\mathbf{B}_s$ as 
\begin{align*}
\mathbf{B}_s & = \int \theta(\ell) I(r=1_d) \left[\prod_{\tau\in\Xi,\tau\neq s} O_{\tau,0}(\ell_\tau)\right]\Delta O_s(\ell_s)p_{0}(\ell,r)d\ell dr\\
& = \underbrace{\int \theta(\ell) I(r=1_d) \left[\prod_{\tau\in\Xi,\tau> s} O_{\tau,0}(\ell_\tau)\right]p_{0}(\ell_{\bar s}, \ell_s,r)d\ell_{\bar s} dr}_{
= \E[\theta(L)I(R=1_d)\prod_{\tau\in\Xi,\tau> s} O_{\tau,0}(L_\tau) |L_s=\ell_s]\cdot p_0(\ell_s)}\\
&\qquad \times \Delta O_s(\ell_s) \left[\prod_{\tau\in\Xi,\tau< s} O_{\tau,0}(\ell_\tau)\right]d\ell_s+O(\epsilon^2).
\end{align*}
Now recall that $m_{\Xi,s}(\ell_s) = \E[\theta(L)I(R=1_d)\prod_{\tau\in\Xi,\tau> s} O_{\tau,0}(L_\tau) |L_s=\ell_s]$
from equation \eqref{eq::EIF::m},
which appears in the first term. 
This, together with equation \eqref{eq::EIF::pf3}, implies 
\begin{align*}
\mathbf{B}_s & = \epsilon\int  \frac{m_{\Xi,s}(\ell_s)}{p_0({\sf PA}_s|\ell_s)} [I(w=s) -O_{s,0}(\ell_s) I(w\in{\sf PA_s})]  \left[\prod_{\tau\in\Xi,\tau< s} O_{\tau,0}(\ell_\tau)\right]\\
&\qquad\times p_0(\ell, w) g(\ell,w)d\ell_s d\ell_{\bar s} dw +O(\epsilon^2).
\end{align*}
Comparing this expression to equation \eqref{eq::EIF::pf1},
we conclude that
$$
\EIF_{\Xi,s}(\ell_s,r) = \frac{m_{\Xi,s}(\ell_s)}{p_0({\sf PA}_s|\ell_s)} 
[I(r=s) -O_{s,0}(\ell_s) I(r\in{\sf PA_s})]\left[\prod_{\tau\in\Xi,\tau< s} O_{\tau,0}(\ell_\tau)\right]
$$
is the EIF from $\mathbf{B}_s$ and is what appears in equation \eqref{eq::EIF::pi_s}.

Recall equation \eqref{eq::EIF::pf2} that 
$\theta_{\Xi,\epsilon} =\theta_{\Xi,0} + \mathbf{A} + \sum_s \mathbf{B}_s +O(\epsilon^2)$, 
so the EIF of the entire path is the EIF from term $\mathbf{A}$ and the EIF of each node $s\in\Xi$ in $\mathbf{B}_s$, leading to 
\begin{align*}
\EIF'_{\Xi}(\ell,r) =  \theta(\ell)I(r=1_d)\left[\prod_{s\in\Xi}O_s(\ell_s)\right] + \sum_{s\in \Xi}\EIF_{\Xi,s}(\ell_s,r).
\end{align*}

Finally, the constraint $\int p_0(\ell,r)g(\ell,r) = 1$ implies that
we can add/subtract any constant to $\EIF'_{\Xi}(\ell,r)$
without affecting the fact that it satisfies equation \eqref{eq::EIF::pf1}. 
To make it the EIF, we need its mean to be $0$.
One can easily show that $\E[\EIF'_{\Xi}(L,R)] = \E\left[\theta(L)I(R=1_d)\left[\prod_{s\in\Xi}O_s(L_s)\right] \right]$,
so the EIF of $\theta_\Xi$ is
$$
\EIF_{\Xi}(\ell,r) =  \sum_{s\in \Xi}\EIF_{\Xi,s}(\ell_s,r),
$$
and the EIF of $\theta$ is $\EIF(\ell,r) = \sum_{r\neq 1_d} \sum_{\Xi\in\Pi_r} \EIF_{\Xi}(\ell,r)$,
which completes the proof.

\end{proof}

\begin{proof}[ of Proposition \ref{prop::mu_RA}]

{\bf Part 1: identification.}
Using the fact that 
$$
O_\tau(\ell_\tau) = \frac{P(R=\tau|\ell_\tau)}{P(R\in{\sf PA}_\tau|\ell_\tau)} = \frac{p(\tau, \ell_\tau)}{p({\sf PA}_\tau, \ell_\tau)},
$$
the regression function we want to identify can be decomposed as
\begin{align*}
\mu_{\Xi,s}(\ell_s) &= \frac{\E[\theta(L)I(R=1_d)\prod_{\tau>s, \tau\in\Xi} O_\tau(L_\tau)|L_s=\ell_s]}{P(R\in {\sf PA}_s|L_s=\ell_s)}\\
&= \int \theta(\ell) I(r=1_d) \frac{p(\ell,r)}{p({\sf PA}_s, \ell_s)} \left[\prod_{\tau>s, \tau\in\Xi} \frac{p(\tau, \ell_\tau)}{p({\sf PA}_\tau, \ell_\tau)}\right] d\ell_{\bar s}dr\\
&= \int \theta(\ell)\frac{p(1_d, \ell)}{p({\sf PA}_s, \ell_s)} \prod_{\tau>s, \tau\in\Xi}  \left[\frac{p(\tau, \ell_\tau)}{p({\sf PA}_\tau, \ell_\tau)}\right] d\ell_{\bar s}.
\end{align*}

We can always write 
$$
p({\sf PA}_\tau, \ell_\tau ) = \sum_{r\in{\sf PA}_\tau} p(\ell_\tau|R=r) P(R=r),
$$
which is identifiable from $\{p(\ell_r|R=r): r\in{\sf PA}_\tau\}$ and ${\sf PA}_\tau$ is  a subset of ${\sf Ans}_s$ (ancestors of $ss$) when $\tau>s$.
Thus, the above equation shows that $\mu_{\Xi,s}(\ell_s)$ can be identifiable 
from
$
\{p(\ell_r|R=r): r \in {\sf Ans}_s\}.
$

{\bf Part 2: the equality $\sum_{\Xi\in\Pi_r} \mu_{\Xi,r}(\ell_r) = m(\ell_r, r)$.}

By Theorem~\ref{thm::Codds} and Proposition \ref{lem::pathG},
we have 
$$
Q_r(L) \equiv \frac{P(R=r|L)}{P(R=1_d|L)} = \sum_{\Xi\in \Pi_r} \prod_{s\in\Xi}O_s(L_s) = O_r(L_r)\sum_{\Xi\in \Pi_r} \prod_{s\in\Xi, s>r}O_s(L_s).
$$
Recall that $\mu_{\Xi,r}(\ell_r)$ is
$$
\mu_{\Xi,r}(\ell_r) = \frac{E(\theta(L)I(R=1_d) \prod_{\tau\in\Xi, s>r}O_s(L_s)|L_r)}{P(R\in{\sf PA}_r|L_r)}.
$$
Thus, using the fact that $Q_r(L)/O_r(L_r) = \sum_{\Xi\in \Pi_r} \prod_{s\in\Xi, s>r}O_s(L_s)$, 
\begin{align*}
\sum_{\Xi\in\Pi_r} \mu_{\Xi,r}(\ell_r)& = \frac{E(\theta(L)I(R=1_d)\sum_{\Xi\in \Pi_r}  \prod_{\tau\in\Xi, s>r}O_\tau(L_s)|L_r=\ell_r)}{P(R\in{\sf PA}_r|\ell_r)}\\
& = \frac{E(\theta(L)I(R=1_d)Q_r(L)/O_r(L_r)|L_r=\ell_r)}{P(R\in{\sf PA}_r|\ell_r)}\\
& = \frac{E(\theta(L)I(R=1_d)Q_r(L)|L_r=\ell_r)}{P(R=r|\ell_r)}\\
& = \int \theta(\ell) \frac{p(r,\ell)}{p(1_d,\ell)} \frac{1}{p(r,\ell_r)}  p(1_d, \ell)d\ell_{\bar r}\\
& = \int \theta(\ell)  p(\ell_{\bar r}|\ell_r,r) d\ell_{\bar r} \\
&= \E(\theta(L)|L_r=  \ell_r, R=r) = m(\ell_r,r).
\end{align*}

{\bf Part 3: the ancestor expression.}
The key to this proof is the following observation. 
For any path $\Xi\in\Pi_r$ and a pattern $s\in \Xi, s>r$, 
the pair $(\Xi, s)$ can be uniquely expressed as 
a pattern $s\in{\sf Ans}_r$ and a path $\Xi$ containing $s$. 
Namely, 
\begin{equation}
\{(\Xi,s): \Xi\in\Pi_r, s\in \Xi, s>r\} \equiv \{(\Xi,s): s\in {\sf Ans}_r, \Xi\in\Pi_r, \Xi\ni s, s>r\}.
\label{eq::path::equiv}
\end{equation}
In the expression of right-handed-side, for a fixed $s$,
we are thinking of paths $\Xi\in\Pi_r$  containing $s$
so any path with this property can be written in the following form:
$$
\Xi = (\overbrace{1_d,\cdots}^{=\Xi_1},
\underbrace{s,\cdots, r}_{=\Xi_2}),
$$
so it can be decomposed as $\Xi = (\Xi_1,\Xi_2)$,
and the second part $\Xi_2\in \Upsilon_{s\rightarrow r}$ (recall that $\Upsilon_{s\rightarrow r}$
is the collection of all paths from $s$ to $r$).
Moreover, one can easily see that for each $s\in{\sf Ans}_r, s>r$, 
\begin{equation}
\{\Xi: \Xi\in \Pi_r, \Xi\ni s\} = \{(\Xi_1,\Xi_2): (\Xi_1,s) \in \Pi_s, \Xi_2 \in \Upsilon_{s,r}\}.
\label{eq::path::equiv2}
\end{equation}

Recall that the path-specific EIF is
\begin{align*}
\EIF_{\Xi,s}&(L_s,R) \\
&= \mu_{\Xi,s}(L_s)\left(I(R=s) - O_s(L_s) I(R\in {\sf PA}_s)\right) \prod_{w\in\Xi, w<s} O_w(L_w)
\end{align*}
and the function
\begin{align*}
\mu_{\Xi,s}(L_s) &= \frac{\E(\theta(L)I(R=1_d)\prod_{\tau>s, \tau\in\Xi}O_\tau(L_\tau)|L_s)}{P(R\in {\sf PA}_s|L_s)}\\
&= \frac{\E(\theta(L)I(R=1_d)\prod_{\tau\in\Xi_1}O_\tau(L_\tau)|L_s)}{P(R\in {\sf PA}_s|L_s)}.
\end{align*}
One may notice that the regression function 
$\mu_{\Xi,s}(L_s)$ only depends on the first part of the path $\Xi_1$
and
is independent of the second part
and 
$$
\mu_{\Xi,s}(L_s) = \mu_{(\Xi_1,s),s}(L_s)
$$
with $(\Xi_1,s)\in\Pi_s$.

Using equations \eqref{eq::path::equiv} and \eqref{eq::path::equiv2}, the EIF of pattern $r$  can be written as
\begin{equation}
\begin{aligned}
\EIF_r(L,R) &= \sum_{\Xi\in\Pi_r} \sum_{s\in \Xi, s>r} \EIF_{\Xi,s}(L_s,R) \\
& = \sum_{s\in{\sf Ans}_r,  s>r}\sum_{\Xi\in\Pi_r, \Xi\ni s} \EIF_{\Xi,s}(L_s,R)\\
& = \sum_{s\in{\sf Ans}_r, s>r} \sum_{\Xi_1: (\Xi_1,s) \in \Pi_s} \sum_{\Xi_2\in \Upsilon_{s,r}}\EIF_{(\Xi_1,\Xi_2),s}(L_s,R)\\
\end{aligned}
\label{eq::path::xi1}
\end{equation}
For a fixed $s$ and $\Xi_2$,
the summation over $\Xi_1$ in the above expression leads to 
\begin{align*}
 \sum_{\Xi_1: (\Xi_1,s) \in \Pi_s} &\EIF_{(\Xi_1,\Xi_2),s}(L_s,R) = \\
 &=   \sum_{\Xi_1: (\Xi_1,s) \in \Pi_s}\mu_{\Xi,s}(L_s)\left(I(R=s) - O_s(L_s) I(R\in {\sf PA}_s)\right) \prod_{w\in\Xi_2, w<s} O_w(L_w)\\
 & = \left[\sum_{\Xi_1: (\Xi_1,s) \in \Pi_s}\mu_{(\Xi,s),s}(L_s)\right]\left(I(R=s) - O_s(L_s) I(R\in {\sf PA}_s)\right) \prod_{w\in\Xi_2, w<s} O_w(L_w)\\
 & = \underbrace{\left[\sum_{\Xi'\in \Pi_s}\mu_{\Xi',s}(L_s)\right]}_{=m(\ell_r,r)}\left(I(R=s) - O_s(L_s) I(R\in {\sf PA}_s)\right) \prod_{w\in\Xi_2, w<s} O_w(L_w)\\
 & = m(\ell_s,s) \left(I(R=s) - O_s(L_s) I(R\in {\sf PA}_s)\right) \prod_{w\in\Xi_2, w<s} O_w(L_w),
\end{align*}
where we use the result in Part 2 in the last equality.
Putting this into equation \eqref{eq::path::xi1}, we conclude that 
\begin{align*}
\EIF_r(L,R)& = \sum_{s\in{\sf Ans}_r, s>r} \sum_{\Xi_1: (\Xi_1,s) \in \Pi_s} \sum_{\Xi_2\in \Upsilon_{s,r}}\EIF_{(\Xi_1,\Xi_2),s}(L_s,R)\\
&= \sum_{s\in{\sf Ans}_r, s>r}m(\ell_s,s) \left(I(R=s) - O_s(L_s) I(R\in {\sf PA}_s)\right)  \sum_{\Xi_2\in \Upsilon_{s,r}}\prod_{w\in\Xi_2, w<s} O_w(L_w),
\end{align*}
which is the desired result.

\end{proof}

\begin{proof}[  of Theorem \ref{thm::MR}]

For simplicity, 
we write $ O^\dagger_s(L_s) =  O_s(L_s;\eta^*)$
and $\mu_{\Xi,s}^\dagger(L_s) = \mu_{\Xi,s} (L_s;\lambda^*)$.
Also, for abbreviatioon, 
we set $\mathcal{L}^\dagger(L,R) = \mathcal{L}_{\sf semi, \Xi}(L,R; \lambda^*,\eta^*)$. 


Consider a pattern $s\in \Xi$. 
It is easy to see that when $O^\dagger_s$ is correctly specified ($O^\dagger_s = O_s$), 
\begin{equation}
\begin{aligned}
\E({\sf EIF}_{\Xi,s}(L_s, R)) &= \E\left(\mu_{\Xi,s}(L_s)(I(R=s) - O^\dagger_s(L_s)I(R\in{\sf PA}_s) ) \prod_{w<s}O^\dagger_s(L_w)\right)\\
& =\E\left(\mu_{\Xi,s}(L_s)\E[(I(R=s) - O^\dagger_s(L_s)I(R\in{\sf PA}_s) )|L_s] \prod_{w<s}O^\dagger_s(L_w)\right)\\
& =\E\left(\mu_{\Xi,s}(L_s)\underbrace{(P(R=s|L_s) - O^\dagger_s(L_s)P(R\in{\sf PA}_s|L_s) )}_{=0} \prod_{w<s}O^\dagger_s(L_w)\right)\\
&=0.
\end{aligned}
\label{eq::SO::correct}
\end{equation}
This holds regardless of other selection odds or regression function $\mu_{\Xi,s}$ being correct or not. 

Thus, when all selection odds are correctly specified, 
clearly 
$$
\E(\underbrace{\theta(L)I(R=1_d) \prod_{r\in \Xi}O^\dagger_r(L_r) + \sum_{s\in \Xi}\EIF_{\Xi,s}(L_s,R)}_{\mathcal{L}_{\Xi}(L,R)}) = \theta_\Xi.
$$
So we consider the case where some selection odds are incorrectly specified,
but the regression function is incorrectly specified.

{\bf Case 1: One selection odds is incorrectly specified.}
Suppose that we have only one an incorrect model $O^\dagger_s(L_s)$
for a selection odds of pattern $s$,
but all other selection odds are correct and the regression function $\mu_{\Xi,s}(L_s)$
is also correct. 
In this case, the quantity $\mathcal{L}^\dagger(L,R) = \mathcal{L}_{\sf semi, \Xi}(L,R; \lambda^*,\eta^*)$ becomes 
\begin{align*}
\mathcal{L}^\dagger_\Xi(L,R) &= \theta(L)I(R=1_d) O^\dagger_s(L_s)\prod_{r\in \Xi, r\neq s}O_r(L_r) \\
&\qquad+\EIF^\dagger_{\Xi,s}(L_s,R)+  \sum_{w\in \Xi, w\neq s}\EIF_{\Xi,w}(L_w,R),
\end{align*}
where 
\begin{equation}
\EIF^\dagger_{\Xi,s}(L_s,R) = \mu_{\Xi,s}(L_s) (I(R=s) - O^\dagger_s(L_s) I(R\in{\sf PA}_s)) \prod_{w\in\Xi, w<s}O_w(L_w).
\label{eq::dagger}
\end{equation}

By equation \eqref{eq::SO::correct},
the last part has mean $0$ (correctly specified EIFs)
so we obtain
\begin{align*}
\E[\mathcal{L}^\dagger_\Xi(L,R)] = \E\left(\underbrace{\theta(L)I(R=1_d) O^\dagger_s(L_s)\prod_{r\neq s}O_r(L_r)}_{(A)} + 
\EIF^\dagger_{\Xi,s}(L_s,R)\right).
\end{align*}
So we just need to prove that the above quantity is $\theta_\Xi$. 

For  part (A), a direct computation shows that 
\begin{equation}
\begin{aligned}
\E((A))& = \E\left(\theta(L)I(R=1_d) O^\dagger_s(L_s)\prod_{r\neq s}O_r(L_r)\right)\\
& = \E\left(\E\left(\theta(L)I(R=1_d) \prod_{\tau> s}O_\tau(L_\tau)|L_s\right) O^\dagger_s(L_s)\prod_{w< s}O_w(L_w)\right)\\
& = \E\left(\mu_{\Xi,s}(L_s) P(R\in{\sf PA}_s|L_s)O^\dagger_s(L_s)\prod_{w< s}O_w(L_w)\right)\\
\end{aligned}
\label{eq::MR::A}
\end{equation}

For the EIF part, its expectation is
\begin{equation*}
\begin{aligned}
\E(\EIF^\dagger_{\Xi,s}(L_s,R))& = \E\left(\mu_{\Xi,s}(L_s) (I(R=s) - O^\dagger_s(L_s) I(R\in{\sf PA}_s)) \prod_{w<s}O_w(L_w)\right)\\
&=\E\left(\mu_{\Xi,s}(L_s) \E(I(R=s) - O^\dagger_s(L_s) I(R\in{\sf PA}_s)|L_s) \prod_{w<s}O_w(L_w)\right)\\
& =\E\left(\mu_{\Xi,s}(L_s) (P(R=s|L_s) - O^\dagger_s(L_s) P(R\in{\sf PA}_s|L_s)) \prod_{w<s}O_w(L_w)\right)\\ 
\end{aligned}
\end{equation*}
The second component of $\E(\EIF^\dagger_{\Xi,s}(L_s,R))$
is identical to $\E((A))$ in equation \eqref{eq::MR::A},
so the summation leads to 
\begin{align*}
\E[\mathcal{L}^\dagger_\Xi(L,R)] &=  \E\left(\mu_{\Xi,s}(L_s) P(R=s|L_s)  \prod_{w<s}O_w(L_w)\right)\\
& = \E\left(\frac{\E(\theta(L)I(R=1_d)\prod_{\tau>s}O_\tau(L_\tau)|L_s)}{P(R\in{\sf PA}_s|L_s)} P(R=s|L_s)  \prod_{w<s}O_w(L_w)\right)\\
& = \E\left(\E\left(\theta(L)I(R=1_d)\prod_{\tau>s}O_\tau(L_\tau)|L_s\right) O_s(L_s)  \prod_{w<s}O_w(L_w)\right)\\
& = \E\left(\theta(L)I(R=1_d)\left[\prod_{\tau>s}O_\tau(L_\tau)\right] O_s(L_s)  \prod_{w<s}O_w(L_w)\right)\\
& = \theta_\Xi.
\end{align*}
Thus, when the selection odds of pattern $s$ is incorrectly specified, 
as long as the regression function $\mu_{\Xi,s}$ is correctly specified,
we still recover the true parameter.

{\bf Case 2: Two or more selection odds are incorrectly specified.}
We prove the case when there are two patterns $s_1>s_2\in \Xi$ that are both mis-specified.
The case of more selection odds being mis-specified can be proved in a similar way. 
In this case, we use two incorrect selection odds $O^\dagger_{s_1}$ and $O^\dagger_{s_2}$,
but the corresponding regression function $\mu_{\Xi,s_1}$ and $\mu_{\Xi,s_2}$ are correct.
In this case, $\mathcal{L}_\Xi(L,R)$ becomes
\begin{align*}
\mathcal{L}^\dagger_\Xi(L,R) &= \underbrace{\theta(L)I(R=1_d) O^\dagger_{s_1}(L_{s_1})O^\dagger_{s_2}(L_{s_2})\prod_{r\in \Xi, r\neq s_1,s_2}O_r(L_r)}_{(C)} \\
&\qquad+\EIF^\dagger_{\Xi,s_1}(L_{s_1},R)+\EIF^\dagger_{\Xi,s_2}(L_{s_2},R)+  \sum_{w\in \Xi, w\neq s_1,s_2}\EIF_{\Xi,w}(L_w,R),
\end{align*}

Similar to the case of one selection odds being mis-specified, 
the component $\sum_{w\in \Xi, w\neq s_1,s_2}\EIF_{\Xi,w}(L_w,R)$ has mean $0$
so we can ignore it.
So we only need to focus on the mean of term (C) and $\EIF^\dagger_{\Xi,s_1}(L_{s_1},R), \EIF^\dagger_{\Xi,s_2}(L_{s_2},R)$.
Because of $s_1>s_2$, $\EIF^\dagger_{\Xi,s_2}(L_{s_2},R)$ does not involve $O^\dagger_{s_1}$
so it is the same as equation \eqref{eq::dagger}. 
However, term $\EIF^\dagger_{\Xi,s_1}(L_{s_1},R)$ involves $O^\dagger_{s_2}$,
and it is 
\begin{align*}
\EIF^\dagger_{\Xi,s_1}(L_{s_1},R) =& \mu_{\Xi,s_1}(L_{s_1}) (I(R=s_1) - O^\dagger_{s_1}(L_{s_1}) I(R\in{\sf PA}_{s_1})) \\
&\qquad\times O^\dagger_{s_2}(L_{s_2})\prod_{ w<s_1, w\neq s_2}O_w(L_w).
\end{align*}

Using a similar derivation as $\E((A))$ in equation \eqref{eq::MR::A}, we have
\begin{align*}
\E((C))& = \E\Bigg(\mu_{\Xi,s_1}(L_{s_1}) (P(R=s_1|L_{s_1}) - O^\dagger_{s_1}(L_{s_1}) P(R\in{\sf PA}_{s_1}|L_{s_1}))\\
&\times O^\dagger_{s_2}(L_{s_2}) \prod_{w<s_1,w\neq s_2}O_w(L_w)\Bigg)
\end{align*}
Now we compute the expectation of $\EIF^\dagger_{\Xi,s_1}(L_{s_1},R)$.
Using the law of total expectation that we condition on $L_{s_1}$ first, one can show that 
\begin{align*}
\E(\EIF^\dagger_{\Xi,s_1}(L_{s_1},R))& = \E\Bigg( \mu_{\Xi,s_1}(L_{s_1}) (I(R=s_1) - O^\dagger_{s_1}(L_{s_1}) I(R\in{\sf PA}_{s_1})) \\
&\qquad\times O^\dagger_{s_2}(L_{s_2})\prod_{ w<s_1, w\neq s_2}O_w(L_w)\Bigg)\\
& = \E\Bigg( \mu_{\Xi,s_1}(L_{s_1}) \E\left(I(R=s_1) - O^\dagger_{s_1}(L_{s_1}) I(R\in{\sf PA}_{s_1})|L_{s_1}\right) \\
&\qquad\times O^\dagger_{s_2}(L_{s_2})\prod_{ w<s_1, w\neq s_2}O_w(L_w)\Bigg)\\
& = \E\Bigg( \mu_{\Xi,s_1}(L_{s_1}) (P(R=s_1|L_{s_1}) - O^\dagger_{s_1}(L_{s_1}) P(R\in{\sf PA}_{s_1}|L_{s_1})) \\
&\qquad\times O^\dagger_{s_2}(L_{s_2})\prod_{ w<s_1, w\neq s_2}O_w(L_w)\Bigg)
\end{align*}
Again, the second term in the above equality (the one involving $O^\dagger_{s_1}(L_{s_1}) P(R\in{\sf PA}_{s_1}|L_{s_1})$)
is identical to  $\E((C))$. 
So we conclude that 
\begin{align*}
\E((C) + \EIF^\dagger_{\Xi,s_1}(L_{s_1},R)) &= \E\Bigg( \mu_{\Xi,s_1}(L_{s_1}) P(R=s_1|L_{s_1})  \\
&\qquad\times O^\dagger_{s_2}(L_{s_2})\prod_{ w<s_1, w\neq s_2}O_w(L_w)\Bigg),
\end{align*}
which is the same result as $\E((A))$ in equation \eqref{eq::MR::A}
with replacing $s$ by $s_2$. 
Thus, the problem  reduces to Case 1: only one selection odds is mis-specified.
By applying the analysis of Case 1,
we conclude that 
$$
\E(\mathcal{L}^\dagger_\Xi(L,R)) = \E((C) + \EIF^\dagger_{\Xi,s_1}(L_{s_1},R)+ \EIF^\dagger_{\Xi,s_2}(L_{s_2},R)) = \theta_\Xi.
$$

One can adapt this procedure to any number of selection odds being mis-specified.
As long as the corresponding regression function $\mu_{\Xi,s}$ is correctly specified,
we recover the same pathwise effect $\theta_\Xi$. 
Thus, we conclude that 
for all $s\in \Xi$, as long as $O_s(\cdot; \eta^*_s)  = O_s(\cdot)$ or $\mu_{\Xi,s}(\cdot;\lambda^*) = \mu_{\Xi,s}(\cdot)$,
$$
\E(\mathcal{L}_{\sf semi, \Xi}(L,R; \lambda^*,\eta^*)) = \theta_\Xi,
$$
which completes the proof.


\end{proof}


\begin{proof}[ of Theorem~\ref{thm::perturb}]
The proof consists of several ``if and only if" statements:
\begin{align*}
&\frac{P(R=r|\ell)}{P(R\in {\sf PA}_r|\ell)} = \frac{P(R=r|\ell_r)}{P(R\in {\sf PA}_r|\ell_r)} \cdot g(\ell_{\bar r})\\
\Leftrightarrow&\frac{P(R=r,\ell)}{P(R\in {\sf PA}_r,\ell)} = \frac{P(R=r,\ell_r)}{P(R\in {\sf PA}_r,\ell_r)} \cdot g(\ell_{\bar r})\\
\Leftrightarrow&\frac{P(R=r,\ell)}{P(R=r,\ell_r)} = \frac{P(R\in {\sf PA}_r,\ell)}{P(R\in {\sf PA}_r,\ell_r)} \cdot g(\ell_{\bar r})\\
\end{align*}
Now using the fact that 
\begin{align*}
p(\ell_{\bar r}|\ell_r,R=r) &= \frac{P(\ell|R=r)}{P(\ell_r|R=r)} = \frac{P(R=r,\ell)}{P(R=r,\ell_r)} \\
p(\ell_{\bar r}|\ell_r,R\in{\sf PA}_r) &= \frac{P(\ell|R\in{\sf PA}_r)}{P(\ell_r|R\in{\sf PA}_r)} = \frac{P(R\in{\sf PA}_r,\ell)}{P(R\in{\sf PA}_r,\ell_r)},
\end{align*}
the above if and only if statement becomes
\begin{align*}
&\frac{P(R=r|\ell)}{P(R\in {\sf PA}_r|\ell)} = \frac{P(R=r|\ell_r)}{P(R\in {\sf PA}_r|\ell_r)} \cdot g(\ell_{\bar r})\\
\Leftrightarrow&\frac{P(R=r,\ell)}{P(R=r,\ell_r)} = \frac{P(R\in {\sf PA}_r,\ell)}{P(R\in {\sf PA}_r,\ell_r)} \cdot g(\ell_{\bar r})\\
\Leftrightarrow&p(\ell_{\bar r}|\ell_r,R=r) = p(\ell_{\bar r}|\ell_r,R\in{\sf PA}_r) \cdot g(\ell_{\bar r}),
\end{align*}
which completes the proof.

\end{proof}

\begin{proof}[ of Proposition~\ref{lem::Lnumber}]
In non-monotone case, there are ${d \choose k}$ distinct missing patterns with $k$ missing variables. 
For a pattern $r$ with $d-|r|$ variables missing, 
there are totally $2^{d-|r|} -1$ patterns in the set $\mathcal{H}_r =\{s: s>r\}$
that can be a parent of $r$. 
Any non-empty subsets of $\mathcal{H}_r$ can be a parent of $r$
so there is a total of $2^{2^{d-|r|}-1}-1$ possible parent sets of pattern $r$.
%
%

To specify an identifying restriction,
we need to specify every parent pattern, and a parent set must be a subset of $\mathcal{H}_r$.
Because parent sets of different patterns can be specified independently,
so the total number is
\begin{align*}
M_d &= (2^{2^d-1}-1)^{d\choose 0}\times (2^{2^{d-1}-1}-1)^{d\choose 1}\times\cdots \underbrace{(2^{2^{d-k}-1}-1)^{d\choose k}}_\text{$k$ variable missing}\cdots \times (2^{2^0-1}-1)^{d\choose d-1}\\
&= \prod_{k=0}^{d-1} (2^{2^{d-k}-1}-1)^{d\choose k}.
\end{align*}

\end{proof}

\begin{proof}[ of Proposition~\ref{lem::Gpm}]

{\bf Case of $\Delta_{+1}$.}
We first prove
\begin{align*}
\{G&\oplus e_{s\rightarrow r}: s>r, s\notin {\sf PA}_r\}\\
&\subset \{G': |G'-G| = 1, \mbox{condition (G1-2) holds for $G'$}, G\subset G'\}
\end{align*}
and then prove the other way around.
Apparently, we only add one arrow so $|G'-G| = 1$ holds. 
Similarly, since we are adding edges, $G\subset G'$. 
Also, it is straightforward that the new graph also satisfies (G1-2) so this inclusion holds.

We now turn to showing that 
\begin{align*}
\{G': |G'-G| = 1, &\mbox{ conditions (G1-2) hold for $G'$}, G\subset G'\}\\
&\subset \{G\oplus e_{s\rightarrow r}: s>r, s\notin {\sf PA}_r\}.
\end{align*}
$G\subset G'$ and $ |G'-G| = 1$ implies that $G'$ has one additional edge compared to $G$.
Let ${s\rightarrow r}$ be the newly added arrow.
Since ${s\rightarrow r}$ has to satisfies (G2), it must satisfy condition $s>r$ and $s\notin{\sf PA}_r$.
Thus, the inclusion condition holds so the two sets are the same.

{\bf Case of $\Delta_{-1}$.}
We first prove 
\begin{align*}
\{G&\ominus e_{s\rightarrow r}: s\in {\sf PA}_r, |{\sf PA}_r|>1\}\\
&\subset  \{G': |G'-G| = 1, \mbox{condition (G1-2) holds for $G'$}, G'\subset G\}
\end{align*}
and then derive the other direction later.
Apparently, 
we are removing one edge so conditions $|G'-G|=1$ and $G'\subset G$ hold
automatically.
Also, since we are deleting an edge, 
the partial ordering condition (G2)
holds for $G'$. 
All we need is to show that the resulting graph still has the unique source $1_d$ (condition (G1)).
Because we are deleting an arrow ${s\rightarrow r}$ with $s\in {\sf PA}_r, |{\sf PA}_r|>1$,
so the node $r$ still has parents. 
Thus, this will not create any new source and the condition (G1) holds, which proves this inclusion direction.

Now we prove 
\begin{align*}
\{G': |G'-G| = 1, &\mbox{condition (G1-2) holds for $G'$}, G'\subset G\}\\
&\subset \{G\ominus e_{s\rightarrow r}: s\in {\sf PA}_r, |{\sf PA}_r|>1\}.
\end{align*}
Conditions $|G'-G|=1$ and $G'\subset G$ implies that we are deleting one edge so  
$G'=G\ominus e_{s\rightarrow r}$ for some $s\in {\sf PA}_r$. 
Thus, we only need to show that we can only delete this edge if $|{\sf PA}_r|>1$.
Note that condition (G2) holds for the new graph $G'$ so they do not provide any additional constraint.
The only constraint we have is condition (G1)--we need to make sure
that the deletion will not create a new source.

We will prove that to satisfy (G1), the arrow ${s\rightarrow r}$ being deleted must satisfies $|{\sf PA}_r|>1$.
We prove this by contradiction. Suppose that we delete an arrow ${s\rightarrow r}$ with $|{\sf PA}_r|=1$.
Then the node $r$ in the graph $G'$ has no parents, so it becomes a source, which contradicts to (G1).
Thus, condition (G1) implies that the arrow ${s\rightarrow r}$ being deleted must satisfies $|{\sf PA}_r|>1$,
and this has proved the inclusion.
As a result, the two sets are the same, and we have complete the proof.



\end{proof}

Before proving Theorem \ref{thm::GPG} and~\ref{thm::equiv} , we first introduce a useful lemma. 

\begin{lemma}[Generation number]
Let $G$ be a DAG with a unique source $s^*$. 
For a node $r$ of $G$,
we define $\Pi_r $ to be the collection of all paths from $s^*$ to $r$. 
For a path $\Xi \in\Pi_r$, let $\|\Xi\|$ be the number of elements in the path. 
We define the generation number
$$
g(r) =\max\{\|\Xi\|: \Xi\in \Pi_r\} - 1
$$
and set $g(s^*)= 0$.
Then
\begin{itemize}
\item[(P1)] for all $s\in{\sf PA}_r$, $g(s)\leq g(r)-1$.
\item[(P2)] there exists $s\in{\sf PA}_r$ such that $g(s) = g(r)-1$.
\end{itemize}
\label{lem::generation}
\end{lemma}
Both statements can be easily proved using the proof by contradiction so we omit the proof.

\begin{proof}[ of Theorem~\ref{thm::GPG}]

{\bf Equivalence between selection odds model and pattern mixture model. }
This proof is essentially the same as the proof of Theorem~\ref{thm::PMM2}.
We can directly apply the proof here because  the proof of Theorem~\ref{thm::PMM2} does not 
use assumption (G2). 

{\bf Identifying property.}
This proof follows from the same idea as the proof of Theorem~\ref{thm::PMM1} (PMMs); we use the proof by induction. 
The only difference is that the order of induction is no longer based on the number 
of observed variables but instead, the order is determined by $g(r)$, the generation number defined in Lemma~\ref{lem::generation}. 

The induction goes from $g(r)=0,1,2,\cdots$. 
In the case of $g(r)=0$, there is only one node with this property: $r=1_d$.
Under assumption (G1), 
the pattern $1_d$ is the unique source. 
So $1_d$ can be treated as the starting point of the induction. 
Clearly, $p(\ell|R=1_d)$ is identifiable.
For the case of $g(r)=1$, since node $1_d$ is identifiable, clearly $r$ is identifiable. 

Now we assume that a pattern $r$ has generation number $g(r)=k$
and for any other patterns with $g(s) <k$, the conclusion holds (i.e., they are identifiable).
Note that property (P2)  in Lemma~\ref{lem::generation} implies that if there is a pattern with $g(r)=k$, there must  be a pattern $q$ with $g(q) = k-1$
so there is no gap in the sequence.
The pattern mixture model formulation shows that 
$$
p(\ell_{\bar r}|\ell_r,R=r) = p(\ell_{\bar r}|\ell_r,R\in {\sf PA}_r).
$$
Because $p(\ell| R=s)$ is identifiable for all $s \in {\sf PA}_r$ due to the induction assumption, 
$p(\ell_{\bar r}|\ell_r,R\in {\sf PA}_r)$ is identifiable, which implies that $p(\ell_{\bar r}|\ell_r,R=r)$
is identifiable.
By induction, the result follows.




\end{proof}

%

\begin{proof}[ of Theorem~\ref{thm::equiv}]

Let $s^*$ be the pattern satisfying the two conditions in the theorem.
We prove this theorem by 
showing that 
\begin{equation}
p(\ell_{\bar r}|\ell_r,r) = p(\ell_{\bar r}|\ell_r,s^*)
\label{eq::pf::equiv1}
\end{equation}
so that we can replace all the arrows to $r$ by a single arrow from $s^*$ to $r$. 

Our strategy is the proof by induction. 
We first construct a subgraph $G^*\subset G$ formed by all the nodes where
they appear in at least one of the path from $s^*$ to $r$.
We only keep the arrows if the arrows are used in a path from $s^*$ to $r$. 

It is easy to see that the resulting graph $G^*$ is still a DAG and has a unique source $s^*$. 
Lemma~\ref{lem::generation} shows that we can label every pattern $s$ in $G^*$ with an integer 
given by the generation number $g(s)$ and $g(s^*) = 0$.

The generation number is the quantity that we use for induction.
We will show that 
\begin{equation}
p(\ell_{\bar r}|\ell_r,s) = p(\ell_{\bar r}|\ell_r,s^*)
\label{eq::pf::equiv2}
\end{equation}
for all $s$ in the graph $G^*$, which implies the desired result (equation \eqref{eq::pf::equiv1}).

{\bf Case $g(s)=0$ and $g(s)=1$.}
The case $g(s) = 0$ occurs only if $s=s^*$ so this is trivially true.
For $g(s)=1$, they only have one parent: $s^*$, so by the pattern mixture model factorization
and the fact that $s<r$ due to the uninformative condition in Theorem~\ref{thm::equiv},
we immediately have equation \eqref{eq::pf::equiv2}. 
Thus, both cases have  been proved. 

{\bf Case $g(s)\leq k$ implies case $g(s)=k+1$.}
Now we assume that for any $s$ such that $g(s)\leq k$, equation \eqref{eq::pf::equiv2} is true.
And our goal is to show that this implies $g(s)=k+1$ is also true. 

Let $s$ be a pattern such that $g(s)= k+1$.
By the uninformative condition, $s<r$ so
using the rule of conditional probability, 
we can decompose 
\begin{align*}
p(\ell_{\bar r}|\ell_r, s) = \frac{p(\ell,s)}{p(\ell_r,s)} = \frac{p(\ell_{\bar s}|\ell_s, s)p(\ell_s,s)}{p(\ell_{r-s}|\ell_s,s)p(\ell_s,s)} = \frac{p(\ell_{\bar s}|\ell_s, s)}{p(\ell_{r-s}|\ell_s,s)},
\end{align*}
where both $p(\ell_{\bar s}|\ell_s, s)$ and $p(\ell_{r-s}|\ell_s,s)$ are from the extrapolation density of pattern $s$.
Thus, by the pattern mixture model factorization in equation \eqref{eq::PMM}, it equals to 
\begin{equation}
p(\ell_{\bar r}|\ell_r, s)=\frac{p(\ell_{\bar s}|\ell_s, s)}{p(\ell_{r-s}|\ell_s,s)}=\frac{p(\ell_{\bar s}|\ell_s, {\sf PA}_s)}{p(\ell_{r-s}|\ell_s,{\sf PA}_s)} = p(\ell_{\bar r}|\ell_r,{\sf PA}_s).
\label{eq::pf::equiv3}
\end{equation}
We can further decompose $p(\ell_{\bar r}|\ell_r,{\sf PA}_s)$ as 
\begin{align*}
p(\ell_{\bar r}|\ell_r,{\sf PA}_s) &= \frac{p(\ell_{\bar r},R\in{\sf PA}_s|\ell_r)}{P(R\in{\sf PA}_s|\ell_r)} \\
&= \frac{\sum_{w\in {\sf PA}_s}p(\ell_{\bar r},R=w|\ell_r)}{P(R\in{\sf PA}_s|\ell_r)}\\
&= \sum_{w\in {\sf PA}_s}p(\ell_{\bar r}|\ell_r,w)\frac{P(R=w|\ell_r)}{P(R\in{\sf PA}_s|\ell_r)}\\
&= \sum_{w\in {\sf PA}_s}p(\ell_{\bar r}|\ell_r,w)P(R=w|R\in{\sf PA}_s,\ell_r).
\end{align*}
For each $w\in {\sf PA}_s$, the generation number $g(w)\leq g(s)-1 = k$ due to Lemma~\ref{lem::generation} so by the assumption in the induction,
$p(\ell_{\bar r}|\ell_r,w) = p(\ell_{\bar r}|\ell_r,s^*)$. 
Thus, the above equality becomes
\begin{align*}
p(\ell_{\bar r}|\ell_r,{\sf PA}_s) &= \sum_{w\in {\sf PA}_s}p(\ell_{\bar r}|\ell_r,w)P(R=w|R\in{\sf PA}_s,\ell_r)\\
&= \sum_{w\in {\sf PA}_s}p(\ell_{\bar r}|\ell_r,s^*)P(R=w|R\in{\sf PA}_s,\ell_r)\\
&=p(\ell_{\bar r}|\ell_r,s^*)\underbrace{\sum_{w\in {\sf PA}_s}P(R=w|R\in{\sf PA}_s,\ell_r)}_{=1}.
\end{align*}
Putting this into equation \eqref{eq::pf::equiv3},
we conclude 
$$
p(\ell_{\bar r}|\ell_r, s) = p(\ell_{\bar r}|\ell_r,{\sf PA}_s) = p(\ell_{\bar r}|\ell_r,s^*),
$$
which proves the case.

Therefore, we have shown that equation \eqref{eq::pf::equiv2} holds for all $s$ in the graph $G^*$, which
proves equation \eqref{eq::pf::equiv1} and completes the proof of this theorem.



\end{proof}

Before we prove Proposition \ref{prop::aug},
we first introduce a lemma to characterize the augmentation space $\mathcal{G}$
under MNAR. 
\begin{lemma}
The space $\mathcal{G}$ can be equivalently expressed as 
\begin{align*}
\mathcal{G} = \Bigg\{\frac{\theta(L)I(R=1_d)}{\pi(L)}+ \sum_{r\neq 1_d}\Big(I(R=r) - {\frac{P(R=r|L)}{P(R=1_d|L)}} I(R=1_d)&\Big) h(L_r,r): \\
&\E(h^2(L_r,r))<\infty \Bigg\}
\end{align*}
\label{lem::augG}
\end{lemma}

Note that this lemma appears in Theorem 10.7 of \cite{tsiatis2007semiparametric}, page 29 of \citealt{malinsky2019semiparametric}),
and was implicitly used in the proof of Theorem 4 of \cite{tchetgen2018discrete} 

\begin{proof}
It is easy to see that the above augmentation term 
$$
g(L_R,R)=  \sum_{r\neq 1_d}\Big(I(R=r) - {\frac{P(R=r|L)}{P(R=1_d|L)}} I(R=1_d)\Big) h(L_r,r)
$$
satisfies $\E(g(L_R,R))= 0$
so it is a subset of $\mathcal{G}$.
Now we show that for any augmentation $w(L_R,R)$ with $\E(w(L_R,R)|R) = 0$, 
it can be written in terms of the above expression.

The equality $\E(w(L_R,R)|R) = 0$ implies that 
$$
\sum_r w(L_r,r) P(R=r|L) = 0.
$$
Thus,
$$
w(L,1_d) = w(L_{1_d},1_d) = -\sum_{r\neq 1_d}\frac{P(R=r|L)}{P(R=1_d|L)} w(L_r,r).
$$
Note that any function $w(L_R,R)$ can be written as 
\begin{align*}
w(L_R,R) &= \sum_{r} w(L_r,r) I(R=r)\\
&= w(L,1_d)I(R=1_d) + \sum_{r\neq 1_d} w(L_r,r) I(R=r)\\
& = -\sum_{r\neq 1_d}\frac{P(R=r|L)}{P(R=1_d|L)} w(L_r,r)I(R=1_d)+w(L_r,r) I(R=r)\\
& = \sum_{r\neq 1_d} \left(I(R=r) - \frac{P(R=r|L)}{P(R=1_d|L)} I(R=1_d)\right) w(L_r,r).
\end{align*}
By identifying $w(L_r,r) = h(L_r,r)$, we have shown that any augmentation can be written
as the expression in $g(L_R,R)$, which completes the proof.

\end{proof}

\begin{proof}[ of Proposition \ref{prop::aug}]

It is easy to see that $\mathcal{F}\subset\mathcal{G}$.
So we focus on showing that $\mathcal{G}\subset\mathcal{F}$.

Consider any augmentation $g(L_R,R)$ in $\mathcal{G}$.
By Lemma~\ref{lem::augG}, we can rewrite the augmentation term as 
\begin{equation}
g(L_R,R) = \sum_{r\neq 1_d}\left(I(R=r) - \underbrace{\frac{P(R=r|L)}{P(R=1_d|L)}}_{Q_r(L)} I(R=1_d)\right) h(L_r,r)
\label{eq::aug::g}
\end{equation}
for some functions $h(L_r,r)$ such that $\E(h^2(L_r,r))<\infty$.


Let ${\sf CH}_r $ be the children node of pattern $r$. 
The augmentation in $\mathcal{F}$ can be written as
\begin{equation}
\begin{aligned}
\sum_{r\neq 1_d}&\left(I(R=r) - O_r(L_r) I(R\in{\sf PA}_r)\right) \Psi_r(L_r)\\
&=\sum_{r\neq 1_d}\left(I(R=r) - O_r(L_r) \sum_{s\in {\sf PA}_r}I(R=s)\right) \Psi_r(L_r)\\
&=\sum_{r\neq 1_d}I(R=r)\left( \Psi_r(L_r) - \sum_{s\in {\sf CH}_r}O_s(L_s) \Psi_s(L_s)\right)\\
& \quad+ I(R=1_d)\sum_{s\in{\sf CH}_{1_d}} O_s(L_s)\Psi_s(L_s)\\
&=\sum_{r\neq 1_d}I(R=r)\Psi_r'(L_r)
+ I(R=1_d)\sum_{s\in{\sf CH}_{1_d}} O_s(L_s)\Psi_s(L_s),
\end{aligned}
\label{eq::aug::f1}
\end{equation}
where $\Psi_r'(L_r) =  \Psi_r(L_r) - \sum_{s\in {\sf CH}_r}O_s(L_s) \Psi_s(L_s)$.

Suppose that 
\begin{equation*}
h(L_r,r) = \Psi_r'(L_r) =  \Psi_r(L_r) - \sum_{s\in {\sf CH}_r}O_s(L_s) \Psi_s(L_s)
\end{equation*}
for each $r$.
It is easy to see that the above equation defines a one-to-one mapping between $\{h(L_r,r):r\in\mathcal{R}\}$
and $\{\Psi_r(L_r):r\in\mathcal{R}\}$ by the Gauss elimination. 
Namely, given $\{h(L_r,r):r\in\mathcal{R}\}$, we can find a unique set of functions $\{\Psi_r(L_r):r\in\mathcal{R}\}$
such that the above equality holds.
With this insight, a sufficient condition that
equation \eqref{eq::aug::g} can be expressed using
equation \eqref{eq::aug::f1} is
\begin{equation}
\begin{aligned}
\sum_{s\in{\sf CH}_{1_d}} &O_s(L_s)\Psi_s(L_s) \\
&= -\sum_{r\neq 1_d} Q_r(L) h(L_r,r)\\
& = -\sum_{r\neq 1_d} Q_r(L) \left( \Psi_r(L_r) - \sum_{s\in {\sf CH}_r}O_s(L_s) \Psi_s(L_s)\right).
\end{aligned}
\label{eq::aug::f2}
\end{equation}
Thus, we focus on deriving equation \eqref{eq::aug::f2}.

Using the fact that (exchanging parents and children)
$$
\sum_{r\neq 1_d}\sum_{s\in {\sf CH}_r} Q_r(L)O_s(L_s) \Psi_s(L_s) 
= \sum_{r}\sum_{s\in {\sf PA}_r, s\neq 1_d} Q_s(L)O_r(L_r) \Psi_r(L_r),
$$
we have
\begin{align*}
\sum_{r\neq 1_d}\sum_{s\in {\sf CH}_r} Q_r(L)O_s(L_s) \Psi_s(L_s)
&= \sum_{r}\sum_{s\in {\sf PA}_r} Q_s(L)O_r(L_r) \Psi_r(L_r)\\
&\quad - I(1_d\in{\sf PA}_r)\underbrace{Q_{1_d}(L)}_{=1} O_r(L_r) \Psi_r(L_r).
\end{align*}
By Theorem~\ref{thm::Codds}, $\sum_{s\in {\sf PA}_r} Q_s(L)O_r(L_r)  = Q_r(L)$
so the above implies 
$$
\sum_{r\neq 1_d}\sum_{s\in {\sf CH}_r} Q_r(L)O_s(L_s) \Psi_s(L_s) 
= \sum_{r\neq 1_d}Q_r(L)\Psi_r(L_r)
- I(1_d\in{\sf PA}_r)O_r(L_r) \Psi_r(L_r).
$$
Putting this into the last quantity in equation \eqref{eq::aug::f2}, we obtain
\begin{align*}
-\sum_{r\neq 1_d} Q_r(L) &\left( \Psi_r(L_r) - \sum_{s\in {\sf CH}_r}O_s(L_s) \Psi_s(L_s)\right)\\
& = \sum_{r\neq 1_d}I(1_d\in{\sf PA}_r)O_r(L_r) \Psi_r(L_r)\\
& = \sum_{r\in {\sf CH}_{1_d}}O_r(L_r) \Psi_r(L_r),
\end{align*}
which is the first quantity in equation \eqref{eq::aug::f2}.
So equation \eqref{eq::aug::f2} holds, implying that $\mathcal{G}\subset\mathcal{F}$
with the choice
\begin{equation*}
h(L_r,r)=  \Psi_r(L_r) - \sum_{s\in {\sf CH}_r}O_s(L_s) \Psi_s(L_s)
\end{equation*}
for each $r$.

\end{proof}

\bibliographystyle{abbrvnat}
\bibliography{PG}

\begin{thebibliography}{49}
\providecommand{\natexlab}[1]{#1}
\providecommand{\url}[1]{\texttt{#1}}
\expandafter\ifx\csname urlstyle\endcsname\relax
  \providecommand{\doi}[1]{doi: #1}\else
  \providecommand{\doi}{doi: \begingroup \urlstyle{rm}\Url}\fi

\bibitem[Ali et~al.(2009)Ali, Richardson, and Spirtes]{ali2009markov}
R.~A. Ali, T.~S. Richardson, and P.~Spirtes.
\newblock Markov equivalence for ancestral graphs.
\newblock \emph{The Annals of Statistics}, 37\penalty0 (5B):\penalty0
  2808--2837, 2009.

\bibitem[Andersson et~al.(1997)Andersson, Madigan, and
  Perlman]{andersson1997characterization}
S.~A. Andersson, D.~Madigan, and M.~D. Perlman.
\newblock A characterization of markov equivalence classes for acyclic
  digraphs.
\newblock \emph{The Annals of Statistics}, 25\penalty0 (2):\penalty0 505--541,
  1997.

\bibitem[Bhattacharya et~al.(2020)Bhattacharya, Malinsky, and
  Shpitser]{bhattacharya2020causal}
R.~Bhattacharya, D.~Malinsky, and I.~Shpitser.
\newblock Causal inference under interference and network uncertainty.
\newblock In \emph{Uncertainty in Artificial Intelligence}, pages 1028--1038.
  PMLR, 2020.

\bibitem[Chen(2020)]{chen2020supp}
Y.-C. Chen.
\newblock Supplementary materials: Pattern graphs: a graphical approach to
  nonmonotone missing data.
\newblock doi: COMPLETED BY THE TYPESETTER, 2020.

\bibitem[Chen and Sadinle(2019)]{chen2019nonparametric}
Y.-C. Chen and M.~Sadinle.
\newblock Nonparametric pattern-mixture models for inference with missing data.
\newblock \emph{arXiv preprint arXiv:1904.11085}, 2019.

\bibitem[Daniels and Hogan(2008)]{DanielsHogan08}
M.~J. Daniels and J.~W. Hogan.
\newblock \emph{{Missing Data in Longitudinal Studies: Strategies for Bayesian
  Modeling and Sensitivity Analysis}}.
\newblock Chapman and Hall/CRC, Boca Raton, 2008.

\bibitem[Diggle et~al.(2002)Diggle, Heagerty, Liang, Heagerty, and
  Zeger]{diggle2002analysis}
P.~J. Diggle, P.~Heagerty, K.-Y. Liang, P.~J. Heagerty, and S.~Zeger.
\newblock \emph{Analysis of longitudinal data}.
\newblock Oxford University Press, 2002.

\bibitem[Efron(1979)]{efron1979}
B.~Efron.
\newblock Bootstrap methods: Another look at the jackknife.
\newblock \emph{Ann. Statist.}, 7\penalty0 (1):\penalty0 1--26, 01 1979.
\newblock \doi{10.1214/aos/1176344552}.
\newblock URL \url{https://doi.org/10.1214/aos/1176344552}.

\bibitem[Efron and Tibshirani(1994)]{efron1994introduction}
B.~Efron and R.~J. Tibshirani.
\newblock \emph{An introduction to the bootstrap}.
\newblock CRC press, 1994.

\bibitem[Friedman et~al.(2001)Friedman, Hastie, and
  Tibshirani]{friedman2001elements}
J.~Friedman, T.~Hastie, and R.~Tibshirani.
\newblock \emph{The elements of statistical learning}, volume~1.
\newblock Springer series in statistics New York, 2001.

\bibitem[Gill et~al.(1997)Gill, {van der Laan}, and Robins]{Gilletal97}
R.~D. Gill, M.~J. {van der Laan}, and J.~M. Robins.
\newblock Coarsening at random: Characterizations, conjectures,
  counter-examples.
\newblock In \emph{Proceedings of the First Seattle Symposium in Biostatistics:
  Survival Analysis}, pages 255--294, 1997.

\bibitem[Gillispie and Perlman(2002)]{gillispie2002size}
S.~B. Gillispie and M.~D. Perlman.
\newblock The size distribution for markov equivalence classes of acyclic
  digraph models.
\newblock \emph{Artificial Intelligence}, 141\penalty0 (1-2):\penalty0
  137--155, 2002.

\bibitem[Hall(2013)]{hall2013bootstrap}
P.~Hall.
\newblock \emph{The bootstrap and Edgeworth expansion}.
\newblock Springer Science \& Business Media, 2013.

\bibitem[Hoeting et~al.(1999)Hoeting, Madigan, Raftery, and
  Volinsky]{hoeting1999bayesian}
J.~A. Hoeting, D.~Madigan, A.~E. Raftery, and C.~T. Volinsky.
\newblock Bayesian model averaging: a tutorial.
\newblock \emph{Statistical science}, pages 382--401, 1999.

\bibitem[Hoonhout and Ridder(2018)]{HoonhoutRidder18}
P.~Hoonhout and G.~Ridder.
\newblock {Nonignorable Attrition in Multi-Period Panels With Refreshment
  Samples}.
\newblock \emph{J. Bus. Econ. Statist.}, Forthcoming, 2018.

\bibitem[Horowitz and Manski(2000)]{horowitz2000nonparametric}
J.~L. Horowitz and C.~F. Manski.
\newblock Nonparametric analysis of randomized experiments with missing
  covariate and outcome data.
\newblock \emph{Journal of the American statistical Association}, 95\penalty0
  (449):\penalty0 77--84, 2000.

\bibitem[Kim and Yu(2011)]{kim2011semiparametric}
J.~K. Kim and C.~L. Yu.
\newblock A semiparametric estimation of mean functionals with nonignorable
  missing data.
\newblock \emph{Journal of the American Statistical Association}, 106\penalty0
  (493):\penalty0 157--165, 2011.

\bibitem[Linero(2017)]{Linero17}
A.~R. Linero.
\newblock {Bayesian nonparametric analysis of longitudinal studies in the
  presence of informative missingness}.
\newblock \emph{Biometrika}, 104\penalty0 (2):\penalty0 327--341, 2017.

\bibitem[Little(1993{\natexlab{a}})]{little1993pattern}
R.~J. Little.
\newblock Pattern-mixture models for multivariate incomplete data.
\newblock \emph{Journal of the American Statistical Association}, 88\penalty0
  (421):\penalty0 125--134, 1993{\natexlab{a}}.

\bibitem[Little et~al.(2012)Little, D'Agostino, Cohen, Dickersin, Emerson,
  Farrar, Frangakis, Hogan, Molenberghs, Murphy, et~al.]{little2012prevention}
R.~J. Little, R.~D'Agostino, M.~L. Cohen, K.~Dickersin, S.~S. Emerson, J.~T.
  Farrar, C.~Frangakis, J.~W. Hogan, G.~Molenberghs, S.~A. Murphy, et~al.
\newblock The prevention and treatment of missing data in clinical trials.
\newblock \emph{New England Journal of Medicine}, 367\penalty0 (14):\penalty0
  1355--1360, 2012.

\bibitem[Little(1993{\natexlab{b}})]{Little93}
R.~J.~A. Little.
\newblock Pattern-mixture models for multivariate incomplete data.
\newblock \emph{J. Am. Statist. Assoc.}, 88\penalty0 (421):\penalty0 125--134,
  1993{\natexlab{b}}.

\bibitem[Little and Rubin(2002)]{LittleRubin02}
R.~J.~A. Little and D.~B. Rubin.
\newblock \emph{{Statistical Analysis with Missing Data}}.
\newblock Wiley, Hoboken, New Jersey, 2nd edition, 2002.

\bibitem[Liu(2008)]{liu2008monte}
J.~S. Liu.
\newblock \emph{Monte Carlo strategies in scientific computing}.
\newblock Springer Science \& Business Media, 2008.

\bibitem[Malinsky et~al.(2019)Malinsky, Shpitser, and
  Tchetgen]{malinsky2019semiparametric}
D.~Malinsky, I.~Shpitser, and E.~J.~T. Tchetgen.
\newblock Semiparametric inference for non-monotone missing-not-at-random data:
  the no self-censoring model.
\newblock \emph{arXiv preprint arXiv:1909.01848}, 2019.

\bibitem[Manski(1990)]{manski1990nonparametric}
C.~F. Manski.
\newblock Nonparametric bounds on treatment effects.
\newblock \emph{The American Economic Review}, 80\penalty0 (2):\penalty0
  319--323, 1990.

\bibitem[Mohan and Pearl(2014)]{mohan2014graphical}
K.~Mohan and J.~Pearl.
\newblock Graphical models for recovering probabilistic and causal queries from
  missing data.
\newblock In \emph{Advances in Neural Information Processing Systems}, pages
  1520--1528, 2014.

\bibitem[Mohan and Pearl(2018)]{mohan2018graphical}
K.~Mohan and J.~Pearl.
\newblock Graphical models for processing missing data.
\newblock \emph{arXiv preprint arXiv:1801.03583}, 2018.

\bibitem[Mohan et~al.(2013)Mohan, Pearl, and Tian]{mohan2013graphical}
K.~Mohan, J.~Pearl, and J.~Tian.
\newblock Graphical models for inference with missing data.
\newblock In \emph{Advances in neural information processing systems}, pages
  1277--1285, 2013.

\bibitem[Molenberghs et~al.(1998)Molenberghs, Michiels, Kenward, and
  Diggle]{ACMV}
G.~Molenberghs, B.~Michiels, M.~G. Kenward, and P.~J. Diggle.
\newblock {Monotone missing data and pattern-mixture models}.
\newblock \emph{Statistica Neerlandica}, 52\penalty0 (2):\penalty0 153--161,
  1998.

\bibitem[Molenberghs et~al.(2014)Molenberghs, Fitzmaurice, Kenward, Tsiatis,
  and Verbeke]{molenberghs2014handbook}
G.~Molenberghs, G.~Fitzmaurice, M.~G. Kenward, A.~Tsiatis, and G.~Verbeke.
\newblock \emph{Handbook of missing data methodology}.
\newblock Chapman and Hall/CRC, 2014.

\bibitem[Nabi et~al.(2020)Nabi, Bhattacharya, and Shpitser]{nabi2020full}
R.~Nabi, R.~Bhattacharya, and I.~Shpitser.
\newblock Full law identification in graphical models of missing data:
  Completeness results.
\newblock \emph{arXiv preprint arXiv:2004.04872}, 2020.

\bibitem[Robins(1997)]{Robins97}
J.~M. Robins.
\newblock Non-response models for the analysis of non-monotone non-ignorable
  missing data.
\newblock \emph{Statist. Med.}, 16\penalty0 (1):\penalty0 21--37, 1997.

\bibitem[Robins and Gill(1997)]{robins1997non}
J.~M. Robins and R.~D. Gill.
\newblock Non-response models for the analysis of non-monotone ignorable
  missing data.
\newblock \emph{Statistics in medicine}, 16\penalty0 (1):\penalty0 39--56,
  1997.

\bibitem[Robins et~al.(2000)Robins, Rotnitzky, and
  Scharfstein]{robins2000sensitivity}
J.~M. Robins, A.~Rotnitzky, and D.~O. Scharfstein.
\newblock Sensitivity analysis for selection bias and unmeasured confounding in
  missing data and causal inference models.
\newblock In \emph{Statistical models in epidemiology, the environment, and
  clinical trials}, pages 1--94. Springer, 2000.

\bibitem[Rubin(2004)]{rubin2004multiple}
D.~B. Rubin.
\newblock \emph{Multiple imputation for nonresponse in surveys}, volume~81.
\newblock John Wiley \& Sons, 2004.

\bibitem[Sadinle and Reiter(2017)]{sadinle2017itemwise}
M.~Sadinle and J.~P. Reiter.
\newblock Itemwise conditionally independent nonresponse modelling for
  incomplete multivariate data.
\newblock \emph{Biometrika}, 104\penalty0 (1):\penalty0 207--220, 2017.

\bibitem[Seaman and Vansteelandt(2018)]{seaman2018introduction}
S.~R. Seaman and S.~Vansteelandt.
\newblock Introduction to double robust methods for incomplete data.
\newblock \emph{Statistical science: a review journal of the Institute of
  Mathematical Statistics}, 33\penalty0 (2):\penalty0 184, 2018.

\bibitem[Shao and Wang(2016)]{shao2016semiparametric}
J.~Shao and L.~Wang.
\newblock Semiparametric inverse propensity weighting for nonignorable missing
  data.
\newblock \emph{Biometrika}, 103\penalty0 (1):\penalty0 175--187, 2016.

\bibitem[Shpitser(2016)]{shpitser2016consistent}
I.~Shpitser.
\newblock Consistent estimation of functions of data missing non-monotonically
  and not at random.
\newblock In \emph{Advances in Neural Information Processing Systems}, pages
  3144--3152, 2016.

\bibitem[Shpitser et~al.(2015)Shpitser, Mohan, and Pearl]{shpitser2015missing}
I.~Shpitser, K.~Mohan, and J.~Pearl.
\newblock Missing data as a causal and probabilistic problem.
\newblock Technical report, CALIFORNIA UNIV LOS ANGELES DEPT OF COMPUTER
  SCIENCE, 2015.

\bibitem[Sun and Tchetgen~Tchetgen(2018)]{sun2018inverse}
B.~Sun and E.~J. Tchetgen~Tchetgen.
\newblock On inverse probability weighting for nonmonotone missing at random
  data.
\newblock \emph{Journal of the American Statistical Association}, 113\penalty0
  (521):\penalty0 369--379, 2018.

\bibitem[Tchetgen et~al.(2018)Tchetgen, Wang, and Sun]{tchetgen2018discrete}
E.~J.~T. Tchetgen, L.~Wang, and B.~Sun.
\newblock Discrete choice models for nonmonotone nonignorable missing data:
  Identification and inference.
\newblock \emph{Statistica Sinica}, 28\penalty0 (4):\penalty0 2069--2088, 2018.

\bibitem[Thijs et~al.(2002)Thijs, Molenberghs, Michiels, Verbeke, and
  Curran]{Thijs}
H.~Thijs, G.~Molenberghs, B.~Michiels, G.~Verbeke, and Curran.
\newblock {Strategies to fit pattern-mixture models}.
\newblock \emph{Biostatistics}, 3\penalty0 (2):\penalty0 245--265, 2002.

\bibitem[Tian(2015)]{tian2015missing}
J.~Tian.
\newblock Missing at random in graphical models.
\newblock In \emph{Artificial Intelligence and Statistics}, pages 977--985,
  2015.

\bibitem[Tsiatis(2007)]{tsiatis2007semiparametric}
A.~Tsiatis.
\newblock \emph{Semiparametric theory and missing data}.
\newblock Springer Science \& Business Media, 2007.

\bibitem[van~der Vaart(1998)]{van1998asymptotic}
A.~W. van~der Vaart.
\newblock \emph{Asymptotic statistics}, volume~3.
\newblock Cambridge university press, 1998.

\bibitem[van~der Vaart and Wellner(1996)]{van1996weak}
A.~W. van~der Vaart and J.~A. Wellner.
\newblock \emph{Weak convergence}.
\newblock Springer, 1996.

\bibitem[Vansteelandt et~al.(2006)Vansteelandt, Goetghebeur, Kenward, and
  Molenberghs]{Vansteelandtetal06}
S.~Vansteelandt, E.~Goetghebeur, M.~G. Kenward, and G.~Molenberghs.
\newblock Ignorance and uncertainty regions as inferential tools in a
  sensitivity analysis.
\newblock \emph{Statist. Sinica}, 16\penalty0 (3):\penalty0 953--979, 2006.

\bibitem[Zhao et~al.(2017)Zhao, Tang, Qu, and Jiang]{zhao2017semiparametric}
P.~Zhao, N.~Tang, A.~Qu, and D.~Jiang.
\newblock Semiparametric estimating equations inference with nonignorable
  missing data.
\newblock \emph{Statistica Sinica}, pages 89--113, 2017.

\end{thebibliography}

\end{document}